\documentclass{aa} 
\usepackage{natbib}
\usepackage{subcaption}
\usepackage{graphicx,textcomp}
\usepackage{placeins}

\newcommand{\mum}{$\mu$m}

\usepackage{txfonts}
\usepackage[colorlinks]{hyperref}
\hypersetup{citecolor = blue}
\hypersetup{linkcolor = blue}

\newcommand{\Mjup}{M$_{\rm{Jup}}$}
\newcommand{\AU}{au} 

\usepackage{threeparttable}
\begin{document} 

   \title{Spatially resolving polycyclic aromatic hydrocarbons in Herbig Ae disks with VISIR-NEAR at the VLT}

   \author{G. Yoffe
          \inst{1}$^,$\inst{16}
          \and R. van Boekel\inst{1}
          \and A. Li \inst{2}
          \and L.B.F.M. Waters\inst{3}$^,$\inst{4}
          \and K. Maaskant
          \and R. Siebenmorgen\inst{5}
          \and M. van den Ancker\inst{5}
          \and D.J.M. Petit dit de la Roche\inst{6}
          \and B. Lopez\inst{7}
          \and A. Matter\inst{7}
          \and J. Varga\inst{8}
          \and M.R. Hogerheijde\inst{9,10}
         \and G. Weigelt\inst{11}
         \and R.D. Oudmaijer\inst{12}
          \and E. Pantin\inst{13}
          \and M. R. Meyer\inst{14}
        \and J.-C. Augereau\inst{15}
          \and Th. Henning\inst{1}
          }

   \institute{$^1$ Max Planck Institute for Astronomy, K\"onigstuhl 17, 69117 Heidelberg, Germany\\
   $^2$ Department of Physics and Astronomy, University of Missouri, Columbia, MO 65211, USA\\
   $^3$ Institute for Mathematics, Astrophysics and Particle Physics, Radboud University, P.O. Box 9010, MC 62, NL-6500 GL Nijmegen, The Netherlands\\
   ${^4}$ SRON, Sorbonnelaan 2, 3484CA Utrecht, The Netherlands\\
   ${^5}$ European Southern Observatory, Karl-Schwarzschild-Str. 2, 85748 Garching, Germany \\
   ${^6}$ Observatoire de l’Universit\'e de Gen\`eve, Chemin Pegasi 51, 1290 Versoix, Switzerland \\
   ${^7}$ Laboratoire Lagrange, UMR7293,
        CNRS,  Observatoire de la Côte d’Azur,
            Nice, France \\
    ${^8}$ Konkoly Observatory, Research Centre for Astronomy and Earth Sciences, Eötvös Loránd Research Network (ELKH), Konkoly-Thege Miklós út 15-17, 1121 Budapest, Hungary \\
    ${^9}$ Leiden Observatory, Leiden University, PO Box 9513, 2300 RA
            Leiden, The Netherlands \\
    ${^{10}}$ Anton Pannekoek Institute for Astronomy, University of Amsterdam, Science Park 904, 1098 XH, Amsterdam, the Netherlands \\
    ${^{11}}$ Max-Planck-Institut für Radioastronomie, Auf dem Hügel 69, 53121 Bonn, Germany \\
    ${^{12}}$ School of Physics and Astronomy, University of Leeds, Leeds LS2 9JT, UK \\
  ${^{13}}$ Centre d’Etudes de Saclay, Gif-sur-Yvette, France \\
   ${^{14}}$ Department of Astronomy, University of Michigan, Ann Arbor, MI 48109, USA \\
  ${^{15}}$ Université Grenoble Alpes, CNRS, IPAG, 38000 Grenoble, France \\
  ${^{16}}$ Now at: Department of Statistics and Data Science, The Hebrew University of Jerusalem, Mount Scopus, 91905, Jerusalem, Israel\\
              \email{gideon.yoffe@mail.huji.ac.il}
             }


 
   \abstract
  {The emission from polycyclic aromatic hydrocarbons (PAHs) arises from the uppermost layers of protoplanetary disks, higher than the optical/near-infrared scattered light and similar to the emission from the highly thick $^{12}$CO millimeter lines. The PAH intensity profiles trace the gas distribution and can constrain the penetration depth of UV radiation.}
  {We aim to constrain the spatial intensity profiles of the four strongest PAH emission features in the telluric N-band spectral region. Thereby, we seek to constrain the dependence of PAH properties on the (radial) location in the disk, such as charge state, the interrelation with the presence and dynamics of small silicate grains, and the correlation of PAH emission with gas or dust.}
   {We used the long-slit spectroscopy mode of the VISIR-NEAR experiment to perform diffraction-limited observations of eight nearby Herbig~Ae protoplanetary disks. We extracted spectra for various locations along the slit with a spectral resolution of $R\approx300$ and performed a compositional fit at each spatial location using spectral templates of silicates and the four PAH bands. This yields the intensity versus location profiles of each species.}
  {We obtained spatially resolved intensity profiles of the PAH emission features in the N band for five objects (AB~Aurigae, HD~97048, HD~100546, HD~163296, and HD~169142). We observe two kinds of PAH emission geometry in our sample: \textbf{centrally peaked} (HD~97048) and \textbf{ring-like} (AB~Aurigae, HD~100546, HD~163296, and potentially HD~169142). Comparing the spatial PAH emission profiles with near-infrared scattered light images, we find a strong correlation in the disk substructure but a difference in radial intensity decay rate. The PAH emission shows a less steep decline with distance from the star. Finally, we find a correlation between the presence of (sub)micron-sized silicate grains and the depletion of PAH emission within the inner regions of the disks.}
   {In this work we find the following: (\textbf{1}) PAH emission traces the extent of Herbig Ae disks to a considerable radial distance. (\textbf{2}) The correlation between the presence of silicate emission within the inner regions of disks and the depletion of PAH emission can result from dust-mixing and PAH coagulation mechanisms and competition over UV photons. (\textbf{3}) For all objects in our sample, PAHs undergo stochastic heating across the entire spatial extent of the disk and are not saturated. (\textbf{4}) The difference in radial intensity decay rates between the PAHs and scattered-light profiles may be attributed to shadowing and dust-settling effects, which impact the scattering grains more so than the PAHs.}


   \maketitle
%
\section{Introduction}\label{Introduction}

\begin{figure*}
   \centering
   \includegraphics[width=\hsize]{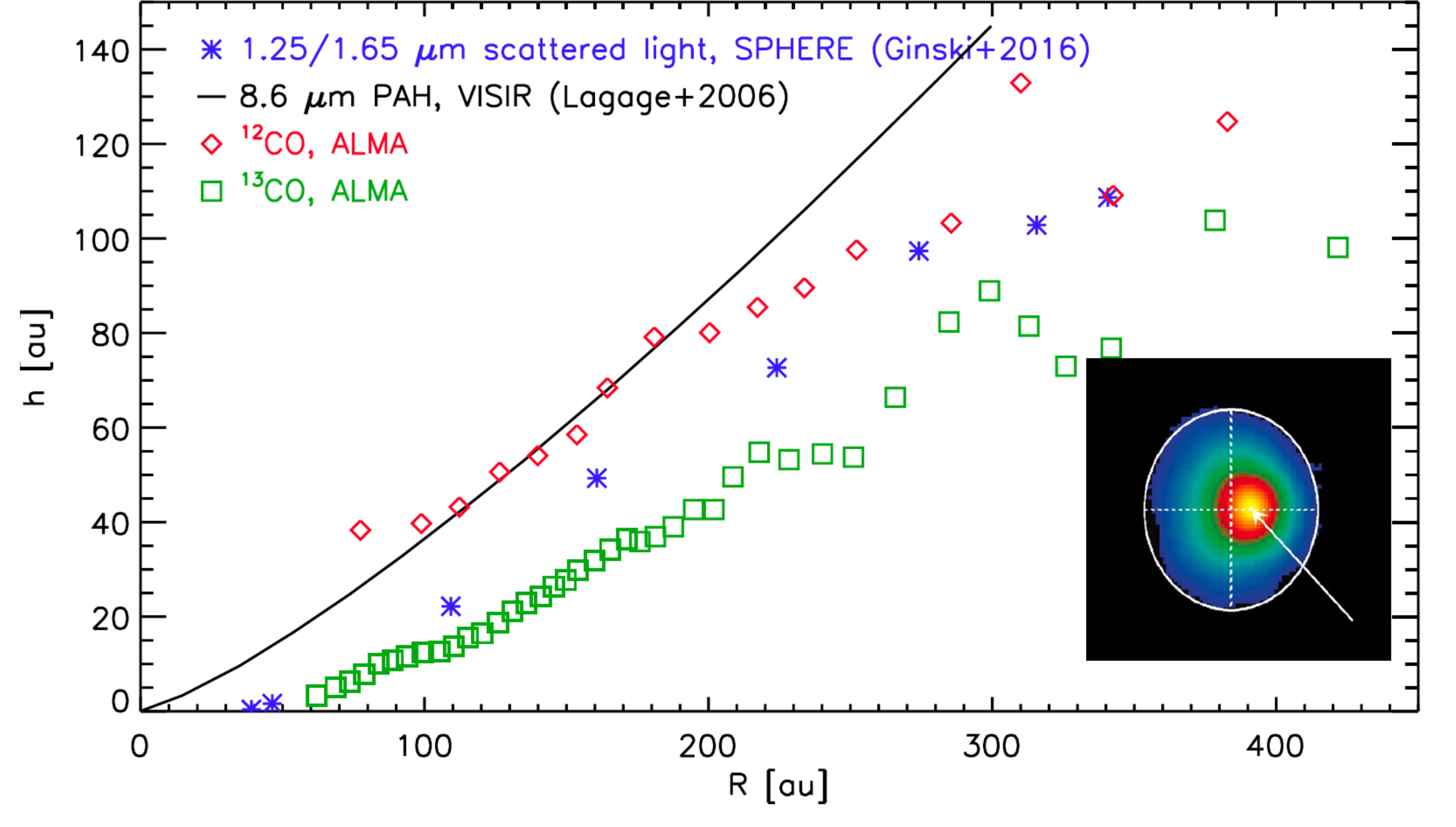}
      \caption{Variations in the effective emission height of different tracers for a 3D disk demonstrated for HD~97048: (\textbf{1}) scattered-light as observed with SPHERE \citep{Ginski2016}, (\textbf{2}) PAH emission as observed with VISIR \citep{Lagage2006}, and (\textbf{3}) $^{13/12}$CO as observed with ALMA \citep{Walsh2016}. The inset on the bottom right is Fig. 4 from \citet{Lagage2006}, demonstrating the 2D spatial distribution of the 8.6~$\mu$m PAH emission intensity of HD~97048.}
         \label{Fig_scaleHeight_tracers}
   \end{figure*}

For most of observational astronomy history, protoplanetary disks had been spatially unresolved, but this has changed drastically -- particularly in the last decade. Nowadays, different facilities provide a wealth of detail by tracing different (vertical) layers of these objects at various wavelengths \citep[e.g.,][]{Menu2015, Ginski2016, Andrews2018, Varga2018, Bertrang2018, perraut2019gravity, Keppler2020, bae2021observational}. The contribution of different but complementary observations is demonstrated in Fig. \ref{Fig_scaleHeight_tracers}. From these observations, we have learned that protoplanetary disks are very rich in structure, such as radial gaps, rings, and spiral arms \citep[i.e., substructure; e.g.,][]{2022arXiv220309991B, Boccaletti2020}. The origination and diversity of substructure in protoplanetary disks may be attributableto different mechanisms, such as the gravitational interaction of the disk's material -- especially forming or embedded planets \citep[e.g.,][]{Mordasini14}, for which compelling evidence exists in some cases \citep{keppler2018discovery}. 


Thermal infrared emission of protoplanetary disks in most cases remains spatially unresolved to 10 m class telescopes, even when they operate at the diffraction limit, because the equilibrium temperature of the disk material is too low to yield significant radiation at, for example, 10~$\mu$m on scales that can be resolved \citep[e.g.,][]{Menu2015}. For spatially resolving the thermal emission, long-baseline interferometry is therefore typically required (cf. GRAVITY \citep{Gillessen2006}, MATISSE \citep{lopez2022matisse}). The situation is different for the emission from polycyclic aromatic hydrocarbon (PAH) particles -- large molecules consisting of multiple benzene rings with hydrogen atoms around the edges. These particles are so small that the absorption of a single stellar ultraviolet (UV) photon provides sufficient heating for 10~$\mu$m emission to occur far into the disk, allowing them to be observed with single-dish telescopes \citep{Geers2007, Visser2007}. \color{black} Such nanoparticles experience stochastic heating followed by a quick de-excitation through infrared emission and then remain cold (and invisible) until they absorb another UV photon\citep{natta1995pah, li2003modeling, siebenmorgen2010destruction} \color{black}. Thus, for PAH emission, the expected emission profile does not follow the (nonlinear) profile resulting from a thermal equilibrium calculation but instead scales directly with the number of UV photons absorbed in a given disk location. It thus approximately follows an inverse square law with respect to the distance to the central star, modulated by the spatial distribution of the PAH molecules as well as the ``flaring profile'' of the disk (effective surface height vs. radius) and substructure, such as radial gaps \citep[e.g.,][]{Doucet2007}.

Contrary to the macroscopic-sized (millimeter-scale) grains that dominate the emission seen in Atacama Large Milimeter Array (ALMA) continuum images \citep[e.g.,][]{Andrews2018}, PAH molecules are strongly coupled to the disk gas \citep{Seok2017}. They are presumably present throughout the disk but are excited and therefore observable only in regions directly exposed to stellar UV radiation and thus probe the highest disk layers \citep{Siebenmorgen2012}. 

Additionally, it is well established that the PAH emission bands show clear variation in wavelength, shape, and both absolute and relative intensities -- not only among different sources but also across spatially resolved sources \citep[e.g.,][]{Lagage2006}. In the N band, it has been shown that the 7.9 and 8.6~$\mu$m bands are tightly correlated with each other, but not with the 11.3~$\mu$m band \citep{Maaskant2014}. Theoretical and laboratory work indicates that the PAH charge state predominantly determines these correlations. While neutral PAHs emit strongly in the 11.3~$\mu$m band, ionized PAHs emit strongly in the 7.9 and 8.6~$\mu$m bands \citep[e.g.,][]{Hudgins2005, Draine2020}. Since the ionization is sensitive to UV fluxes \citep{Maaskant2014}, spatial tracking of the distinctly ionized and neutral PAH features can provide insight into the optical thickness profile of the disk in which they are embedded.

 In early observations performed with 4--8 meter-range telescopes operating in seeing-limited mode, some of the largest disks could be spatially resolved (e.g., Fig. \ref{Fig_scaleHeight_tracers}). Long-slit spectroscopy aided in this, allowing robust relative measurements between the continuum and the PAH features. However, without adaptive optics (AO) correction, the spatial resolution was significantly lower than the diffraction limit. The most prominent example is the disk of HD~97048 \citep{Doucet2007}, which was already highly resolved with Thermal Infrared MultiMode Instrument (TIMMI2) on the European Southern Observatory (ESO) 3.6 m telescope \citep{vanBoekel2004}. Furthermore, the seeing-limited VLT Imager and Spectrometer for mid-Infrared (VISIR) was able to secure a high-quality image of it, which showed the disk surface and allowed the vertical disk height versus radius to be directly (geometrically) measured for the first time \citep{Lagage2006}. Follow-up studies \citep[e.g.,][]{Pinte2019} showed that, in some objects, the PAH emission is dominated by the dense disk regions. In contrast, in other cases, the emission arises primarily in much less dense "gap" regions also prevalent in ALMA/Spectro-Polarimetric High-contrast Exoplanet Research (SPHERE) observations. In these observations, the emission is dominated by ionized PAH molecules, whereas the PAHs in the dense regions are mostly neutral \citep{Maaskant2014}.

We report on observations performed on eight Herbig Ae disks with the Very Large Telescope (VLT) spectrometer and imager for the VISIR mid-infrared New Earths in the Alpha Cen Region (NEAR) as a long-slit spectrometer with a diffraction-limited point spread function (PSF). We obtained spatially resolved N-band spectra (8--13$\mu$m) with a spectral resolution of $\approx$300 and spatial sampling of $\approx$1.1$\lambda/D$, which corresponds to several tens of astronomical units for our targets. Using a nonlinear Markov chain Monte Carlo (MCMC) optimization technique, we fit the reduced spectra with a parametric model consisting of a decoupled blackbody component, an additional blackbody component that is coupled to opacity curves of ten silicate species, and four prominent PAH emission features in the N band, roughly centered at 8.6, 11.3, and 12.6~$\mu$m, following, for example, \cite{vanBoekel2005}. We also include the "red" tail of the 7.9~$\mu$m emission feature as a power law.  Finally, we produce radial profiles of the PAH emission, with which we explore the physical properties of the disks in our sample, such as the projected geometry of their photospheres, the distribution of PAHs of different charge states across the disk, and the dependence thereof on the intrinsic structure. In Sect. \ref{Target_Sample} we present our target sample; in Sect. \ref{Observations} we provide details about the performed observations and subsequent data processing; and in Sect. \ref{Modeling} we present our modeling methodology. In Sects. \ref{Results} and \ref{Discussion}, we present our results and discuss them, respectively, and in Sect. \ref{Summary} we summarize and conclude.
 
\section{Target sample} \label{Target_Sample}

\color{black}  We selected several nearby disks based on the presence of PAH features and indications of extended emission in previous VISIR observations \citep{Maaskant2014}. These are: AB Aurigae, HD~95881, HD~97048, HD~100453, HD~100546, HD~163296, and HD~169142. In some sources, the continuum emission may be extended \citep[e.g.,][]{vanBoekel2004,Lagage2006}. Furthermore, sufficient optical brightness for the NEAR wavefront sensor was required. Of the resulting sample, eight targets were observed during the VISIR-NEAR science verification time runs; all these targets are known to have disk gaps or ring-like disk structures in ALMA continuum data, scattered light observations, or infrared interferometry.

\begin{figure*}
\centering
\includegraphics[width=0.7\textwidth,trim=80 170 110 220,clip]{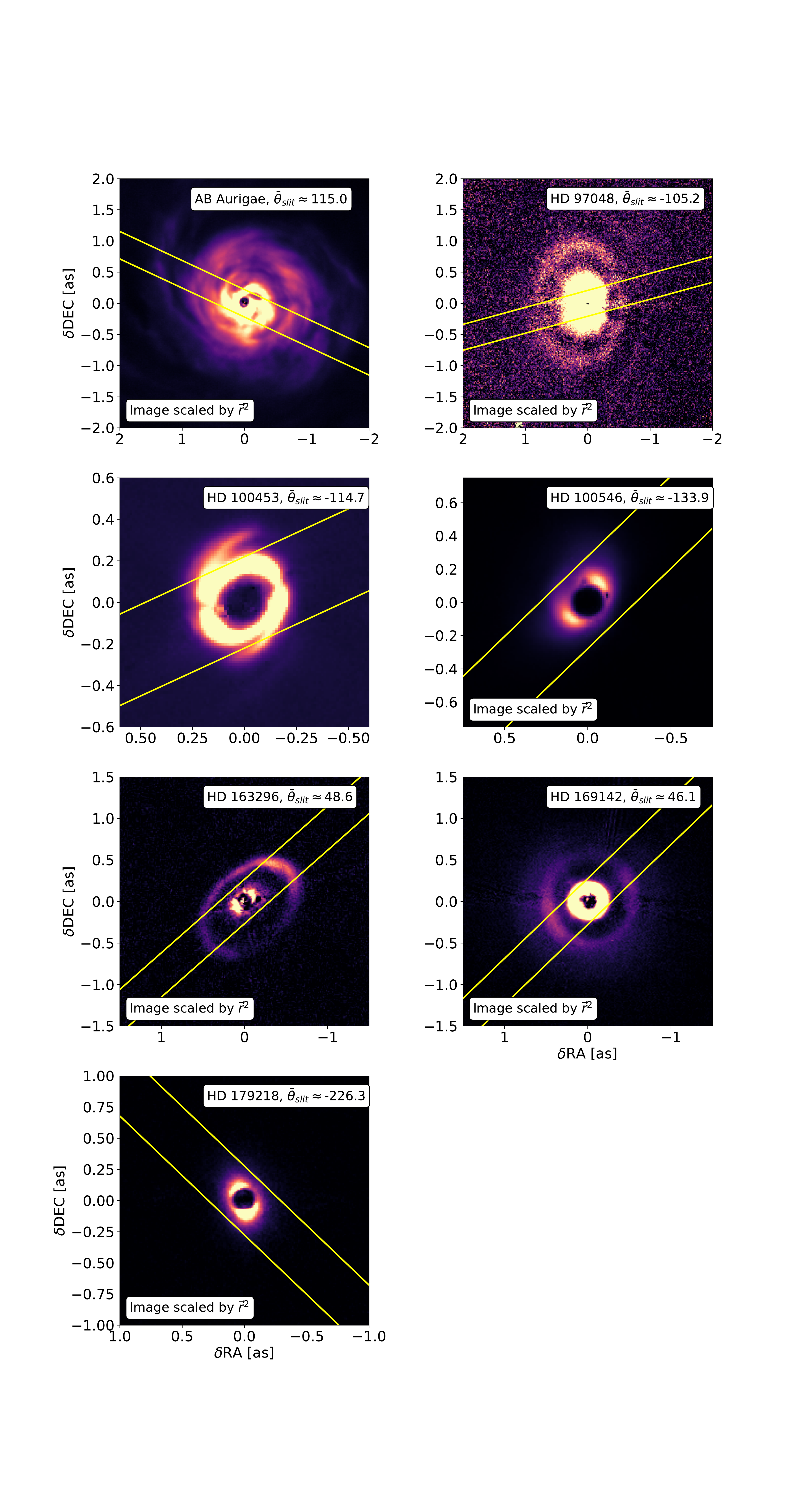}
  \caption{Mean slit orientation (yellow \color{black} rectangle) overlaid on scattered-light images observed with SPHERE for all targets in our sample except HD~95881.}
     \label{Fig_SPHERE_allSources}
\end{figure*}

Table \ref{Tab_Targets} lists our target sample and some fundamental properties. In Appendix~\ref{app_targets}, we provide background information on the individual targets. In Fig. \ref{Fig_SPHERE_allSources}, we present reduced SPHERE polarized scattered-light images for all targets in our sample, except HD~95881, which was not observed in this mode. In addition, we over-plot the VISIR long-slit width and mean orientation during observation.
In Appendix \ref{app_targets}, we individually introduce each target in our sample.

\begin{table*}
\begin{threeparttable}
\small
    \centering
    \begin{tabular}{c  c  c  c  c  c  c  c } 
    \hline\hline
    Object & $d$ [pc] & T$_\star$ [K] & Log(L$_\star$) [L$_\odot$] & M$_\star$ [M$_\odot$] & t$_\star$ [Myr] &  FWHM$_{\rm{PSF,8.6}}$ [au] & FWHM$_{{PSF,11.3}}$ [au] \\ 
    \hline
    AB Aurigae & 155$^{+0.90}_{-0.89}$ & 9000$\pm$125 & 1.66$\pm$0.01 & 2.36$^{+0.04}_{-0.05}$ & 4.14$^{+0.20}_{-0.13}$ & 32.87 & 42.34  \\
    
    HD~95881 & 1098.03$^{+24.34}_{-23.30}$ & 9750$\pm$125 & 2.97$\pm$0.02 & 6.40$^{+0.07}_{-0.19}$ & 0.23$^{+0.04}_{-0.03}$ & 233.65 & 300.41  \\
    
    HD~97048 & 184.11$^{+0.85}_{-0.84}$ & 11000$\pm$125  & 2.22$\pm$0.06 & 2.80$^{+0.03}_{-0.03}$ & 3.90$^{+0.07}_{-0.60}$ & 43.68 & 55.31  \\
    
    HD~100453 & 103.61$^{+0.24}_{-0.24}$ & 7250$\pm$125 & 0.79$\pm$0.01 & 1.60$^{+0.05}_{-0.04}$ & 19.28$^{+0.70}_{-0.68}$ & 26.21 & 33.74  \\
    
    HD~100546 & 107.97$^{+0.44}_{-0.44}$ & 9250$\pm$125 & 1.34$\pm$0.01 & 2.10$^{+0.05}_{-0.03}$ & 7.67$^{+0.36}_{-0.67}$ & 22.89 & 28.35  \\
    
    HD~163296 & 100.57$^{+0.41}_{-0.41}$ & 8750$\pm$125 & 1.19$^{+0.04}_{-0.05}$ & 1.91$^{+0.12}_{-0.00}$ & 10.00$^{+9.50}_{-2.00}$ & 28.80 & 36.97  \\
    
    HD~169142 & 114.42$^{+0.35}_{-0.35}$ & 7250$\pm$125 & 0.76$\pm$0.01 & 1.55$^{+0.03}_{-0.00}$ & $<$20 & 27.62 & 36.24 \\
    
    HD~179218 & 257.95$^{+2.21}_{-2.17}$ & 9750$\pm$125 & 2.02$\pm$0.01 & 2.99$^{+0.01}_{-0.04}$ & 2.35$^{+0.19}_{-0.16}$ & 47.45 & 61.32  \\
    \hline
    \end{tabular}
    \caption{Target sample and physical parameters.}
    \begin{tablenotes}
        \item Note 1: Physical parameters are adopted from \citet{guzman2021homogeneous}.
        \item Note 2: The FWHM of the PSF (i.e., limits of spatial resolution), expressed in au, at the two prominent PAH emission bands -- 8.6 and 11.3~$\mu$m are listed in the ninth and tenth columns, respectively. 
    \end{tablenotes}
    \label{Tab_Targets}
\end{threeparttable}
\end{table*}

\color{black}

\section{Observations and processing} \label{Observations}

\color{black} We used VISIR-NEAR in long-slit spectroscopy configuration, whose spectral resolution is $\approx$300 and spatial resolution is 0\farcs076/pixel. The slit width is 0\farcs4. Our observations were conducted from 15-17 September and 12-19 December 2019. An overview of the observations is listed in Table~\ref{Tab_ObsLog}. Observing setup and data processing stages are discussed in Sects. \ref{Observing_Setup} and \ref{Data_Processing}, respectively. 


\begin{table*}
\begin{threeparttable}
\small
    \centering
    \begin{tabular}{c  c  c  c  c  c  c  c   c} 
    \hline\hline
    Object & Date & Obs. Start & Duration [hr]  & Airmass & Calibrator & Calib. Obs. Start [hr] & Calib. Airmass & Slit Angle [deg] \\ 
    \hline
    AB Aurigae & 19.12.2019 & 04:10:27 & 0.44 & 1.77-1.83 & HD~39045 & 04:55:35 & 1.83 & \color{black} 111.1 -- 115.9 \color{black}\\
    HD~95881 & 16.12.2019 & 07:36:04 & 0.40 & 1.54-1.58  & HD~98292 & 07:30:00, 08:20:45 & 1.46-1.56 & \color{black} -87.8 -- -94.7 \color{black}\\
    HD~97048 & 16.12.2019 & 06:36:30 & 0.40 & 1.83-1.88 & HD~98292 & 07:17:28 & 1.57 & \color{black} -102.0 -- -108.4  \color{black} \\
    HD~100453 & 17.12.2019 & 07:26:21& 0.40 & 1.32-1.39 & HD~102461 & 08:11:59 & 1.35 & \color{black} -111.3 -- -118.0 \color{black} \\
    HD~100546 & 13.12.2019 & 05:38:53 & 0.56 & 1.95-2.12 & HD~98292 & 06:23:41 & 1.79 & \color{black} -129.9 -- -138.1 \color{black}\\
    HD~163296 & 15.9.2019 & 00:57:20 & 0.34 & 1.12-1.16  & HD~169420 & 01:19:15 & 1.12 & \color{black} 48.4 -- 48.6   \color{black}\\
    HD~169142 & 16.9.2019 & 03:06:09 & 0.20 & 1.56-1.65 & HD~171115 & 03:04:17 & 1.47 & \color{black} 45.5 -- 46.7  \color{black}\\
    HD~179218 & 16.9.2019 & 23:35:14 & 0.33 & 1.31-1.33 & HD~185622 & 00:02:54 & 1.34 & \color{black} -222.8 -- 229.8   \color{black}\\

    \hline
    \end{tabular}
    \caption{Log of the VISIR-NEAR observations.}
    \begin{tablenotes}
        \item[] Note: We list the observing date, time (UT), duration of the observation, and airmass thereof (Cols. 2–5). The calibrators used for the atmospheric correction are also given, with the calibrator star, time, and airmass of the measurements (Cols. 6–8) and the range of orientation angles of the slit for all observations \color{black} (clockwise with respect to east. The mean orientation for each object ($\bar{\theta}_{slit}$) is listed atop each panel in Fig. \ref{Fig_SPHERE_allSources}) (Col. 9) \color{black}.
    \end{tablenotes}
    \label{Tab_ObsLog}
\end{threeparttable}
\end{table*}

\begin{figure}
\centering
\includegraphics[scale = 0.65]{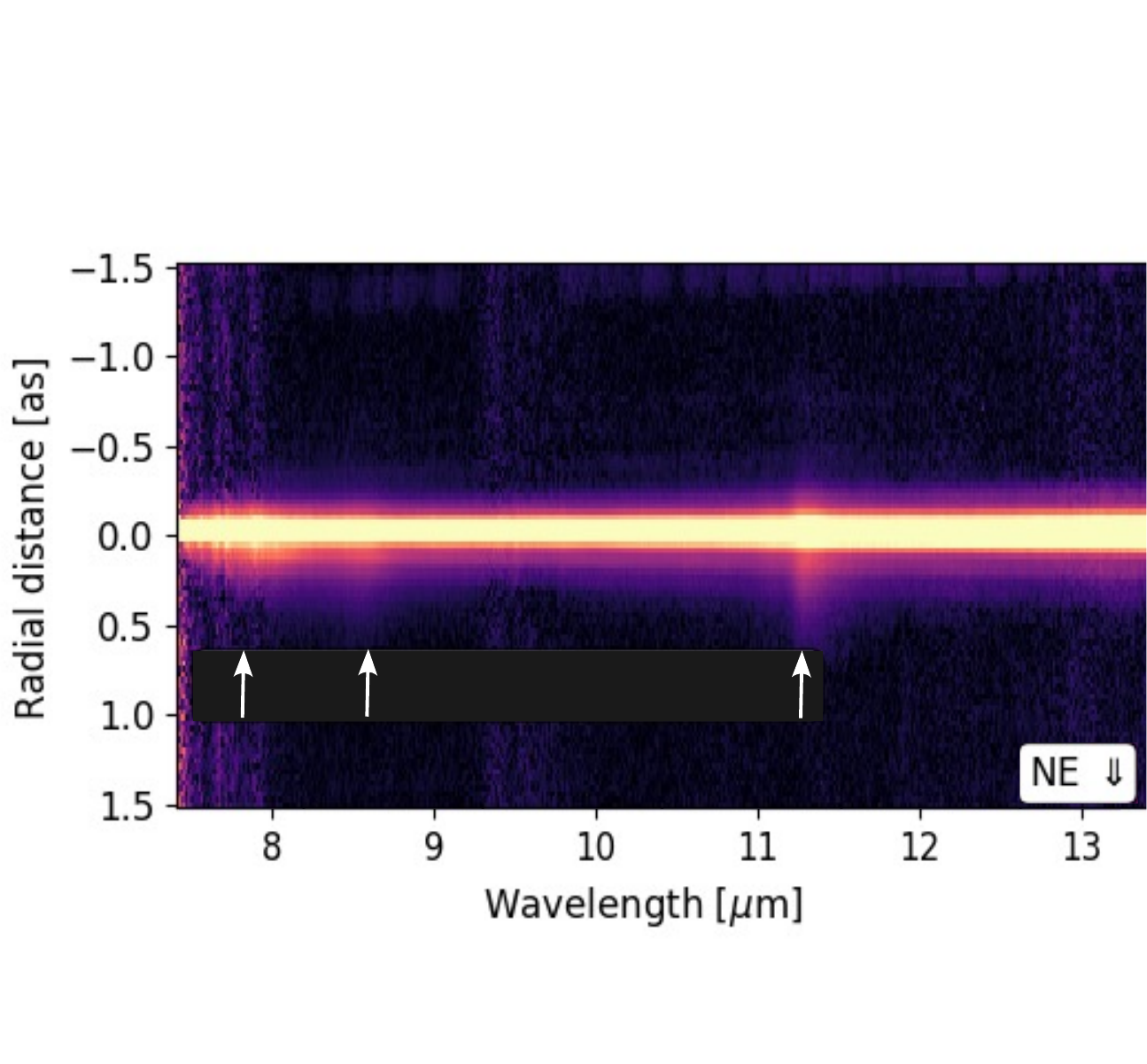}
  \caption{Calibrated 2D spectrum of HD~97048. The x-axis is the spectral dimension perpendicular to the long axis of the slit, and the y-axis is the spatial axis parallel thereto. The source is positioned in the center of the y-axis. Spatially resolved emission in the PAH bands is highlighted with white arrows. {Note} the image was square-root-stretched for demonstration purposes.}
     \label{Fig_2D_Spectrum}
\end{figure}

\subsection{Observing setup} \label{Observing_Setup}

The long-slit spectrograph of VISIR covers the full N band in one setting at a spectral resolution of approximately 300 (see the \href{https://www.eso.org/sci/facilities/paranal/instruments/visir/doc/VLT-MAN-ESO-14300-3514_2020-03-03.pdf}{VISIR user manual}).

Observations with NEAR could be performed only in pupil-tracking mode, with a predetermined instrument rotation angle that cannot be changed (the NEAR Lyot stop and VLT pupil are then aligned, and the AO control matrix has been calculated only for this angle). As a result, the field rotated during observations. This was uncritical for our observations, as they were relatively short, and the field typically rotated by only a few degrees during an observation. Moreover, knowing the field orientation during each integration allowed us to extract subsets of exposures if the amount of field rotation was large enough to cause significant spatial smearing for some targets. For a given target, we achieved this by stacking only the observations for which field rotation caused a spatial offset at the spatial edges (which we limited to 1\farcs5 from the central source on either side of the slit, as for all sources except for HD~97048 -- we detected no significant spatially resolved emission with the analysis presented in Sect. \ref{Modeling_FluxExcess}, when applied to the average of all available exposures of each object,) of the spectra by less than the median full-width half maximum (FWHM) of the 7.9~$\mu$m PSF (the narrowest PSF of a significant spectral feature) from the mean slit orientation angle.





We used chopping with the deformable secondary mirror at the maximum throw of 5\farcs5. The stability of the PSF provided by NEAR ensured operational consistency near the diffraction limit. Through Gaussian fitting, we measured the average FWHM of the PSF at 7.9~$\mu$m (the shortest wavelength in the analyzed spectra) for all observed calibrators (163 exposures) to be $\approx1.15 \pm 0.04$ $\lambda/D$. At longer wavelengths, the PSF quality of science target observations cannot be directly measured in general since the disks that dominate the N-band emission of the science targets may be somewhat spatially resolved in the continuum and more so in the PAH bands. Since the NEAR AO wavefront sensor was specifically designed for a planet detection experiment in the $\alpha$~Cen system, it was not optimized for performance on much fainter targets. While our calibrator stars are much brighter than the adopted brightness limit for the AO system ($I\leq8.5$~mag) and photon shot noise is thus not a limiting factor in the AO performance, the science targets' brightnesses are much closer to the magnitude limit and may be in the regime where the limited number of photons starts to degrade the S/N of the wavefront signal and hence the quality of the PSF. Thus, a marginally resolved disk source observed with a perfect PSF may look very similar to an unresolved disk source observed with a slightly poorer PSF -- and the two scenarios cannot be easily distinguished\footnote{One way to assess the performance of the AO system toward fainter optical magnitudes would be to observe fainter standard stars. However, stars in the optical brightness range of our science targets are very faint in the N band and would yield prohibitively low signal-to-noise N-band data.}.

We assessed the PSF quality at fainter magnitudes with HD~95881 -- which is known to have a very compact N-band continuum emission unresolved with VISIR. At a brightness of $I\approx8.0$~mag, the source is indeed near the limit for the AO wavefront sensor. The FWHM of the PSF ranges between $\approx$1.2 $\lambda/D$ at 7.9~$\mu$m and $\approx$1.05 $\lambda/D$ at 14 $\mu$m. Therefore, we find that the performance of VISIR-NEAR remains close to diffraction-limited over the brightness range of our science targets (the faintest of which is $I\approx8.1$~mag).

\subsection{Data processing} \label{Data_Processing}

We implemented spectral calibration for our targets by observing spectrophotometric standard stars (Table \ref{Tab_ObsLog}), expressing an observed spectrum (of either a calibrator or the science target) as
\begin{equation}
S(\lambda) = I_\star(\lambda)A(\lambda)R(\lambda),
\label{Eq_IncSpectrum}
\end{equation}
where $S(\lambda)$ denotes the raw extracted signal from the source, $I_\star(\lambda)$ is the intrinsic spectrum of the observed object, $A(\lambda)$ is the atmospheric extinction profile (which is a strong function of, e.g., integrated water-vapor column and airmass), and $R(\lambda)$ is the spectral instrument response curve. Since $A(\lambda)$  and $R(\lambda)$ are similar for both the calibrator and the science object (up to a difference in observing conditions), and a simulated stellar spectrum exists for the calibrator star, a calibrated science object spectrum is achieved straightforwardly:
\begin{equation}
I_{\rm science}(\lambda) = \frac{S_{\rm science}(\lambda)}{S_{\rm calib}(\lambda)}\cdot I_{\rm \star, int}(\lambda),
\label{Eq_CalibSpectrum}
\end{equation}
where $I_{\rm science}(\lambda)$ is the calibrated spectrum of the science object and $I_{\rm\star, int}(\lambda)$ is the intrinsic spectrum of the calibrator. However, the science object and its calibrator are not observed under identical atmospheric conditions. The resulting difference between the calibrator's atmospheric extinction profile $A(\lambda)$ should also be considered. We compute this difference in the following manner: we simulate two telluric absorption profiles using \texttt{SkyCalc}\footnote{\url{https://www.eso.org/observing/etc/bin/gen/form?INS.MODE=swspectr+INS.NAME=SKYCALC}}, Sky Model Calculator by ESO, each with the respective atmospheric conditions measured during the observation of the science object and calibrator (which vary as a function of airmass and integrated water-vapor column). The ratio thereof makes the correction we introduce into Eq. \ref{Eq_CalibSpectrum} in order to account for the difference between the $A(\lambda)$ of the science object and the calibrator. We used the spectral templates presented in \citet{Cohen1999} for the intrinsic spectra of the calibrators.

Finally, we superimposed all snapshot exposures of each object (Sect. \ref{Observing_Setup}) to receive a final calibrated 2D spectrum, where one axis represents the radial dimension (along the long axis of the slit) and the other the spectral dimension. In Fig.  \ref{Fig_2D_Spectrum} we present a calibrated 2D spectrum of HD97048, where spatially resolved emission in the PAH bands can be noticed.


\color{black}

\section{Modeling} \label{Modeling}

\subsection{Preliminary excess flux detection in the 8.6 and 11.3~$\mu$m PAH bands} \label{Modeling_FluxExcess}

\begin{figure}
\centering
\includegraphics[scale = 0.5]{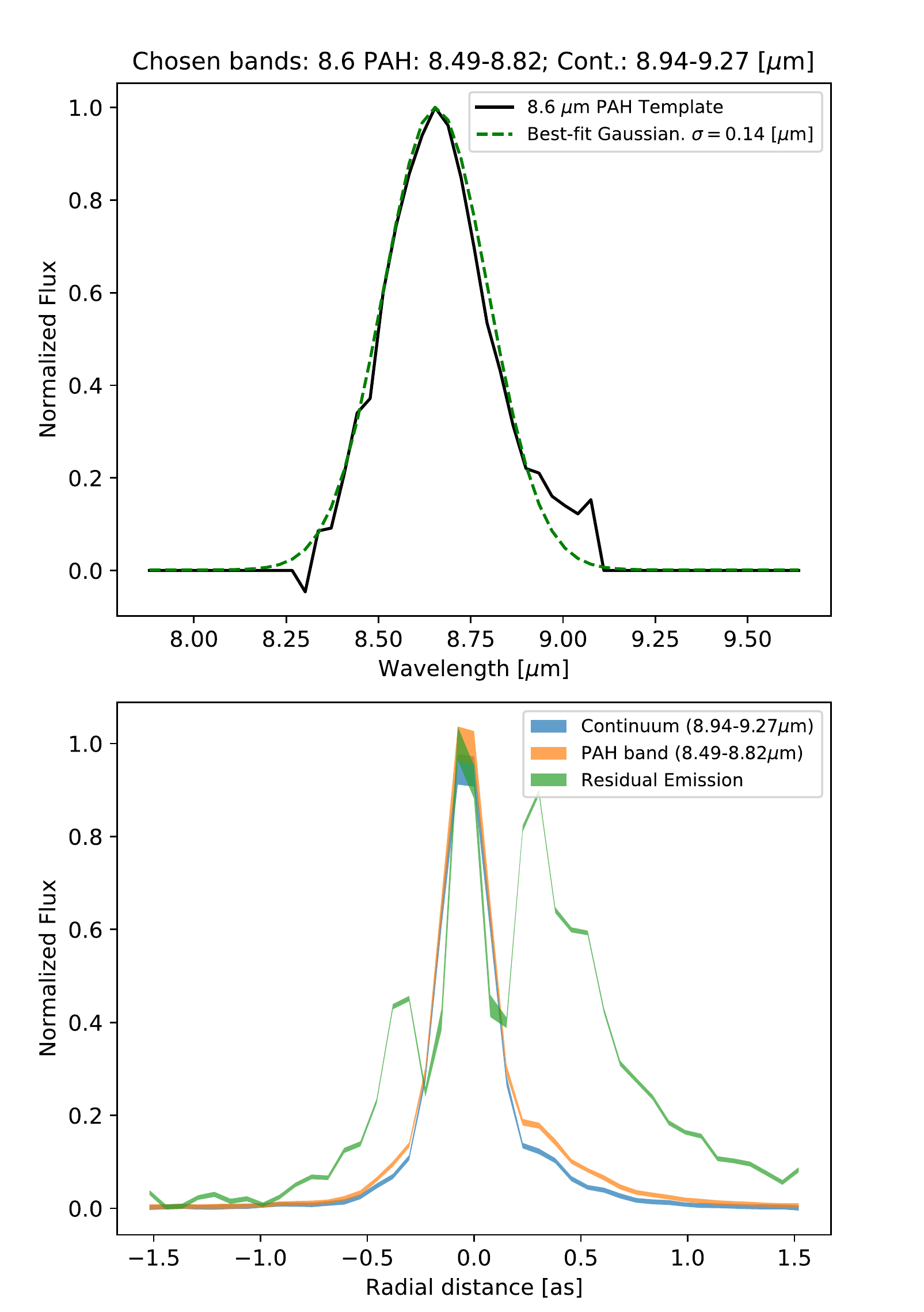}
 \caption{\color{black} Excess flux analysis of the 8.6~\mum~PAH band of HD~97048. \textbf{Top Panel}: Gaussian fitting of the 8.6 \mum~PAH template, with a resulting standard deviation of $\sigma_{template} \approx 0.14$ \mum. The aperture of the PAH spectral window is defined as ±1.17 $\cdot \sigma_{template}$, and the continuum aperture is set to be two FWHMs down the red part of the spectrum (the chosen widths for each are plotted atop the panel). \textbf{Bottom Panel}: Fluxes integrated over the continuum and PAH apertures, and the computed excess flux (blue, orange, and green curves, respectively) and their respective $\pm1\sigma$ uncertainties (see Appendix \ref{APP_Stat_Model}), indicated by curve widths. \color{black}}
    \label{Fig_contFitting}
\end{figure}

\begin{figure*}[h!]
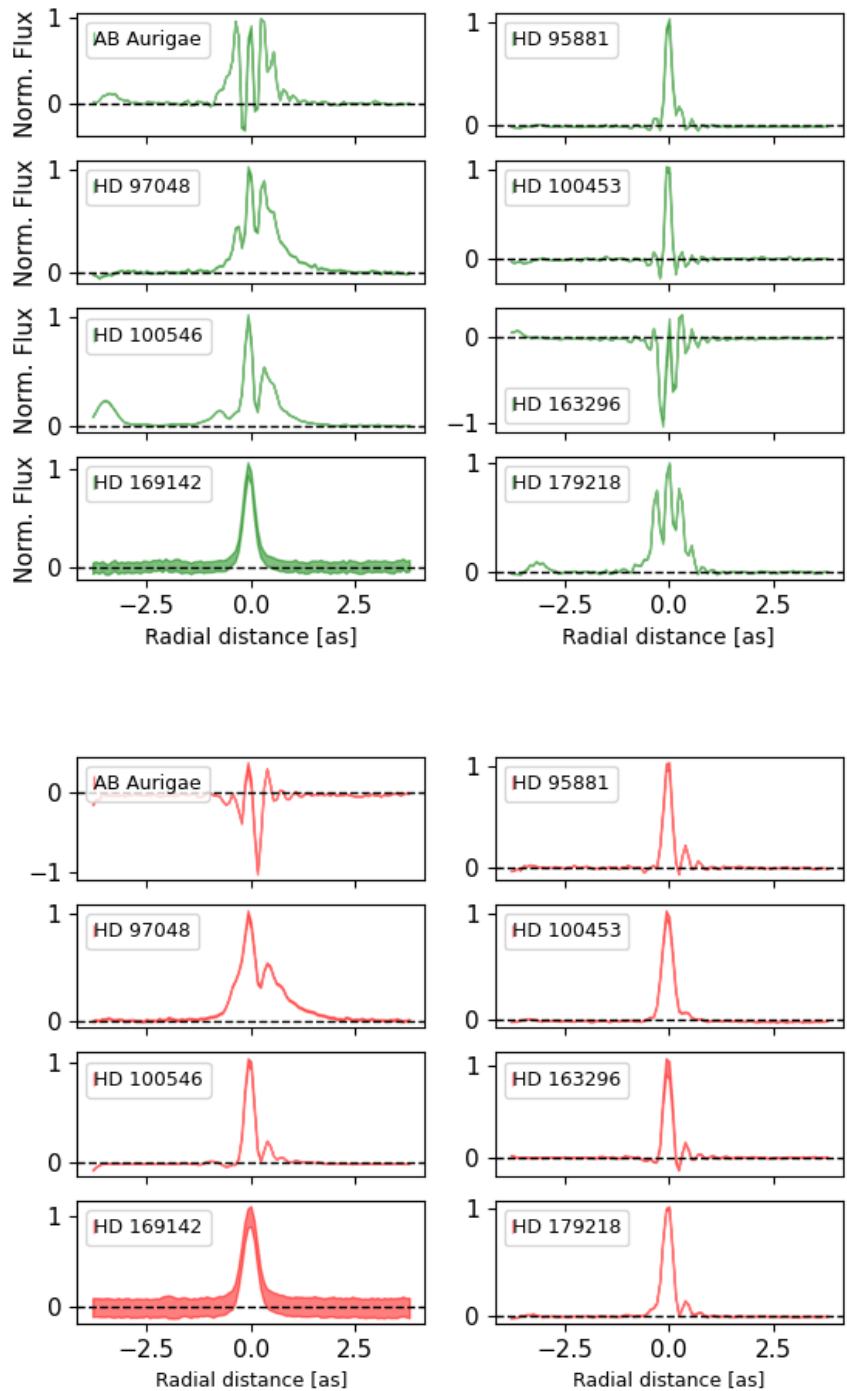

   \centering
    \begin{tabular}{cc}
        \hspace{-1.0cm}
        \includegraphics[scale = 0.8]{fluxExcess_allObjects_86.pdf} \\
        \hspace{-1.0cm}
        \includegraphics[scale = 0.8]{fluxExcess_allObjects_11_3.pdf} \\
    \end{tabular}
    \caption{Normalized flux excess in the 8.6~$\mu$m and 11.3 $\mu$m PAH bands (top and bottom panels, respectively). The width of the lines represents the $\pm$1$\sigma$ residuals of the PAH and continuum windows (see Sect. \ref{Modeling_FluxExcess}). The x-axis is parallel to the slit (i.e., its spatial dimension), and the zero point is the object's center (the star). Note that negative values indicate that the flux in the continuum windows is greater than that in the PAH window. This is caused by silicate emission in the continuum window.}
    \label{Fig_fluxExcess_allObjects}
\end{figure*}

We performed a preliminary analysis of the spatially resolved spectra for all sources in our sample in an attempt to detect spatially resolved excess flux in two prominent PAH emission-bearing spectral bands -- 8.6 and 11.3~$\mu$m \citep[e.g.,][]{vanBoekel2005, Lagage2006, Maaskant2014} -- by defining two spectral windows, each centered on the emission peak of the 8.6, and 11.3~$\mu$m PAH templates, respectively, and whose spectral width is the FWHM of a Gaussian fitted to each template. We then define reference (continuum) spectral windows. In the case of the 8.6~$\mu$m aperture, the continuum spectral aperture is of similar width. It is located 2~FWHM (usually approximately 0.5~$\mu$m) down the red part of the spectrum (since the 7.9~$\mu$m PAH feature often dominates the blue part). The procedure is similar in the case of the 11.3~$\mu$m aperture, except that we average two reference spectra on either spectral side of the PAH aperture. Each spectral aperture is then summed over the spectral dimension and normalized to obtain a 1D normalized flux profile, representing the spatially resolved emission at these spectral windows. Finally, we subtract the reference flux profile from the PAH aperture. This procedure yields positive residuals if the PAH emission is more spatially extended than the emission at the reference wavelengths. We note that this qualitative analysis remains insensitive to flux differences in the vicinity around the {center} of the object because the difference between the two bands there is obscured by the fact that the shape of the PSF dominates the central emission. \color{black} In Fig.~\ref{Fig_contFitting}, we plot an example of this procedure applied to the 8.6~\mum~PAH band of HD~97048, and in Fig.~\ref{Fig_fluxExcess_allObjects} we plot the result of this analysis on all object and both bands. \color{black}

\subsection{Fitting spatially resolved spectra} \label{modeling_Optimization}


\subsubsection{Rationale} \label{optim_rationale}

Following our preliminary analysis of the spatial extent of PAH emission at the 8.6~$\mu$m and 11.3~$\mu$m bands (Sect. \ref{Modeling_FluxExcess}), we analyze the spatial dependence of the spectral shape in more detail for selected objects. We extract spectra at different distances from the star in the along-slit direction, which are the rows of our 2D spectra (see Sect. \ref{Observations}). For each object, we consider $2n$+1 spectra, where $n$ is the number of spectra sampled on each star side in the along-slit direction, and one additional spectrum samples the on-star location. We then perform a spectral decomposition and fitting analysis of each spectrum. Because the telescope PSF is sampled by multiple pixels, the spectra of adjacent rows are correlated. Since the $\approx$diffraction-limited PSF width scales $\approx$linearly with wavelength, the spatial correlation length is longer on the red side of the N band than on the blue side.

In the fitting procedure, we followed the reasoning and methodology of \citet{vanBoekel2005}. We performed a decomposition of the observed spectra into the opacity curves of five common mineral dust species, with two grain sizes for each component, plus an empirical template for each of the significant PAH bands and a gray-body continuum component. The chosen mineral species represent the most abundant dust species in circumstellar material that exhibit spectral structure in the N band \citep[e.g.,][]{Bouwman2001}. The adopted grain sizes of the mineral components have volume equivalent radii of 0.1 $\mu$m ("small") and 1.5 $\mu$m ("big"). This choice of dust species and grain sizes was found to provide a suitable set of base functions for reproducing the observed range of shapes of the $\sim$10 $\mu$m silicate features \citep[][]{vanBoekel2005}.



\citet{vanBoekel2005} demonstrated that such a comparatively simple model is adequate for a limited spectral range (e.g., only the N band, as in the current study); in this case, more sophisticated models such as that by \citet{Juhsz2010}, containing considerably more degrees of freedom describing the disk structure, are insufficiently constrained due to the small spectral coverage and would constitute an over-fitting of the data.


Unlike the single N-band template for all the PAH emissions, which was used for the modeling of \citet{vanBoekel2005}, we modeled four PAH emission bands independently rather than as a single template. We generated the PAH templates for each band as follows: {7.9}~$\mu$m was fitted as a power-law for each source independently since the peak of its emission is not within the modeled spectral range. For {8.6}~$\mu$m, we generated an individual high-S/N template of this band for the following sources: HD\,97048, HD\,100453, and HD\,169142. For the other sources, we assumed a template that is the average. For {11.3}~$\mu$m, we repeated a process similar to that of the 8.6~$\mu$m band. The {12.6}~$\mu$m band exhibits the weakest signal, for which it was impossible to generate a satisfactory high-S/N template from our data. Instead, we compared the highest-S/N instance of this emission band (in HD\,100453) to two functions with optimized parameters: a Gaussian and a Lorentzian, and used the Kolmogorov-Smirnov (K-S) test to determine which of the two functions resembles it more. We found that the Lorentzian (with optimized parameters $df$ and $\gamma$) scored higher on the K-S test and was subsequently adopted as our template for this emission feature.

\subsubsection{Parametric model} \label{Param_Model}

\color{black} For each model N-band spectrum, we fit 18 free parameters, accounting for the following: (\textbf{1}) a decoupled continuum component; and (\textbf{2}) an additional continuum component that is coupled to ten silicate opacity curves of five silicate species, each of two different grain sizes, and four PAH emission components. The model is described analytically as


\begin{equation}
y(\lambda) = D \cdot B_\nu(T_1, \lambda) + C \cdot B_\nu(T_2, \lambda) \cdot \sum_{i = 1}^{10}w_i \cdot S_i(\lambda) + \sum_{j = 1}^4 A_j \cdot P_j(\lambda),
\label{Eq_Model}
\end{equation}

where $\lambda$ is the range of modeled wavelengths ($\approx$8-13 $\mu$m), $y$ is the simulated model. The degrees of freedom are color-coded in orange and are as follows: $D$ and $C$ are the amplitudes of the decoupled and coupled continuum components, respectively, $T_1$ and $T_2$ are the effective temperatures thereof, $w_i$ is the relative abundance of the $i${th} silicate component opacity curve ($S_i$), with ($\sum_i$\,$w_i$ \,$=1$), and $A_j$ is the amplitude of the $j^{\rm{th}}$ PAH component ($P_j$).

We plot the constituents of our parametric model in Appendix \ref{APP_Param_Model}. \color{black} 



\subsubsection{Optimization With \texttt{MultiNest}} \label{Stat_Model}

\color{black} For each N-band spectrum, we perform nonlinear optimization of a parametric spectral model and derive its constituents' uncertainties using \texttt{MultiNest} \citep{Multinest}, a multimodal MCMC algorithm with implementation in Python \citep{Buchner14}. \texttt{MultiNest} implements nested sampling \citep{Skilling04}, a Monte Carlo method targeted at the efficient calculation of the Bayesian evidence. Using \texttt{MultiNest} allows for exploring regions of interest in the parameter space with a reduced risk of being trapped in local likelihood maxima. 

We assume uniform, hardbound priors for all parameters. Priors of the fractional amplitudes of the silicate opacity curves ($w_i$) are always normalized by the sum thereof, the priors of each of the PAH template amplitudes ($A_j$) are bound to a range of 0-100, and the priors of the effective temperatures of the Planckians ($T_1$, $T_2$) span across a range of 50-600K.

That said, from examining the posterior distributions of our spectral fitting procedure (see Appendix \ref{APP_posteriors}), we find that degeneracies between any combination of parameters are insignificant, in agreement with \citet{vanBoekel2005}, who found the silicate components to be largely uncorrelated.

We describe the statistical model used for the MCMC optimization routine in Appendix \ref{APP_Stat_Model}. \color{black}

\section{Results} \label{Results}

\subsection{Fitting unresolved (integrated) spectra} \label{Results_1D}

\begin{figure*}[] 
\centering
    \includegraphics[scale = 1.]{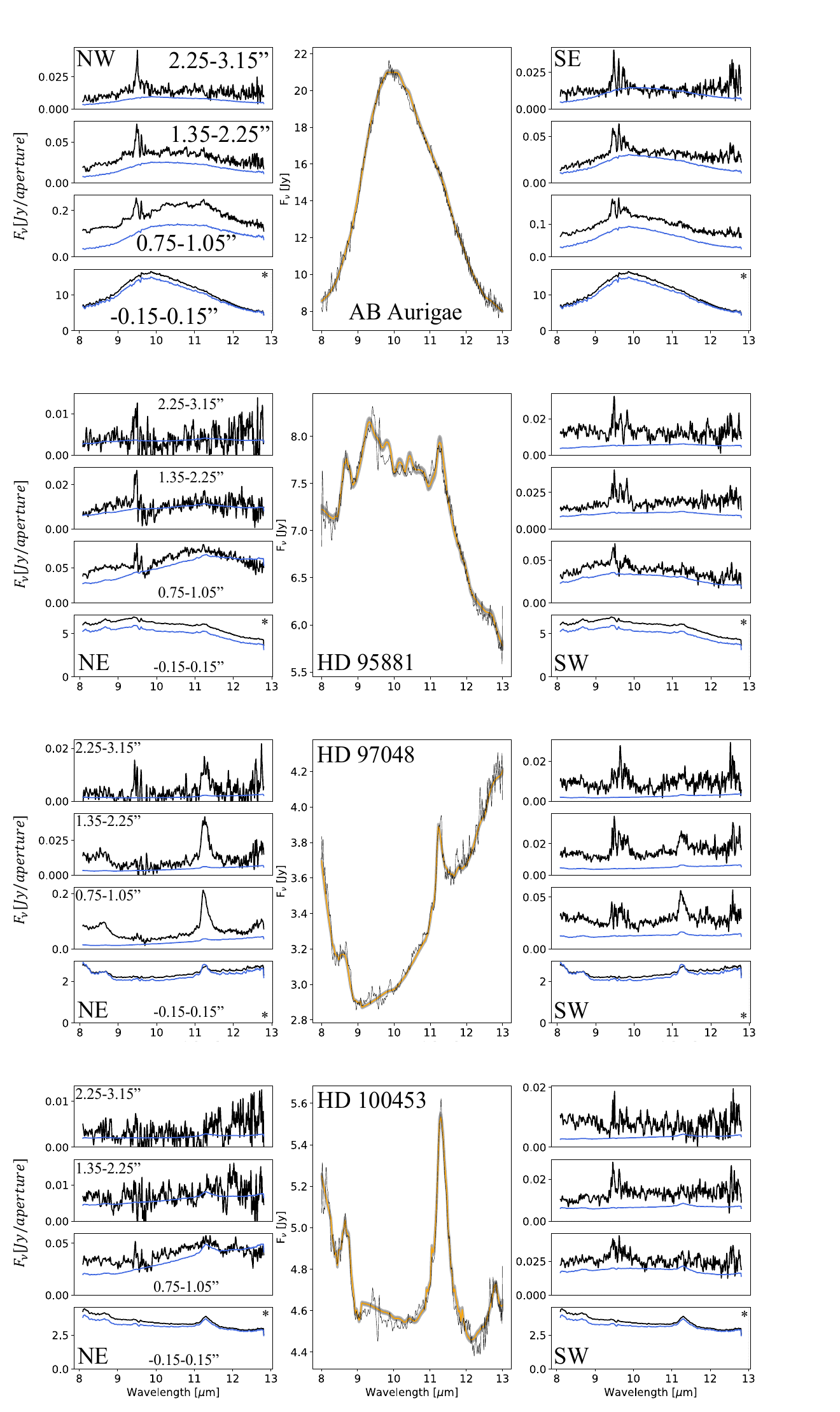}
    \caption{Calibrated long-slit spectra of our targets. \textbf{Left Panel}: Spatially resolved spectra. The black curves are down-slit spatially resolved spectra. Each spectrum represents an average of a $\approx\lambda/D$-wide spectral bin (every three rows of the reduced 2D spectrum, whose centroids are separated by 0\farcs228 and are arranged bottom up, where the bottom spectrum is that of the central object). The uppermost spectrum, farthest from the star, is averaged over three such apertures. The blue curves are the expected flux contamination coming from the unresolved central source, extracted as described in Sect. \ref{Results_1D}. The off-axis direction is indicated by the two letters in the top or bottom panels (e.g., NE for northeast). \textbf{Middle Panel}: Fitted integrated spectra. The solid black line is the observed spectrum, the thicker orange line is the best-fit spectrum, and the filled gray area is the $\pm$1$\sigma$ error-range, where $\sigma$ is the flux-wise standard deviation of spectra generated with all best-fit parameters within the 50\,$\pm$\,34 percentiles range. \textbf{Right Panel}: Similar to the left panel, but on the opposite side of the star with respect to the slit axis. Note that spectra marked with an asterisk are identical in both the left and right panels. Note also that the spectral artifact visible at $\approx$9.6 $\mu$u is caused by the atmospheric ozone absorption band. Continued on the next page.}
    \label{Fig_1D_Spectra}
\end{figure*}

\addtocounter{figure}{-1}
\begin{figure*}
\centering
    \includegraphics[height=1.00\textheight, trim=0 0 0 0]{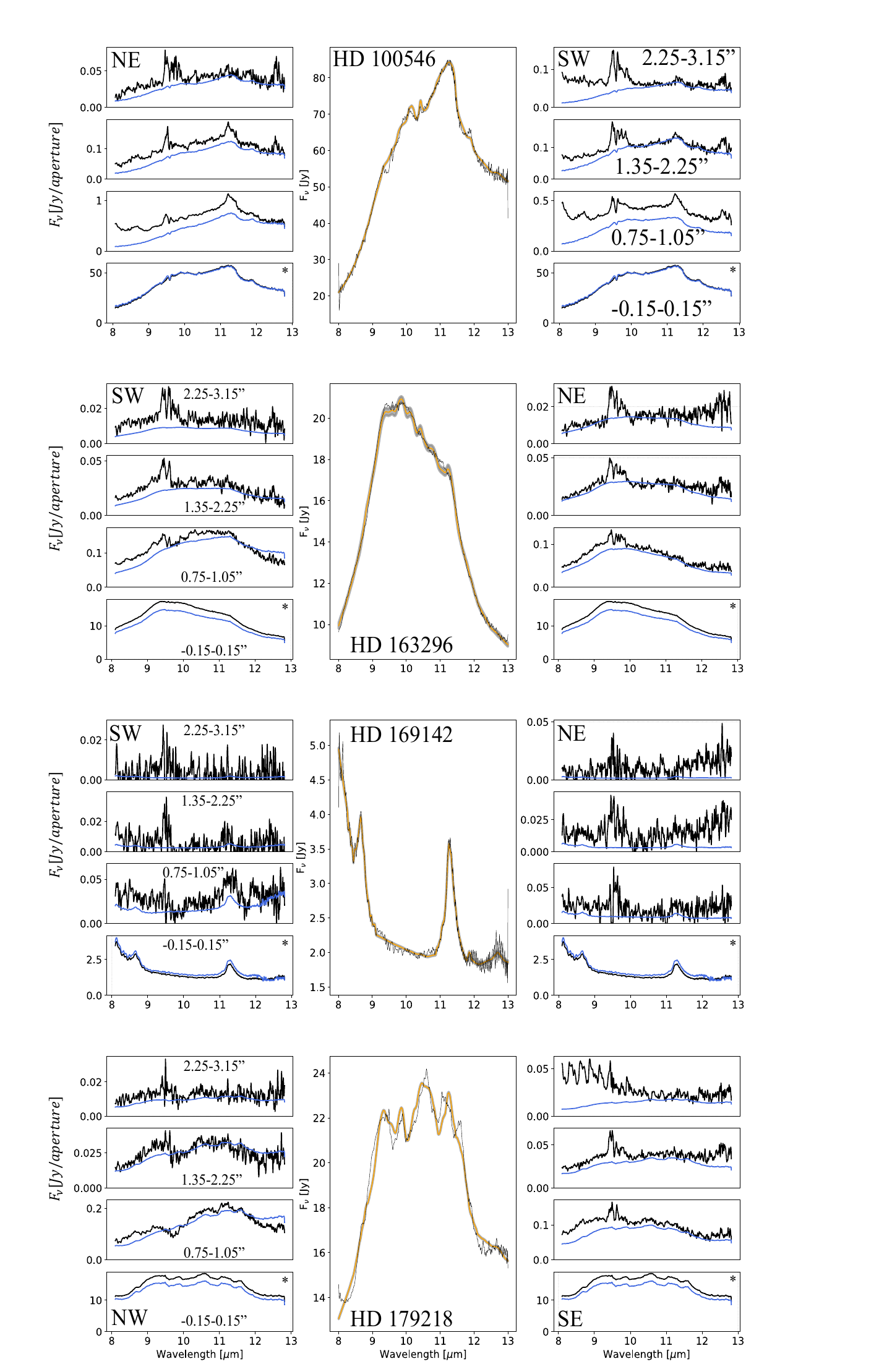}
    \caption{Continued.}
\end{figure*}

\begin{table*}
 \begin{threeparttable}
\small
    \centering
    \begin{tabular}{c c c c c c c c c} 
    \hline\hline
    Parameter & AB Aurigae & HD 95881 & HD 97048 & HD 100453 & HD 100546 & HD 163296 & HD 169142 & HD 179218 \\
    \hline
    $T_{\rm{cont}}$ [K] & 335$^{+8}_{-10}$ & 290$^{+9}_{-8}$ & 176$^{+3}_{-4}$ & 322$^{+8}_{-7}$ & 217$^{+7}_{-6}$ & 297$^{+10}_{-13}$ & 423$^{+28}_{-25}$ & 196$^{+5}_{-4}$ \\
    $T_{\rm{opac}}$ [K] & 215$^{+13}_{-10}$ & 217$^{+29}_{-24}$ & 642$^{+6}_{-13}$ & 57$^{+4}_{-4}$ & 160$^{+10}_{-8}$ & 297$^{+42}_{-30}$ & 62$^{+4}_{-5}$ & 229$^{+4}_{-5}$ \\
    $C$ & 17.83$^{+1.79}_{-1.59}$ & 3.29$^{+0.68}_{-0.62}$ & 1.76$^{+0.17}_{-0.12}$ & 10.57$^{+1.41}_{-1.41}$ & 118.19$^{+16.23}_{-13.82}$ & 15.22$^{+1.31}_{-0.85}$ & 5.91$^{+0.83}_{-0.78}$ &  23.84$^{+0.55}_{-0.47}$ \\
    $D$ & 7.97$^{+0.17}_{-0.34}$ & 6.25$^{+0.21}_{-0.16}$ & 4.06$^{+0.02}_{-0.02}$ & 4.61$^{+0.02}_{-0.01}$ & 36.36$^{+1.47}_{-1.77}$ & 9.02$^{+0.29}_{-0.43}$ & 2.52$^{+0.07}_{-0.79}$ &  12.34$^{+0.10}_{-0.10}$ \\
    PAH$_{\rm{7.9}}$ & 0.30$^{+0.58}_{-0.23}$ & 2.00$^{+0.40}_{-0.46}$ & 33.80$^{+0.82}_{-0.93}$ & 10.49$^{+1.11}_{-1.07}$ & 2.91$^{+1.41}_{-1.36}$ & 0.67$^{+0.91}_{-0.49}$ & 7.16$^{+0.28}_{-0.25}$ & 10.43$^{+0.34}_{-0.40}$ \\
    PAH$_{\rm{8.6}}$ & 0.64$^{+0.42}_{-0.34}$ & 1.83$^{+0.22}_{-0.25}$ & 19.72$^{+0.86}_{-0.70}$ & 6.82$^{+0.92}_{-0.88}$ & 3.53$^{+0.64}_{-0.62}$ & 0.70$^{+0.91}_{-0.48}$ & 7.90$^{+0.25}_{-0.24}$ & 4.75$^{+0.14}_{-0.15}$ \\
    PAH$_{\rm{11.3}}$ & 0.14$^{+0.16}_{-0.10}$ & 0.64$^{+0.11}_{-0.10}$ & 6.68$^{+0.33}_{-0.36}$ & 7.16$^{+0.70}_{-0.63}$ & 3.76$^{+0.96}_{-1.09}$ & 0.49$^{+0.37}_{-0.30}$ & 8.97$^{+0.33}_{-0.29}$ & 2.04$^{+0.05}_{-0.05}$ \\
    PAH$_{\rm{12.6}}$ & 0.41$^{+0.49}_{-0.28}$ & 0.59$^{+0.33}_{-0.30}$ & 21.15$^{+5.68}_{-5.52}$ & 17.28$^{+5.29}_{-4.72}$ & 4.63$^{+3.15}_{-2.68}$ & 0.95$^{+0.86}_{-0.63}$ & 23.78$^{+5.89}_{-5.51}$ & 3.19$^{+0.41}_{-0.41}$ \\
    
    olivine$_{\rm{small}}$ & 0.93$^{+0.05}_{-0.09}$ & 0.08$^{+0.16}_{-0.06}$ & 0.00$^{+0.01}_{-0.00}$ & 0.63$^{+0.25}_{-0.36}$ & 0.58$^{+0.24}_{-0.24}$ & 0.69$^{+0.20}_{-0.28}$ & 0.61$^{+0.27}_{-0.36}$ & 0.00$^{+0.00}_{-0.00}$ \\
    olivine$_{\rm{big}}$ & 0.08$^{+0.10}_{-0.06}$ & 0.37$^{+0.32}_{-0.24}$ & 0.00$^{+0.01}_{-0.00}$ & 0.47$^{+0.32}_{-0.30}$ & 0.77$^{+0.16}_{-0.23}$ & 0.74$^{+0.17}_{-0.29}$ & 0.52$^{+0.31}_{-0.33}$ & 0.00$^{+0.00}_{-0.00}$ \\
    pyroxene$_{\rm{small}}$ & 0.03$^{+0.05}_{-0.02}$ & 0.40$^{+0.36}_{-0.26}$ & 0.26$^{+0.47}_{-0.21}$ & 0.67$^{+0.23}_{-0.36}$ & 0.36$^{+0.38}_{-0.27}$ & 0.47$^{+0.32}_{-0.30}$ & 0.61$^{+0.27}_{-0.36}$ & 0.22$^{+0.07}_{-0.05}$ \\
    pyroxene$_{\rm{big}}$ & 0.14$^{+0.07}_{-0.07}$ & 0.54$^{+0.29}_{-0.34}$ & 0.19$^{+0.25}_{-0.13}$ & 0.63$^{+0.25}_{-0.35}$ & 0.88$^{+0.08}_{-0.17}$ & 0.84$^{+0.11}_{-0.22}$ & 0.58$^{+0.28}_{-0.36}$ & 0.93$^{+0.04}_{-0.09}$ \\
    forsterite$_{\rm{small}}$ & 0.02$^{+0.01}_{-0.01}$ & 0.21$^{+0.12}_{-0.11}$ & 0.01$^{+0.02}_{-0.01}$ & 0.61$^{+0.24}_{-0.27}$ & 0.65$^{+0.14}_{-0.14}$ & 0.52$^{+0.19}_{-0.15}$ & 0.60$^{+0.25}_{-0.27}$ & 0.001$^{+0.00}_{-0.00}$ \\
    forsterite$_{\rm{big}}$ & 0.02$^{+0.01}_{-0.01}$ & 0.26$^{+0.11}_{-0.10}$ & 0.02$^{+0.03}_{-0.01}$ & 0.10$^{+0.15}_{-0.07}$ & 0.01$^{+0.02}_{-0.01}$ & 0.11$^{+0.13}_{-0.07}$ & 0.12$^{+0.18}_{-0.08}$ & 0.11$^{+0.01}_{-0.01}$ \\
    silica$_{\rm{small}}$ & 0.01$^{+0.02}_{-0.00}$ & 0.79$^{+0.14}_{-0.25}$ & 0.19$^{+0.24}_{-0.13}$ & 0.48$^{+0.30}_{-0.30}$ & 0.39$^{+0.28}_{-0.26}$ & 0.17$^{+0.17}_{-0.11}$ & 0.58$^{+0.27}_{-0.33}$ &  0.32$^{+0.21}_{-0.21}$ \\
    silica$_{\rm{big}}$ & 0.01$^{+0.02}_{-0.01}$ & 0.79$^{+0.14}_{-0.23}$ & 0.00$^{+0.01}_{-0.00}$ & 0.49$^{+0.30}_{-0.30}$ & 0.39$^{+0.29}_{-0.25}$ & 0.16$^{+0.18}_{-0.11}$ & 0.58$^{+0.27}_{-0.34}$ & 0.31$^{+0.22}_{-0.21}$ \\
    enstatite$_{\rm{small}}$ & 0.01$^{+0.01}_{-0.00}$ & 0.11$^{+0.20}_{-0.08}$ & 0.00$^{+0.01}_{-0.00}$ & 0.38$^{+0.33}_{-0.25}$ & 0.02$^{+0.03}_{-0.01}$ & 0.25$^{+0.21}_{-0.14}$ & 0.18$^{+0.32}_{-0.14}$ & 0.09$^{+0.01}_{-0.01}$ \\
    enstatite$_{\rm{big}}$ & 0.01$^{+0.01}_{-0.00}$ & 0.75$^{+0.15}_{-0.20}$ & 0.92$^{+0.05}_{-0.10}$ & 0.59$^{+0.25}_{-0.32}$ & 0.24$^{+0.11}_{-0.09}$ & 0.36$^{+0.28}_{-0.18}$ & 0.61$^{+0.24}_{-0.29}$ &  0.41$^{+0.03}_{-0.04}$ \\
     \hline
    \end{tabular}
    \caption{Best-fit values of the model parameters (see Sect. \ref{Param_Model}) for the integrated spectra fits of all objects.}
    \begin{tablenotes}
      \item[] Note: The best-fit values are the median values of the posterior distribution, and the uncertainties are the $50\pm 34.1$ percentiles ranges of the posterior distribution, computed after removal of the MCMC "burn-in" phase (points with relative likelihood $<10^{-3}$ times that of the best-fit).
    \end{tablenotes}
    \label{Tab_BestFit_1D}
     \end{threeparttable}
\end{table*}

In Fig.~\ref{Fig_1D_Spectra} we show the calibrated long-slit spectra of our targets, and in Table \ref{Tab_BestFit_1D} we list the median values and $\pm 50\pm 34.1$ percentiles of each constituent of our parametric model (see Appendix \ref{APP_Param_Model}) for each target. The large central panels show the total spectra ({black} curves), summed over the spatial dimension. The smaller panels on either side show the spectra at different spatial positions along the spatial dimension of the slit. From bottom to top we show spectra extracted at the central source position, 0\farcs9 away from the center, 2\farcs1 away from the center (all extracted over aperture approximately one FWHM of the PSF in size), and integrated over 2\farcs7 to 3\farcs3 (an aperture approximately three PSFs in size), as labeled in the left (or right) column of the object at the top of the figure. We plot the posterior distributions of the model parameters for each object in Appendix \ref{APP_posteriors}.

The total emission is strongly dominated by the flux from the central positions. We do see emission at the off-axis positions, in spectral shape often reminiscent of the spectrum at the central location, but with notably bright PAH emission in some of the targets (HD~97048, HD~100546, HD~169142). For the off-axis apertures, the expected temperatures for dust in thermal equilibrium are typically too low to lead to significant thermal emission; the off-axis continuum emission may therefore be dominated by "contamination" from the central source, that is, light emerging from within the central disk regions that is redistributed in the focal plane by the PSF. We investigate this hypothesis in the following.

We first created a PSF model as described in Appendix~\ref{sec:PSF_model}. We then estimated the flux from the inner disk regions by performing 1D Gaussian fits to the spatial profile at each wavelength, where the flux in each spectral bin was taken to be the volume of the Gaussian fit. We then assumed this light to come from a spatially unresolved source and convolved it with our PSF model to estimate the contamination from the inner disk regions at all spatial locations, and extracted the expected contamination from the respective apertures; the resulting spectra are shown in blue \color{black} in Fig.~\ref{Fig_1D_Spectra}.
We observe the following.

For most off-axis apertures, the contamination from the central source (blue lines) accounts for much or all of the observed flux, except for the PAH bands. We note that the calibrators are optically much brighter than the science targets. Hence, the AO system always operates in a regime where the PSF quality is not limited by photon noise from the star. The science targets are closer to the limiting magnitude of the AO system ($I\approx8.5$~mag), and therefore they will be in the range where the PSF quality depends on the target brightness in the optical. The PSF quality may be somewhat degraded for the science targets compared to the calibration observations from which the PSF model was created.
   
   For AB~Aurigae and HD~100546, we see that in the apertures above and below the central source, there is somewhat more flux than the redistribution model predicts; these systems are dominated by (silicate) dust in thermal equilibrium, with a transition-disk geometry. At larger radii, the observed fluxes approach those predicted by the contamination model again. Thus, the silicate emission appears slightly resolved in the VISIR-NEAR observations.

For HD~97048, the continuum emission in the off-axis apertures is substantially higher than predicted for contamination by the central source, particularly on the southwest side. The large spatial extent in the continuum was already noted by \cite{vanBoekel2004}; this emission stems from nonthermal equilibrium dust.

The PAH bands in the off-axis apertures for HD~97048 are much stronger than can be explained by contamination from the central source; the off-axis equivalent widths of the PAH bands are much larger than at the central position. For HD~100546, the PAH emission is also clearly stronger off-axis than in the central aperture. For HD~169142, this may be marginally the case.

Finally, for HD~169142, the central peak of the spatial profiles is shaped somewhat differently from the other sources. Whereas in the other sources, the continuum profile is consistent with strongly centrally peaked emission broadened by the PSF, in HD~169142, the observed profile is broader and more triangular. We qualitatively discuss this in Sect.~\ref{Disc_dist_HD169142}.

\subsection{Object-wise spatial  intensity distribution of PAHs} \label{Results_dist_objectwise}

We performed a nonlinear optimization analysis (Sect. \ref{modeling_Optimization}) of the spectra of five targets that exhibit spatially resolved PAH emission: AB Aurigae, HD~97048, HD~100546, HD~163296, and HD~169142. The remaining objects were not analyzed for the following reasons: HD~100453 did not display any spatially resolved PAH structure across either band, and HD~95881 and HD~179218 displayed complex spectral structures of silicate emission that could not be accurately modeled with our templates.

\begin{figure*}[] 
    \centering
    \includegraphics[width=1.0\textwidth, trim = 150 10 655 30, angle = 0, clip]{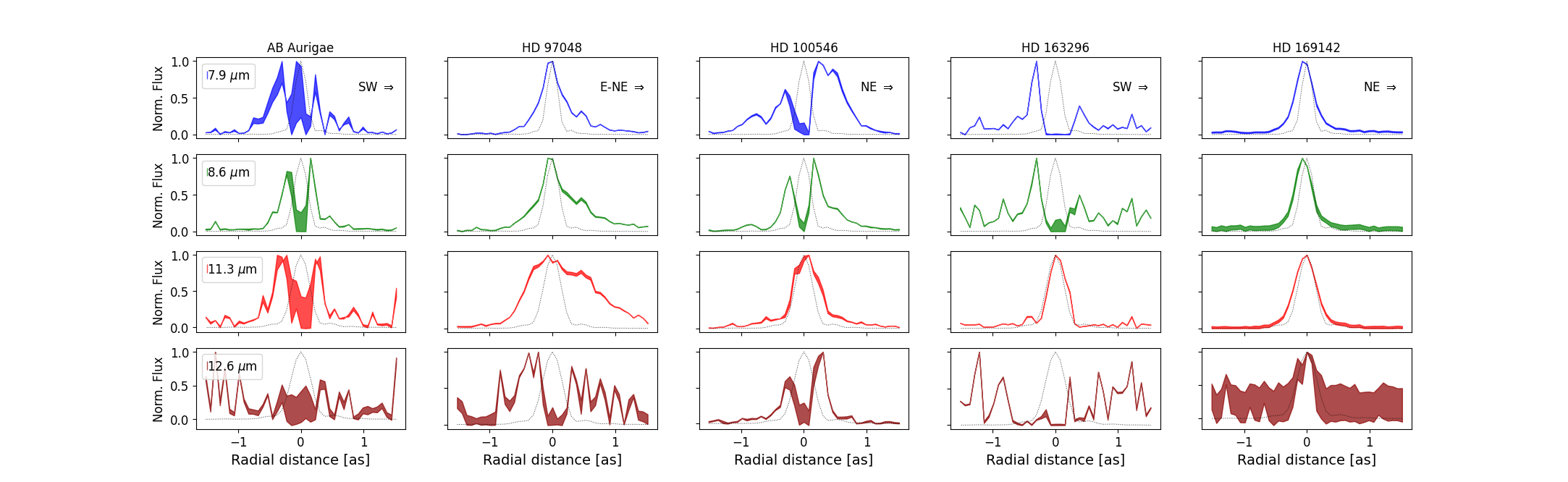} 
    
    \includegraphics[width=0.64\textwidth, trim = 930 10 155 30, angle = 0, clip]{PAH_Profiles_2D_selectSources.pdf}

    \caption{Radial profile of the normalized PAH emission in four spectral bands for select sources. Sources are ordered column-wise, and the PAH bands are ordered row-wise and are color-coded (from 7.9~$\mu$m in the uppermost row to 12.6~$\mu$m in the bottom row). The filled area represents the $\pm$1$\sigma$ error range around the median value of the emission amplitude posterior distribution of the relevant PAH template (see Appendix \ref{APP_Stat_Model}). The orientation of the spatial dimension (parallel to the long axis of the slit) is denoted in the upper-right corner of the uppermost panel. Additionally, the dashed gray line is the high-S/N synthesized PSF (Appendix \ref{sec:PSF_model}), plotted for reference.}
         \label{Fig_2D_selectSources}
\end{figure*}

In Fig. \ref{Fig_2D_selectSources} we present our results of spatially resolved PAH emission of four spectral bands of five select targets. In the following, we discuss the results of each object individually.

\subsubsection{AB~Aurigae} \label{Disc_dist_ABAUR}

From Fig. \ref{Fig_1D_Spectra}, it is evident that PAHs across all bands contribute very little to the overall emission and that silicates (primarily olivine and pyroxene) dominate the emission, in agreement with \citet{Bouwman2000}, who used an Infrared Space Observatory (ISO) spectrum to determine the composition of the circumstellar material. Indeed, the silicates are so dominant in the integrated spectra that the PAHs are hardly seen, except for a slight peak at 11.3 $\mu$m. However, the spatially resolved spectra show weak PAH emission (primarily also of the 11.3 $\mu$m band) that peaks at some distance from the star on either side. 


The PAH intensity (Fig. \ref{Fig_2D_selectSources}) shows that its distribution across all bands exhibits a ring-like structure peaking relatively uniformly at $\pm$0\farcs3 and $\pm$0\farcs4 at the NE/SW sides, respectively, with varying widths. However, when observing the integrated spectrum (Fig. \ref{Fig_1D_Spectra}), the marginal flux of the PAH bands cannot be significantly determined due to the much-brighter silicate emission. This is in agreement with \citet{vanBoekel2005}, who reported that no PAH emission was detected in the integrated N-band spectrum of this object, as was observed with TIMMI2 on the 3.6 m telescope. As we are aware, we present the first-ever detected PAH emission from within this object.
Amorphous silicate emission is dominant in the central region close to the star, and no PAH emission is detected. Unlike in the case of HD~100546 (see Sect. \ref{Disc_dist_HD100546}), we detect no degeneracy between the amorphous silicate templates and the PAH templates (where here the dominant species are olivines).

\subsubsection{HD~97048} \label{Disc_dist_HD97048}

As the spectrum of this object is dominated by a featureless continuum with strong PAH emission on top, discerning this emission is not as challenging as it is for objects with substantial silicate emission (e.g., AB Aurigae, HD100546, and HD163296). Here, a few interesting features of the PAH emission are evident: (\textbf{1}) for all three bands of significant S/N (that excludes the 12.6~$\mu$m band), the PAH emission exhibits radial asymmetry, consistent with the SPHERE intensity profile and the derived projected geometry of the disk \citep{Lagage2006, doucet2007hd, vanderPlas2016}. (\textbf{2}) the intensity profiles of all bands detected at high S/N (mainly the 8.6~$\mu$m and 11.3~$\mu$m) exhibit substructure that is reminiscent of that seen in the SPHERE intensity profile, indicating that the PAH emission follows the gas density in the high disk regions as traced by the small dust grains, well coupled to the gas, that is seen in scattered light. (\textbf{3}) when considering Fig. \ref{Fig_1D_Spectra} and Fig. \ref{Fig_2D_selectSources}, it is evident that for this object, the 11.3~$\mu$m feature becomes relatively strong compared to the 7.9~$\mu$m and 8.3 $\mu$m bands at larger distances from the star. This, in turn, implies that the fraction of PAHs with neutral charge increases with increasing radial distance; that is, the ionization fraction is the highest close to the star; this is consistent with the observations and analysis by \citet{Maaskant2014}.

\subsubsection{HD~100546} \label{Disc_dist_HD100546}

In this object, we observe clear differences in the spatial intensity profiles between the different PAH bands. The most prominent example is the 11.3~\mum \ feature, which is centrally peaked whereas the other three bands show a ring-like profile, albeit with different radial decay rates (see Fig.~\ref{Fig_2D_selectSources}). At first sight, this is puzzling, and cannot be explained with, for example, a specific spatial distribution of neutral and ionized PAHs: the emission in both the 11.3 and 12.6~\mum \ band is dominated by neutral PAHs, so they should show a similar behavior if this is the underlying effect governing the observed profiles. However, the 12.6~\mum \ band shows a ring-like structure, as do the 7.9 and 8.6~\mum \ bands.


We suspect that the apparent deviant behavior seen in the 11.3~\mum \ band occurs due to methodological rather than astrophysical reasons. The 11.3~\mum \ band spectrally overlaps with a strong emission feature of small forsterite grains that is both of similar wavelength and spectral width. If the template of forsterite we used in our spectral decomposition is imperfect (i.e., its spectral shape does not exactly match that of the astrophysical grains), the minimization procedure may compensate for this by adding 11.3~\mum \ PAH emission. We have two sources in our sample in which forsterite is abundant, HD~100546 and HD~163296. In both these (and only in these), our spectral decomposition yields a strongly centrally peaked 11.3~\mum \ PAH profile, whereas all other PAH bands show a ring-like structure and are depleted near the central star. While from such a small number of stars, no hard conclusion can be drawn, this observation is consistent with our hypothesis that in disks rich in forsterite, our spectral decomposition mistakes the crystalline silicate emission close to the star for 11.3~\mum \ PAH emission.




In their analysis of properties of forsterite in HD~100546, \citet{Mulders2011} find forsterite to be the most abundant of the crystalline silicates in HD~100546 and find it to be primarily concentrated in the central $\sim$20~au vicinity around the star. In addition, they find a grain-size distribution of 0.1-1.5~$\mu$m to fit the observed emission features optimally. More importantly for our discussion, \citet{Juhsz2010} and \citet{Mulders2011} show that the 
central wavelength
The emission feature at 11.3~$\mu$m exhibits a shift of several percent as a function of grain-size distribution and can thus degenerate with the 11.3~$\mu$m PAH feature. They furthermore demonstrate this dependence on the central wavelength 69~$\mu$m emission feature (e.g., iron content and temperature); however, in the case of extensive laboratory experiments, \citet{Zeidler2015}, who explored the temperature dependence of such emission spectra, demonstrate that they exhibit very little spectral shift for the temperatures that we expect (on the order of 200 K).
In the integrated spectrum, we find that the contribution of the 11.3~$\mu$m PAH emission band is approximately 2\% in the relevant spectral window. Thus, a systematic discrepancy in the fitting procedure of the 11.3~$\mu$m PAH feature might have occurred because our forsterite templates fail to reproduce the correct shape of its emission since our model considers only two distinct grain sizes. In contrast, more should be considered for a robust fit \citep{Mulders2011}. This would then lead our fitter to systematically compensate the difference in the spectral shapes of the forsterite emission and our templates by increasing the 11.3~$\mu$m PAH emission -- which is unlikely to be detected in the posterior distribution analysis.
Thus, we decide not to consider the 11.3~$\mu$m PAH contribution in the unresolved center of the object as a solid detection.


The three remaining bands do not exhibit such proximity to crystalline emission features and exhibit a relatively consistent asymmetric ring-like structure of the spatial PAH intensities. In   Sect. \ref{Disc_SPHERE_HD100546}, we compare our derived PAH intensity profiles with that of the SPHERE image to characterize it further. Here, the brighter side in these bands coincides with the far side of the disk, as expected from the larger projected disk surface area visible in the SPHERE images (Fig.~\ref{Fig_SPHERE_allSources}). 

Finally, in their work, \citet{Maaskant2014} find that gas flowing through disk gaps can contribute significantly to the observed ionized PAH emission. This could manifest as an increase in emission in the 7.9~\mum \ and 8.6~\mum \ (approximately ionized) bands, relative to the 11.3~\mum \ and 12.6~\mum \ ($\sim$neutral) bands, corresponding to the angular size of a gap. Here, we find that while the two bluer PAH bands exhibit considerably different spatial distribution, they are wider than the 12.6~\mum \ bands -- the only one probing the predominantly neutral PAH emission. That said, their respective widths also vary significantly and do not seem to represent the distribution of ionized PAHs identically. A more informative parameterization of the radial behavior of the charge state of PAHs may be through {ratios} of PAH band intensities, as discussed in \citet{Maaskant2014}, rather than the individual intensities themselves. We hope to address the probing of the spatial distribution of the charge state of PAHs in future work.

\subsubsection{HD~163296} \label{Disc_dist_HD163296}

In this object, we notice a discrepancy similar to that in HD~100546 (Sect. \ref{Disc_dist_HD100546}), where the 11.3~\mum \ PAH band profile is centrally peaked, whereas the two asymmetric bluer bands exhibit a ring-like shape. However, the bluer bands in HD~163296 exhibit a roughly similar radial decay.
The total flux of the PAH bands is considerably lower (by a factor of $\approx$4) than that of HD~100546, and the profiles are noisier, especially the 12.6~\mum \ band that is of too low S/N to constrain significantly. We suspect that here, too, the centrally peaked shape of the 11.3~\mum \ PAH profile is caused by a degeneracy with the forsterite templates and does not represent the real intensity profile.
In the case of the two bluer bands, the asymmetry in the ring-like lobes of the intensity distribution is in agreement with the projected geometry seen in the SPHERE image (Fig.~\ref{Fig_SPHERE_allSources}). 

This is interesting in light of recent work by \citet{dit2021new}, who had also observed this object with VISIR-NEAR (albeit in the imaging mode) through two different N-band filters, where one corresponds to the 8.6~\mum \ PAH band (PAH1 filter, centered at $\lambda_0 = 8.58$~\mum \ with $\Delta \lambda = 0.41$~\mum) and the other contains the 11.3~\mum \ PAH band (NEAR filter, centered at $\lambda_0 = 11.25$~\mum \ with $\Delta \lambda = 2.25$~\mum). They found, however, that in both cases--the bulk of the intensity is contained within an unresolved inner region of the disk. This is probably because of the much brighter silicate emission originating from this unresolved inner region, which also dominates much of the N-band spectral range (see Fig. \ref{Fig_1D_Spectra}) and is difficult to decouple from the PAH emission without spectral decomposition. In our work, the indication that the silicate emission originates in an unresolved inner disk region is also manifested in the aforementioned degeneracy with the 11.3~\mum \ PAH band. Thus, by probing the individual PAH bands, we can resolve the disk and trace prominent substructure features better spatially.



\subsubsection{HD~169142} \label{Disc_dist_HD169142}



In most sources, the emission ("along-slit" intensity) profile in the continuum is consistent with strongly centrally peaked emission broadened by the PSF. The observed profile of HD~169142 is broader and more triangular. It suggests overall more spatially extended but less centrally peaked emission, marginally resolved and broadened by our PSF. We qualitatively test this scenario, inspired by the SPHERE image that shows a bright ring of $\approx$0\farcs17 in radius \citep{Bertrang2018}, and with a radial width of roughly one-quarter of the radius.

\begin{figure}
  \centering
  \includegraphics[angle = 0, width=0.48\textwidth,trim=35 15 18 5,clip]{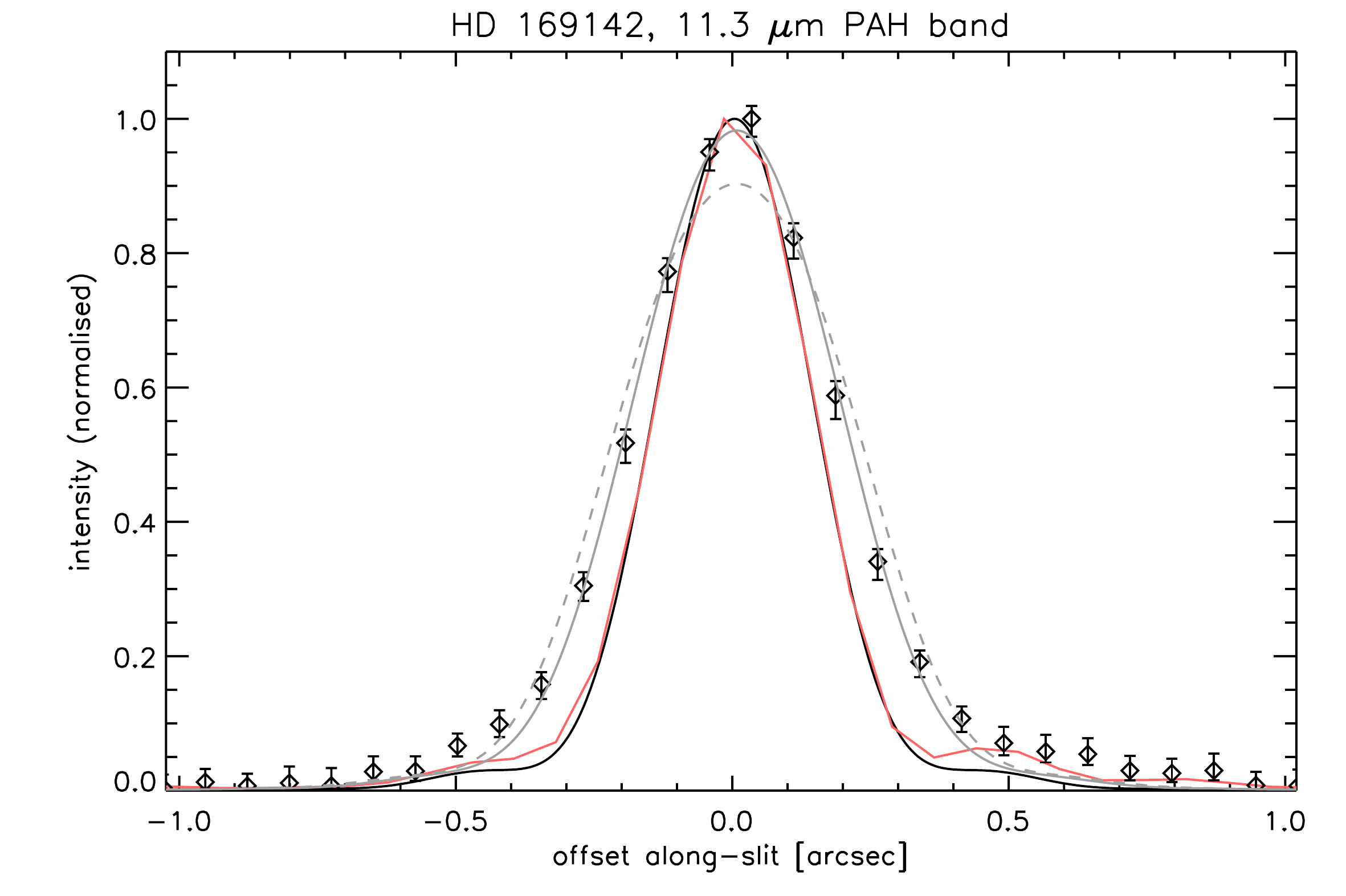}
    \caption{Spatial emission profile of HD~169142 in the 11.3~$\mu$m PAH band. The red \color{black} curve shows the observed PSF profile (see Appendix~\ref{sec:PSF_model}), and the {black} curve shows the employed approximation thereof used in the 2D convolution of the model (see Sect. \ref{Disc_dist_HD169142}). The  {solid gray} \color{black} curve shows the ring model that best fits the observed central peak (ring radius 0\farcs14). The  {dashed gray} \color{black} curve shows the best-fit model with the ring radius fixed to the value observed in the SPHERE data (0\farcs17).}
    \label{Fig_HD169142_Ring}
\end{figure}

We model the intrinsic emission profile as a simple ring of radius $R$ and a radial width of $R/4$ of uniform intensity. This mimics the profile observed with SPHERE and is a reasonable choice for the functional form of the intensity distribution. Still, we do not spatially resolve the width of the ring in the VISIR data, and other choices of parametrization may be equally valid. The situation is simplified by the system's nearly perfectly face-on orientation; therefore, we do not incline the ring. We convolve the model image with a model PSF and project it onto the 0\farcs4 VISIR slit. Then, we integrate over the across-slit dimension to obtain a spatial, along-slit profile of our model that can be combined directly with the observed spatial profile. We then optimize the ring radius $R$ to match the observed profile best. This is illustrated in Fig.~\ref{Fig_HD169142_Ring}. As a model PSF, we use the following approximation: we take a perfect diffraction-limited PSF of a circular aperture with a diameter of 8.2~m and a central obscuration of 1.2~m, following the geometry of the VLT unit telescopes. We slightly broaden and "soften" the PSF by convolving it with a Gaussian kernel of width $\sigma=$~0\farcs075 to account for the PSF being close to but not quite diffraction-limited. The kernel width was chosen so that the synthetic PSF profile (black curve in Fig.~\ref{Fig_HD169142_Ring}) best matches the core of the observed PSF profile (red curve in Fig.~\ref{Fig_HD169142_Ring}). We see that the central core of the model PSF matches the observed one nearly perfectly (note that the model curve has a much denser spatial sampling), but the in the wings, the observed curve has slightly more power. We use this approximation because we need to convolve the 2D model image with the PSF before projecting it onto the slit. Still, we do not directly measure the PSF in our long-slit spectroscopy data. Since the current discussion is qualitative only, we deem this sufficient.

The ring radius for which the modeled profile best matches the observations is 0\farcs14 (gray curve in Fig.~\ref{Fig_HD169142_Ring}); here shown for the 11.3~$\mu$m \ PAH band. The model curve reproduces the observed profile within the observational uncertainties out to a radius of $\approx$0\farcs35 on either side of the center. In contrast, at larger radii, the observed emission levels are somewhat higher than those given by the model. This is partly due to the true PSF having more power than the PSF model and, in part, to the source showing low-intensity PAH emission beyond the ring in our simplified model. 

This experiment shows that the observed PAH emission profile in HD~169142 can indeed have a marginally resolved ring-like shape, plus some extra lower intensity emission at larger radii. There is no dominant contribution from a centrally peaked source. Interestingly, the best-fit ring radius of $R=$~0\farcs14 is smaller than the one observed in near-infrared scattered light with SPHERE ($\approx$0\farcs17) and also smaller than the continuum size of $\approx$0\farcs16 estimated from Q-band observations (18.8 and 24.5~$\mu$m) by \cite{2012ApJ...752..143H}. In Fig.~\ref{Fig_HD169142_Ring}, we also show the best-fit profile for a ring with $R=$~0\farcs17 with the gray, dashed curve. This model yields a worse match and is not consistent with the observations. 

Thus, the emerging picture is that the central disk gap in HD~169142 is mostly devoid of PAH emission. The transition from the gap to the optically thick disk region is smooth, with gradually increasing gas density. In such a configuration, the PAH emission traces the lowest density surface layers of the inner wall of the optically thick disk, analogous to the profiles shown in Fig. \ref{Fig_scaleHeight_tracers}. We expect the PAH emission to trace layers similar to the highly optically thick tracer $^{12}$CO. Indeed, recent ALMA observations show a $^{12}$CO profile peaking around 0\farcs12 \citep{2021ApJ...920L..33Y}, which is at a slightly smaller radius than our best-fit value for the PAH ring.

This picture agrees qualitatively with the study's results by \cite{2022A&A...663A.151D}, who analyzed archival observations of HD\,169142 taken with VLT-Nasmyth Adaptive Optics System - Near-Infrared Imager and Spectrograph (NACO)(AO-supported) and VLT-VISIR (seeing-limited but with image quality close to diffraction-limited). Using radiative transfer modeling, they find that the PAH intensity must be strongly suppressed close to the central star. Their observations can both be reproduced with an inner cavity (radius 0\farcs17 $=$ 20~au) that is entirely devoid of PAH emission (i.e., all PAH emission arises from the optically thick disk part beyond 0\farcs17) and with a model where PAHs are still present in the 5$-$20~au region but at a much-reduced abundance compared to the region beyond 20~au. Our simple model, which has all PAH emission confined to a narrow radial region by design, would favor the latter option since a model in which the PAH intensity inside 20~au is zero would yield a broader profile than observed (see Fig.~\ref{Fig_HD169142_Ring}). We also observe PAH emission at much larger radii, which is not included in our simple model, as is evident in Fig. \ref{Fig_HD169142_Ring} where the measured intensities are higher than the modeled ones.

Finally, \cite{dit2021new} compared the radial intensity profiles of this object, as observed through the PAH1 and NEAR spectral filters (see Sect. \ref{Disc_dist_HD163296}) with ProDiMo-generated models at these bands \citep{woitke2009radiation}. While the ProDiMo model predicted a roughly similar apparent size for both bands (43 and 45 AU, respectively), they find that the object's apparent size, as observed through the PAH1 filter, is considerably smaller (24$\pm$1 AU). This agrees with observations by \citet{Okamoto2017}, who attribute $>$9~\mum \ emissions to the inner wall of the disk, rendering the apparent size larger. They suggest a potential explanation for this discrepancy based on the results of \citet{Maaskant2014} (see Sect. \ref{Disc_dist_HD100546}): it is possible that gas flowing through the disk gap could manifest in increased emission at the 8.6~\mum \ PAH band relative to the 12.6~\mum\ PAH band, corresponding to the angular size of the gap. Assuming that the neutral PAH emission is centered in the gap wall, one would expect a more compact apparent size seen through the PAH1 filter instead of the NEAR filter. Since the ProDiMo model does not account for this difference, this leads to the aforementioned discrepancy. 

In this work, however, we find that the object's apparent size is identical across all four PAH bands to within $1\sigma$, such that we do not seem to identify the different spatial distributions of PAHs as a function of their charge state. Therefore, we suspect that the effect detected by \citet{dit2021new} may reflect a difference in continuum emission (as suggested by \citet{Okamoto2017}), implicitly integrated within the filter apertures (especially the wider NEAR filter), rather than a difference in the distribution of ionized versus neutral PAHs.

\FloatBarrier

\section{Discussion}
\label{Discussion}



\subsection{Spatial PAH intensity profiles: Ring-like versus centrally peaked}
\label{Disc_SpatialDist}


Of the five sources in Fig. \ref{Fig_2D_selectSources}, three (AB~Aurigae, HD~100546, HD~163296) have a well-resolved ring-like geometry in the PAH emission. Interestingly, these all have strong emission features from (sub)micron-sized silicate grains. Two sources (HD~97048, HD~169142) show more centrally peaked PAH emission, and these both do not show silicate emission; in HD~169142, the PAH emission may also have a ring-like geometry but with a much smaller radius than in the "silicate" sources (Sect. \ref{Disc_dist_HD169142}). While the number of sources is too small (and their diversity in properties too large) to draw any firm conclusions, we may establish that: (\textbf{1}) the spatial structure of the disks, in particular radial gaps, causes differences in the emergent spectra by excluding specific temperature regimes (see Fig. \ref{Fig_2D_selectSources}). Our targets have diverse geometries, and all but HD~163296 have large gaps in their disks' inner $\approx$20~\AU \ (see Fig. \ref{Fig_SPHERE_allSources}). However, substructure alone cannot account for the observed dichotomy of spatial PAH emission profile regimes, as, for example, HD~97048 and HD~100546 are fairly similar in disk substructure (i.e., both have an inner disk then a large radial gap and an outer disk; e.g., \citealt {Ginski2016, Mendiguta2017}, respectively) but exhibit the opposite behavior of the PAH profiles. (\textbf{2}) The correlation between the geometry of the PAH emission and the presence of small silicate grains is suggestive. With that in mind, potential explanations are as follows.

First, small silicate grains are absent in the surface layers of HD~97048 and HD~169142 due to grain growth and settling. This has two observational consequences: (\textbf{1}) spectral features of small silicates naturally disappear, and (\textbf{2}) PAHs can absorb more UV photons for lack of competition from small silicate grains, thus increasing the brightness of PAH emission features \citep{dullemond2004effect}. In the other three targets, small silicate grains remain in the inner disk regions, absorbing stellar radiation and providing much surface area for PAHs to coagulate onto, thereby dimming PAH emission from the inner disk regions. Why the small silicate grains have disappeared from the inner disk regions in two sources but not in the other three remains an open question.

Second, the composition of the inner disks of HD~97048 and HD~169142 may be intrinsically different (more carbon-rich; \citealt{van2002nanodiamonds, devinat2022radial}) from the other objects and as such may contain predominantly carbonaceous dust and little or no silicates. However, the required depletion of oxygen as a leading-order explanation for the lack of silicate emission within the inner regions of disks where we detect a ring-like PAH emission profile appears chemically unlikely. The bulk composition of the disk material is unknown, but the starting material (interstellar medium) is expected to be oxygen-rich \citep[C/O\,$\approx$\,0.52][]{anders1989abundances}. The photospheres of both stars do not suggest a strong depletion in oxygen in the accreted material (for HD~169142,  C/O\,$\approx$\,0.68 \citep{folsom2012chemical}, for HD~97048 we could not find a carbon abundance in the literature, but the oxygen abundance is roughly solar \citep{acke2004chemical}. If anything, carbon rather than oxygen is expected to be depleted in the solid phase in the hot inner disks, as was strongly the case in the solar nebula. 
 
 We note that in the sample of \cite{folsom2012chemical} HD~169142 has a somewhat higher than average photospheric C/O ratio, but that of AB~Aurigae is even higher (approaching unity). That disk has one of the strongest 10~micron silicate features known. In summary, attributing the absence of a silicate feature to an intrinsically oxygen-depleted dust population containing little or no silicates appears currently not supported by observations.


\color{black}


Finally, PAHs may be depleted in the inner disks due to coagulation onto other dust. In their work, \citet{geers2009lack} find that the absence of PAH emission can be explained by the depletion of PAHs in the early phases of star formation, by being trapped in ices and or coagulating with dust or other PAHs. This process likely continues well into the class II phase, where an abundance of small dust grains in the disk to coagulate exists. Therefore, one would expect such depletion to manifest in inner regions of disks (where dust densities are highest), where small silicate grain emission is observed.

\subsection{Comparison with SPHERE intensities} \label{Disc_SPHERE}

We then performed a qualitative comparison of the derived PAH intensities for select sources (Sect. \ref{Results_dist_objectwise}) with a 1D intensity profile of the SPHERE images (Fig.~\ref{Fig_SPHERE_allSources}). This comparison is worthwhile because both these observations probe the uppermost vertical layer of the disk essentially, albeit different components thereof, which might adhere to different physics \citep[e.g.,][]{Lagage2006, Ginski2016}. Another critical difference between the two is the inherent modulation of the intensity distribution of polarized radiation due to the projected geometry of the disk. Similarly, the PAH intensity profile is expected to scale with the amount of UV radiation reaching the disk surface; this would also yield a $1/r^2$ intensity profile, modulated by disk structure effects and spatial variations in the abundance and molecular properties of the PAHs.

To compare the two, we generated 1D intensity profiles from the SPHERE images as follows: first, we overplotted the proper dimensions and orientation of the VISIR long-slit as illustrated in Fig.~\ref{Fig_SPHERE_allSources}. We proceeded to average the intensity of all pixels within the slit, parallel to its long-axis, to receive the average intensity of the SPHERE image along the long-axis of the slit, corresponding to the spatial dimension of our 2D spectra (Sect. \ref{Data_Processing}). We then normalized and convolved the resulting intensity profile with a high-S/N VISIR PSF (see Appendix \ref{sec:PSF_model}) at the wavelengths of the respective PAH bands for comparison. Finally, we scaled the intensity profile with $\vec{r}^2$, assuming that scattered-light intensity decays as an inverse square law (except for the polarized light intensity modulation caused by projected geometry) with the stellar radiation field. This scaling is only approximate as it ignores inclination effects and is strictly accurate only for disks seen close to face-on. We note that the SPHERE observations have been performed with an apodized Lyot coronagraph with a radius of 93~mas. Therefore, any scattered light arising at smaller radii will be strongly suppressed in the SPHERE data. Hence, we may underestimate the scattered light intensities within approximately the central resolution element of our VISIR data for the 7.9 and 8.6~\mum \ PAH bands and about the central 2/3 of a resolution element for the 11.3 and 12.6~\mum \ bands.


In Figs. \ref{Fig_PAHvsSPHERE_ABAUR}-\ref{Fig_PAHvsSPHERE_HD169142} it is evident that overall--PAH intensities decrease less steeply with radius than the scattered light profiles. In their work on HD~163296, \cite{muro2018dust} compare the strength of scattered light emission of the rings to ALMA images, and find that its outer ring is fainter than expected, and potentially attribute it to shadowing or settling. This also suggests that the scattered light intensity might show a different radial dependence than the diluted stellar radiation field. This observation, coupled with the results of \cite{dullemond2004effect}, may indicate that in all objects in our sample, dust-settling or shadowing effects may cause the discrepancy in the spatial scattered light and PAH intensities.

Another potential explanation for this discrepancy is that in regions where the different PAH/scattered-light profiles decay rates are observed, PAH molecules are saturated by incoming exciting stellar radiation (i.e., their cooling timescales are comparable to- or longer than- the photon absorption rate), in which case the PAH emission is not coupled to the diluted stellar radiation field, but instead maintains its decay power-law \citep{Draine2020}. However, if the PAH excitation is in the single photon regime (i.e., the average time between the absorption of individual exciting stellar photons is long compared to the time over which the PAH molecule cools through the emission of infrared radiation), then the intensity is expected to be directly proportional to the local stellar radiation field, and the spectral shape is the same everywhere (potentially modulated by disk structure and spatial variations in PAH abundance and molecular properties) \citep{Draine2001, Li2001}. 

To probe the relevance of this effect to the observed discrepancy, we model the heating and cooling processes of C$_{60}$H$_{20}$ and its cation (C$_{60}$H$_{20}^{+}$) in the disk around AB Aurigae at different radial distances from the central star. The specific PAH molecule C$_{60}$H$_{20}$ is selected since it represents the typical PAH in protoplanetary disks around Herbig Ae/Be stars \citep{Seok2017}. In Fig.~\ref{Fig_Li_PAH_ABAUR} we plot the temperature probability distribution functions and emission spectra of C$_{60}$H$_{20}$ and C$_{60}$H$_{20}^{+}$ for two nominal projected radial distances from the central star of AB Aurigae, which are approximately the innermost and outermost disk regions of each object, probed by VISIR-NEAR. In Sect. \ref{Disc_SPHERE_ABAUR}-\ref{Disc_SPHERE_HD169142}, we discuss the comparison between the PAH and scattered-light profiles, as well as the result of this simulation for each object individually. \color{black} We plot the temperature probability and emission spectra for the remaining four objects in Appendix \ref{APP_Li plots}. \color{black}

We note that while additional considerations are likely necessary to compare the two thoroughly, our purpose here is to qualitatively examine whether or not prominent morphological features (e.g., projected asymmetry, substructure) can be identified in both and address large-scale discrepancies, as both diagnostics trace similarly high vertical layers in the disk. We find the morphological comparison of the scattered light and PAH intensity profiles to be visually most evident if we scale the former by $\vec{r}^2$, but not the PAH profiles, though the visual match between the profiles is not equally good in all sources.

\subsubsection{AB Aurigae} \label{Disc_SPHERE_ABAUR}

The SPHERE image of this object (Fig.~\ref{Fig_SPHERE_allSources}) makes it readily clear that its circumstellar disk is extremely rich in substructure, spanning across a range of different scales, and is also azimuthally asymmetric. Therefore, a 1D representation of its intensity distribution does not represent its complexity, and indeed blurs some of the subtler spatial features that fell within the slit. The rather small inclination of the disk (30$^\circ$ according to \citet{Boccaletti2020}) negates the differences in the polarized intensity (PI). During observation, the slit was positioned roughly along the minor axis of the disk (i.e., the axis of minimal PI but maximal asymmetry).

\begin{figure*}[htbp]
\begin{subfigure}[a]{\textwidth}
\centering
\includegraphics[scale = 0.8]{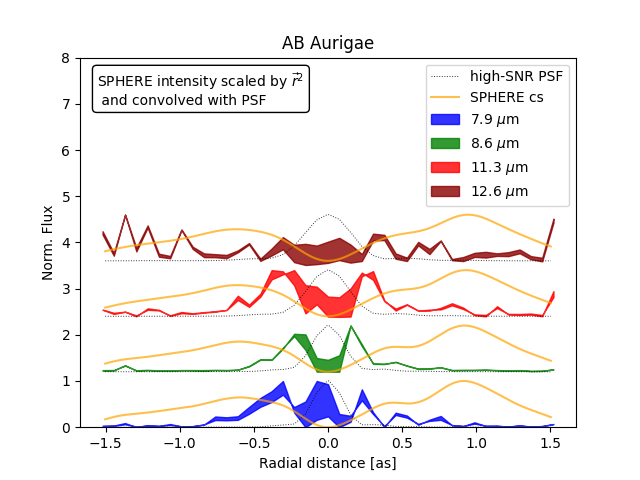}
\caption{}
\label{Fig_PAHvsSPHERE_ABAUR}
\end{subfigure} \\
\begin{subfigure}[b]{\textwidth}
\centering
\includegraphics[scale = 0.8]{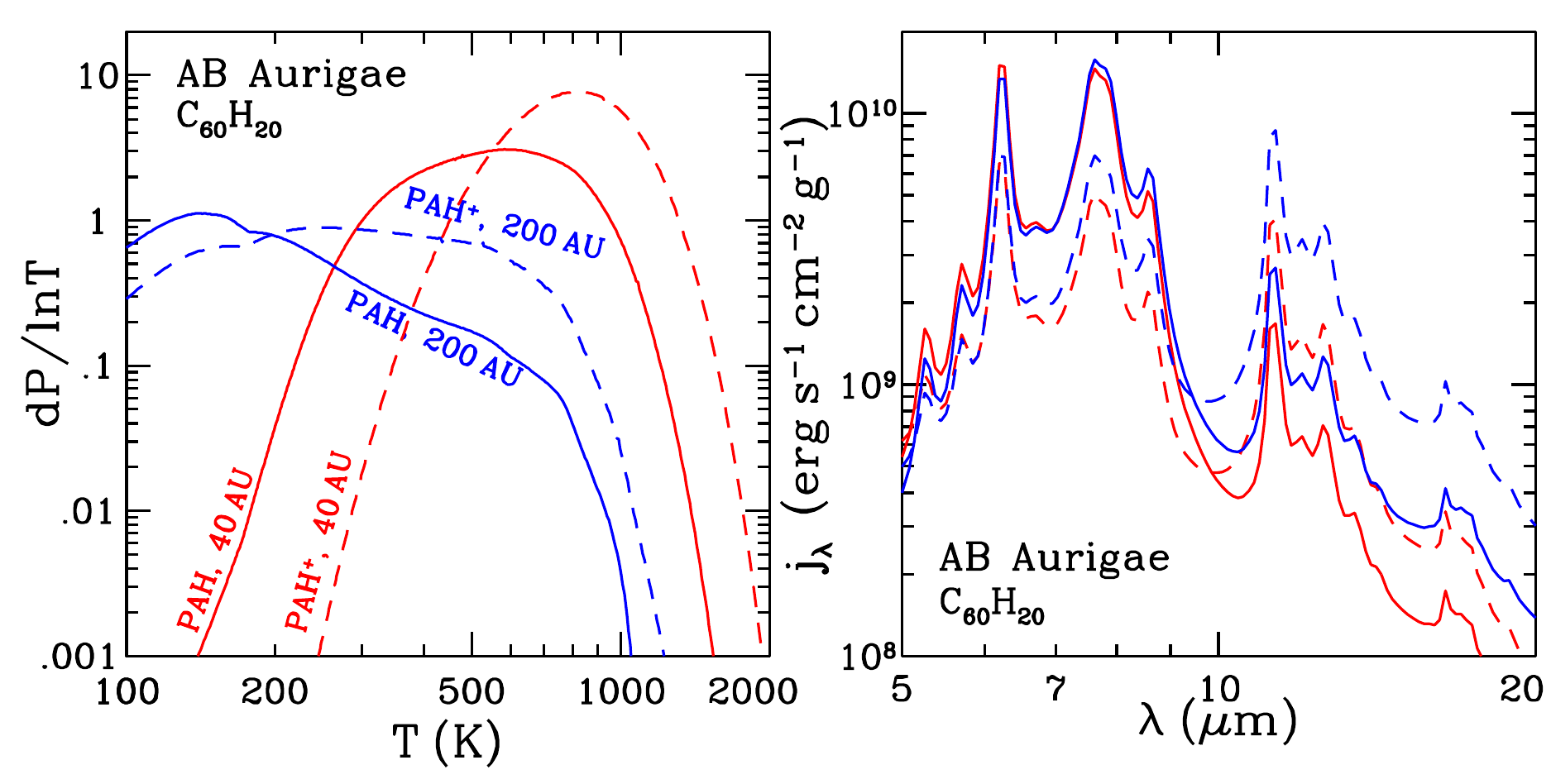}
\caption{ }
\label{Fig_Li_PAH_ABAUR}
\end{subfigure}
\caption{PAH intensity profiles and model calculations for AB~Aurigae. Panel (a): Comparison of the normalized radial profiles of the three prominent PAH bands (7.9~$\mu$m, 8.6~$\mu$m \color{black} and 11.3 $\mu$m, \color{black} and 12.6~$\mu$m \color{black}, from bottom to top, respectively) with a radial intensity profile of the reduced SPHERE image (Fig. \ref{Fig_SPHERE_allSources}). Panel (b): Temperature probability distributions (left panel) and emission spectra (right panel) of neutral C$_{60}$H$_{20}$ (solid lines; labeled ``PAH'') and its cation (dashed lines; labeled ``PAH$^{+}$'') for two nominal distances: 0\farcs228, which corresponds to $\approx$40AU from the central star (red lines), and 1\farcs14, which corresponds to $\approx$200AU from the central star (blue lines), respectively, in the innermost and outermost disk regions probed by VISIR-NEAR. The temperature probability distribution functions for C$_{60}$H$_{20}$ and C$_{60}$H$_{20}^{+}$ are broad even for the innermost disk region, indicating that PAHs undergo stochastic heating by individual stellar photons even in the innermost disk region, where the photon absorption rate is $\sim$25 times more frequent compared to that in the outermost disk region. The stochastic-heating nature of PAHs in the disk around AB Aurigae is also indicated by the close resemblance of the model emission spectra calculated for C$_{60}$H$_{20}$ and C$_{60}$H$_{20}^{+}$ at 40AU (solid and dashed red lines) to that at 200AU, when scaled to the same radial distance, (solid and dashed blue lines).}
\end{figure*}

\begin{figure*}[htbp]
\centering
\includegraphics[scale = 0.8]{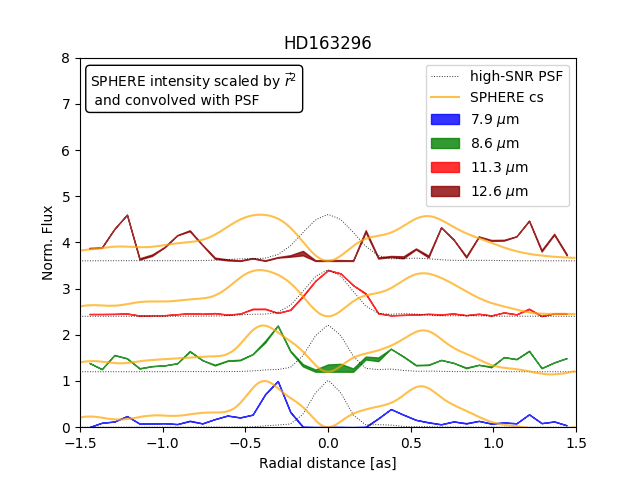}
\caption{Comparison of the normalized radial profiles of the three prominent PAH bands of HD~163296 (similar to Fig. \ref{Fig_PAHvsSPHERE_ABAUR}).}
\label{Fig_PAHvsSPHERE_HD163296}
\end{figure*}

In Fig.~\ref{Fig_PAHvsSPHERE_ABAUR} we compare all PAH bands with a $r^2$-scaled and convolved 1-D profile of the SPHERE intensity distribution contained within the slit (see Sect. \ref{Disc_SPHERE_ABAUR}). 
Unlike in cases where a comparison in this configuration yields the greatest similarity between the two intensity distributions (Sect. \ref{Disc_SPHERE_HD100546}), here the two are not alike. The ring-like PAH distribution evident in the three bluer bands is considerably more compact than the substructure manifested in the SPHERE profile. The slight asymmetry, regarding which both the PAH and the SPHERE profiles were consistent for other objects (Sects. \ref{Disc_SPHERE_HD97048}, \ref{Disc_SPHERE_HD100546}, and \ref{Disc_SPHERE_HD163296}), is equivocal in this case (despite the asymmetry being the smallest with respect to the aforementioned cases, with the exception of the nearly pole-on HD~169142).

As shown in Fig.~\ref{Fig_Li_PAH_ABAUR}, PAHs will be stochastically heated if present in the disk around AB Aurigae. Therefore, the observed discrepancy between the two profiles cannot be explained by a potential saturation state of the emitting PAHs and remains an open question. Comparison between scattered light and millimeter continuum imaging of AB Aurigae is made in \citet{tang2017planet} and \citet{currie2022images}.

\subsubsection{HD~97048} \label{Disc_SPHERE_HD97048}

The PAH emission of this object serves as the optimal case of our analysis. It is very well spatially resolved, its spectral quality is high and there is no silicate emission present to potentially hamper the robustness of the constraints placed on the PAH emission.

Here, the slit was once again positioned almost in parallel to the short axis of the rather inclined object ($\approx$41$^\circ$ according to \citealt{vanderPlas2016}). Such a positioning of the slit, while incidental, is optimal for our purposes, as it captures a considerable measure of the projected asymmetry of the object, thus allowing a comparison thereof between the SPHERE intensity profile and that of the PAHs.

Indeed, all bands (with the exception of the insignificantly constrained 12.6~$\mu$m band) exhibit a consistent behavior of centrally peaked emission with considerable spatial extent. Additionally, the spatial asymmetry and the consistent bump(s) located at $\approx$0\farcs5 (0\farcs75) in all three bands, are features to look for in the SPHERE intensity profile as well. We find PAHs to undergo stochastic heating in the disk around HD~97048, and therefore the radial profiles of the PAH emission bands (after being scaled by $r^{-2}$) directly reflect the spatial distribution of PAHs in the disk (see Appendix \ref{APP_Li plots}).

When considering Fig. \ref{Fig_PAHvsSPHERE_HD97048}, two important observations are made: (\textbf{1}) Both intensity profiles display identical asymmetry, where the right (far) disk side exhibits more prominent emission that decays more slowly. (\textbf{2}) The bump visible in the PAH profiles at $\approx$0\farcs5 is also present in the SPHERE profile, roughly at the same location, and represents an increase in  the surface density of the dust. One can argue, therefore, that here, PAH molecules' column density is coupled with that of the dust. The keen-eyed reader may notice that a second, albeit weaker, bump at $\approx$0\farcs75 may also be correlated, particularly visible at the stronger, 11.3~$\mu$m band.

This object presents an outstanding potential for a detailed analysis of the properties of PAHs (e.g., size distribution and charge state) as a function of distance from the star. We hope to explore this in future work. A comparison between scattered light and millimeter continuum imaging of HD~97048 is made in \citet[][]{van2017cavity}.

%



\begin{figure*}[htbp]
\centering
\includegraphics[scale = 0.8]{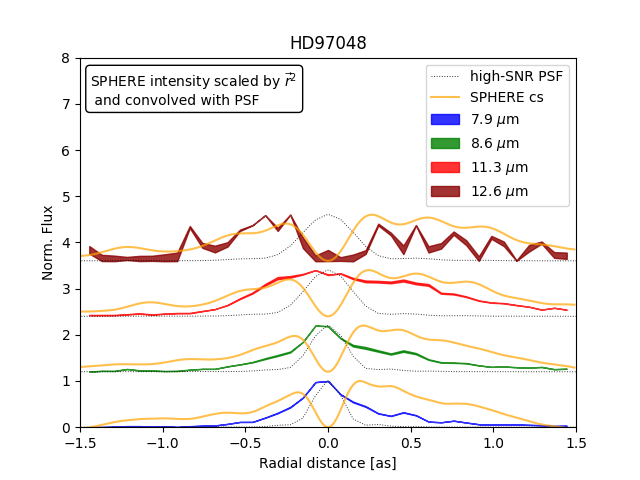}
\caption{Comparison of the normalized radial profiles of the three prominent PAH bands of HD~97048 (similar to Fig. \ref{Fig_PAHvsSPHERE_ABAUR}).}
\label{Fig_PAHvsSPHERE_HD97048}
\end{figure*}


\subsubsection{HD~100546} \label{Disc_SPHERE_HD100546}

This object displays the best agreement between the PAH intensity profiles and that of the $r^2$ -scaled SPHERE image (with the exception of the 11.3~$\mu$m band for reasons explained in 
Sect. \ref{Disc_dist_HD100546}).

Here too, similarly to AB Aurigae and HD~97048 (Sect. \ref{Disc_SPHERE_ABAUR}, \ref{Disc_SPHERE_HD97048}), the slit is positioned roughly in parallel to the minor axis of the inclined circumstellar disk ($\approx$41$^\circ$ according to \citet{Ginski2016}), and the minimal PI is compensated by the fact that the minor axis is in fact that of most interest to this comparison, as it potentially exhibits the greatest asymmetry of the projected disk.

\begin{figure*}[htbp]
\centering
\includegraphics[scale = 0.8]{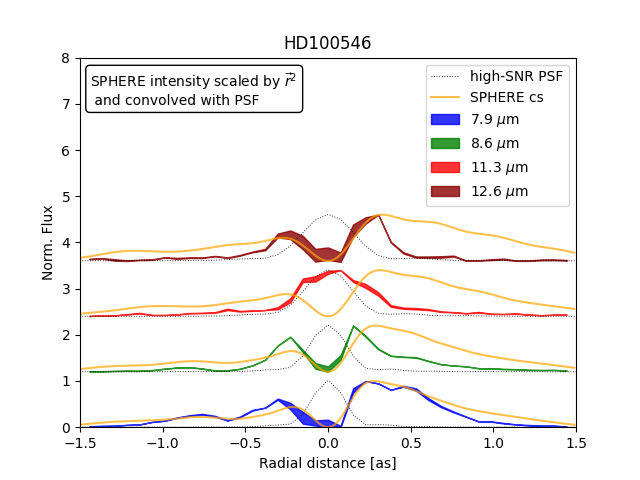}
\caption{Comparison of the normalized radial profiles of the three prominent PAH bands of HD~100546 (similar to Fig. \ref{Fig_PAHvsSPHERE_ABAUR}).}
\label{Fig_PAHvsSPHERE_HD100546}
\end{figure*}

When considering Fig.~\ref{Fig_PAHvsSPHERE_HD100546}, it is evident that the similarity between the PAHs and the SPHERE intensity is considerably greater than for AB~Aurigae and HD~97048. Indeed, the 7.9~$\mu$m nearly completely overlaps the SPHERE profile and reproduces the two smeared bumps extending along the left lobe of the profile until $\approx$-1\arcsec (albeit not perfectly). Surprisingly, it is also considerably more spatially extended than the 8.6~$\mu$m and 12.6~$\mu$m profiles, which one might expect to follow a similar spatial pattern (as seen in the previous two cases). The 8.6~$\mu$m and 12.6~$\mu$m bands, while more centrally centered and compact, still roughly match the inner region of the SPHERE intensity profile, although they decay spatially much faster.

This behavior may be reminiscent, for example, of the measure of coupling of a dominant charge state to the dust. Exploring this possibility further is, unfortunately, out of the scope of this study.

This object emphasizes best the peculiar observation that maximum similarity between the two intensity profiles is reached when the SPHERE profile is scaled by $\vec{r}^2$ while the PAHs are not (esp. for the 7.9~$\mu$m band). A comparison between scattered light and millimeter continuum imaging of HD~100546 is made in \citet[][]{pineda2019high}.



\subsubsection{HD~163296} \label{Disc_SPHERE_HD163296}

 The PAH profiles of this object are of relatively low S/N and allow only a limited qualitative comparison with the SPHERE profile. In this comparison, we only consider the 7.9~$\mu$m and 8.6~$\mu$m PAH bands, as the 12.6~$\mu$m is of too low S/N and is not constrained, and the 11.3~$\mu$m band may experience the same bias as in HD~100546 (see Sect. \ref{Disc_dist_HD163296}). 
 
To our advantage serves, again, the fact that the object is inclined by $\approx$46$^\circ$ with a position angle of $\approx$-88.8$^\circ$ \citep{diep2019}.\ As such, the slit was positioned nearly parallel to the minor axis of the circumstellar disk during observation, thus allowing us to probe the axis of maximal asymmetry. 


When considering Fig.~\ref{Disc_dist_HD163296} it is evident that while the PAH profiles are devoid of much detail, the asymmetry is detectable in both profiles, and is roughly in agreement concerning the position of the dominant ring-like structure. However, it is also evident that the SPHERE image is much richer in substructure extending farther from the star, whereas the PAH emission decays abruptly and we cannot detect spatial features beyond $\approx$0\farcs5 from the star. Unfortunately, the S/N of the PAH profiles does not allow for comparison with the scattered light profiles or investigation of spatial variations in the PAH properties; the two bluest bands, which we detect significantly, represent the same dominant charge state. As for the other stars, multiple-photon excitation is not important on the scales resolved by the VISIR-NEAR observations.

\color{black} A comparison between scattered light and millimeter continuum imaging of HD~163296 is made in \citet[][]{muro2018dust}. \color{black}



\subsubsection{HD~169142} \label{Disc_SPHERE_HD169142}

The PAH emission in this object appears centrally peaked close to the star, though the spatial profile is notably wider than the VISIR-NEAR PSF and has a distinct shape. In Sect. \ref{Disc_dist_HD169142} we showed that the PAH profile is consistent with a ring-like emission, analogous to that seen in scattered light, but with a smaller ring radius (see Figs.~\ref{Fig_SPHERE_allSources} and \ref{Fig_HD169142_Ring}).  

\begin{figure*}[htbp]
\centering
\includegraphics[scale = 0.8]{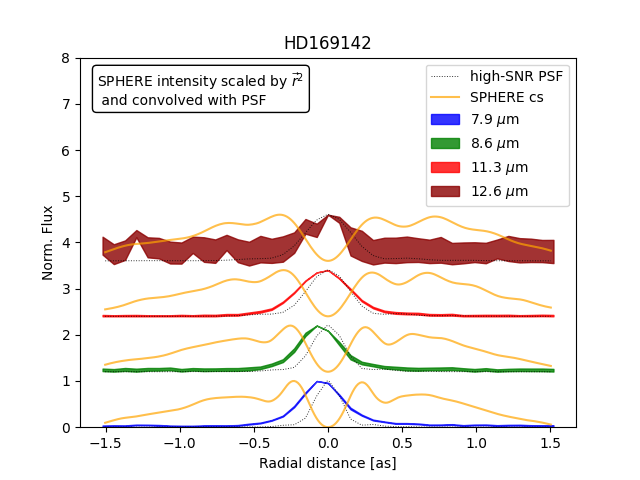}
\caption{Comparison of the normalized radial profiles of the three prominent PAH bands of HD~169142 (similar to Fig. \ref{Fig_PAHvsSPHERE_ABAUR}).}
\label{Fig_PAHvsSPHERE_HD169142}
\end{figure*}

Here, the circumstellar disk is nearly pole-on (with an inclination of $\approx$13$^\circ$ \citep{Macas2019}, so we expect essentially no asymmetry to be detected. Indeed, in Fig.~\ref{Fig_PAHvsSPHERE_HD169142}, we see that nearly all PAH emission ($>1\sigma$) is contained within the innermost cavity of the circumstellar disk. We find PAHs to be stochastically heated in the inner disk and emit prominently in the 6.2, 7.7, 8.6, and 11.3\,$\mu$m bands (see Appendix \ref{APP_Li plots}). 

\color{black} A comparison between scattered light and millimeter continuum imaging of HD~169142 is made in \citet[][]{Macas2019, Prez2020}. \color{black}




\section{Summary} \label{Summary}

We conducted an AO-supported N-band long-slit spectroscopy survey of eight protoplanetary disks and spatially resolved the PAH emission in five objects: AB~Aurigae, HD~97048, HD~100546, HD~163296, and HD~169142. Using a nonlinear optimization routine, we fit the spatially resolved spectra, which we parameterized with a continuum component, five silicate species of two nominal grain sizes, and four PAH bands (7.9, 8.6, 11.3, and 12.6~$\mu$m). For AB~Aurigae, the three bluer bands exhibit a ring-like emission with a radius of roughly 0\farcs3. For HD~97048, we find an asymmetric extended, centrally peaked emission with varying decay rates for each of the three bluer PAH bands. For HD~100546, we find a central cavity in the emission profiles of all PAH bands except for the 11.3~\mum \ band, where a degeneracy exists between the emission in the PAH feature and that of the abundant crystalline forsterite, leading to a false-positive centrally peaked intensity profile. For HD~163296, we similarly find that the extracted 11.3~$\mu$m PAH profile is centrally peaked and possibly confused with forsterite emission, whereas the two bluest bands exhibit a roughly identical ring-like structure. For HD~169142, we observe a widened centrally peaked uniform PAH emission across all four bands, which is consistent with a ring-like geometry with a radius that is somewhat smaller ($\approx$0\farcs41) than that of the ring observed in scattered light (0\farcs17).


Within our small sample, we find that in sources with strong silicate features, the PAH emission appears to be ring-like (AB~Aurigae, HD~100546, and HD~163296), whereas, in sources without silicate emission, the emission tends to be centrally peaked (most notably in HD~97048), which can be due to the competition of the silicate grains and PAHs over UV photons (and, subsequently, a considerably stronger emission feature of the first), dust-settling and shadowing effects, and/or the coagulation of PAHs on the small dust grains that are abundant in the inner regions of disks with strong silicate emission.

Finally, we compared the extracted spatial profiles of the PAH emission of each object with a 1D intensity profile of a scattered light image obtained with SPHERE, as the two probe the highest vertical layers of the disk and are therefore worthwhile to compare. The main conclusions of this analysis are as follows:

\begin{itemize}
    \item PAH emission arises from the highest disk surface layer (similar to $^{12}$CO). Scattered light observations trace a somewhat deeper layer, but one that is still spatially close to the PAH emitting layer.
    \item Naively, we expect both the scattered light and PAH intensity profiles to dilute approximately as $r^{-2}$ with the stellar radiation field. This is modulated by inclination, disk substructure, and spatially variable dust-settling and shadowing effects.
    \item We find that the PAH intensities decrease radially less steeply than the scattered-light profiles. We considered two potential reasons for this: (\textbf{1}) We investigated through modeling whether it could be due to the PAH excitation being saturated in the inner disk regions.\ However, we find that, on the scales probed by our observations, PAHs undergo stochastic heating by individual stellar photons even in the innermost disk regions in all objects in our sample. (\textbf{2}) Dust-settling and shadowing effects impact dust grains more so than PAHs, thus causing a steeper radial intensity decay of the former. Objects with a ring-like PAH intensity profile exhibit richness in silicate emission, whereas the centrally peaked HD~97048 lacks significant silicate emission. 
    \item Despite the difference in intensity decay, steepness, scattered-light, and PAH intensity profiles agree on prominent projected and intrinsic geometric features, such as inclination effects, significant gaps, and rings.
\end{itemize}

\FloatBarrier

 \begin{acknowledgements}

We thank Dr. Christian Ginski for the SPHERE images and the anonymous referee for offering valuable input concerning the editing process of this work.

 \end{acknowledgements}
 

%
  \bibliographystyle{aa} 
  \bibliography{myref} 

\begin{thebibliography}{123}
\expandafter\ifx\csname natexlab\endcsname\relax\def\natexlab#1{#1}\fi

\bibitem[{Acke \& Waelkens(2004)}]{acke2004chemical}
Acke, B. \& Waelkens, C. 2004, Astronomy \& Astrophysics, 427, 1009

\bibitem[{Anders \& Grevesse(1989)}]{anders1989abundances}
Anders, E. \& Grevesse, N. 1989, Geochimica et Cosmochimica acta, 53, 197

\bibitem[{Andrews {et~al.}(2018)Andrews, Huang, P{\'{e}}rez, Isella, Dullemond,
  Kurtovic, Guzm{\'{a}}n, Carpenter, Wilner, Zhang, Zhu, Birnstiel, Bai,
  Benisty, Hughes, \"{O}berg, \& Ricci}]{Andrews2018}
Andrews, S.~M., Huang, J., P{\'{e}}rez, L.~M., {et~al.} 2018, \apj, 869, L41

\bibitem[{Bae {et~al.}(2021)Bae, Teague, \& Zhu}]{bae2021observational}
Bae, J., Teague, R., \& Zhu, Z. 2021, The Astrophysical Journal, 912, 56

\bibitem[{{Benisty} {et~al.}(2022){Benisty}, {Dominik}, {Follette}, {Garufi},
  {Ginski}, {Hashimoto}, {Keppler}, {Kley}, \& {Monnier}}]{2022arXiv220309991B}
{Benisty}, M., {Dominik}, C., {Follette}, K., {et~al.} 2022, arXiv e-prints,
  arXiv:2203.09991

\bibitem[{Benisty {et~al.}(2016)Benisty, Stolker, Pohl, de~Boer, Lesur,
  Dominik, Dullemond, Langlois, Min, Wagner, Henning, Juhasz, Pinilla,
  Facchini, Apai, van Boekel, Garufi, Ginski, M{\'{e}}nard, Pinte, Quanz,
  Zurlo, Boccaletti, Bonnefoy, Beuzit, Chauvin, Cudel, Desidera, Feldt,
  Fontanive, Gratton, Kasper, Lagrange, LeCoroller, Mouillet, Mesa, Sissa,
  Vigan, Antichi, Buey, Fusco, Gisler, Llored, Magnard, Moeller-Nilsson, Pragt,
  Roelfsema, Sauvage, \& Wildi}]{Benisty2016}
Benisty, M., Stolker, T., Pohl, A., {et~al.} 2016, \aap, 597, A42

\bibitem[{Bertrang \& Avenhaus(2018)}]{Bertrang2018}
Bertrang, G. H.-M. \& Avenhaus, H. 2018, Proceedings of the International
  Astronomical Union, 14, 241

\bibitem[{Biller {et~al.}(2014)Biller, Males, Rodigas, Morzinski, Close,
  Juh{\'{a}}sz, Follette, Lacour, Benisty, Sicilia-Aguilar, Hinz, Weinberger,
  Henning, Pott, Bonnefoy, \& K\"{o}hler}]{Biller2014}
Biller, B.~A., Males, J., Rodigas, T., {et~al.} 2014, \apj, 792, L22

\bibitem[{Boccaletti {et~al.}(2020)Boccaletti, Folco, Pantin, Dutrey,
  Guilloteau, Tang, Pi{\'{e}}tu, Habart, Milli, Beck, \&
  Maire}]{Boccaletti2020}
Boccaletti, A., Folco, E.~D., Pantin, E., {et~al.} 2020, \aap, 637, L5

\bibitem[{Boccaletti {et~al.}(2013)Boccaletti, Lagrange, Pantin, Augereau,
  Quanz, \& Meheut}]{Boccaletti2013}
Boccaletti, A., Lagrange, A.-M., Pantin, E., {et~al.} 2013, Proceedings of the
  International Astronomical Union, 8, 208

\bibitem[{Bohn {et~al.}(2022)Bohn, Benisty, Perraut, Van Der~Marel, W{\"o}lfer,
  Van~Dishoeck, Facchini, Manara, Teague, Francis, {et~al.}}]{bohn2022probing}
Bohn, A., Benisty, M., Perraut, K., {et~al.} 2022, Astronomy \& Astrophysics,
  658, A183

\bibitem[{Booth {et~al.}(2019)Booth, Walsh, \& Ilee}]{Booth2019}
Booth, A.~S., Walsh, C., \& Ilee, J.~D. 2019, \aap, 629, A75

\bibitem[{Bouwman {et~al.}(2003)Bouwman, de~Koter, Dominik, \&
  Waters}]{Bouwman2003}
Bouwman, J., de~Koter, A., Dominik, C., \& Waters, L. B. F.~M. 2003, \aap, 401,
  577

\bibitem[{{Bouwman} {et~al.}(2000){Bouwman}, {de Koter}, {van den Ancker}, \&
  {Waters}}]{Bouwman2000}
{Bouwman}, J., {de Koter}, A., {van den Ancker}, M.~E., \& {Waters},
  L.~B.~F.~M. 2000, \aap, 360, 213

\bibitem[{Bouwman {et~al.}(2001)Bouwman, Meeus, de~Koter, Hony, Dominik, \&
  Waters}]{Bouwman2001}
Bouwman, J., Meeus, G., de~Koter, A., {et~al.} 2001, \aap, 375, 950

\bibitem[{Brittain {et~al.}(2014)Brittain, Carr, Najita, Quanz, \&
  Meyer}]{Brittain2014}
Brittain, S.~D., Carr, J.~S., Najita, J.~R., Quanz, S.~P., \& Meyer, M.~R.
  2014, \apj, 791, 136

\bibitem[{Buchner {et~al.}(2014)Buchner, Georgakakis, Nandra, Hsu, Rangel,
  Brightman, Merloni, Salvato, Donley, \& Kocevski}]{Buchner14}
Buchner, J., Georgakakis, A., Nandra, K., {et~al.} 2014, \aap, 564, A125

\bibitem[{Casassus \& P{\'{e}}rez(2019)}]{Casassus2019}
Casassus, S. \& P{\'{e}}rez, S. 2019, \apj, 883, L41

\bibitem[{Chen {et~al.}(2018)Chen, K{\'{o}}sp{\'{a}}l, {\'{A}}brah{\'{a}}m,
  Kreplin, Matter, \& Weigelt}]{Chen2018}
Chen, L., K{\'{o}}sp{\'{a}}l, {\'{A}}., {\'{A}}brah{\'{a}}m, P., {et~al.} 2018,
  \aap, 609, A45

\bibitem[{Chen {et~al.}(2019)Chen, Mo{\'{o}}r, Kreplin, K{\'{o}}sp{\'{a}}l,
  {\'{A}}brah{\'{a}}m, Matter, Carmona, Hofmann, Schertl, \&
  Weigelt}]{Chen2019}
Chen, L., Mo{\'{o}}r, A., Kreplin, A., {et~al.} 2019, \apj, 887, L32

\bibitem[{Cohen {et~al.}(1999)Cohen, Walker, Carter, Hammersley, Kidger, \&
  Noguchi}]{Cohen1999}
Cohen, M., Walker, R.~G., Carter, B., {et~al.} 1999, \apj, 117, 1864

\bibitem[{Cugno {et~al.}(2019)Cugno, Quanz, Hunziker, Stolker, Schmid,
  Avenhaus, Baudoz, Bohn, Bonnefoy, Buenzli, Chauvin, Cheetham, Desidera,
  Dominik, Feautrier, Feldt, Ginski, Girard, Gratton, Hagelberg, Hugot, Janson,
  Lagrange, Langlois, Magnard, Maire, Menard, Meyer, Milli, Mordasini, Pinte,
  Pragt, Roelfsema, Rigal, Szul{\'{a}}gyi, van Boekel, van~der Plas, Vigan,
  Wahhaj, \& Zurlo}]{Cugno2019}
Cugno, G., Quanz, S.~P., Hunziker, S., {et~al.} 2019, \aap, 622, A156

\bibitem[{Currie {et~al.}(2022)Currie, Lawson, Schneider, Lyra, Wisniewski,
  Grady, Guyon, Tamura, Kotani, Kawahara, {et~al.}}]{currie2022images}
Currie, T., Lawson, K., Schneider, G., {et~al.} 2022, Nature Astronomy, 6, 751

\bibitem[{Devinat {et~al.}(2022{\natexlab{a}})Devinat, Habart, Pantin, Ysard,
  Jones, Labadie, \& Di~Folco}]{2022A&A...663A.151D}
Devinat, M., Habart, {\'E}., Pantin, {\'E}., {et~al.} 2022{\natexlab{a}},
  Astronomy \& Astrophysics, 663, A151

\bibitem[{Devinat {et~al.}(2022{\natexlab{b}})Devinat, Habart, Pantin, Ysard,
  Jones, Labadie, \& Di~Folco}]{devinat2022radial}
Devinat, M., Habart, {\'E}., Pantin, {\'E}., {et~al.} 2022{\natexlab{b}},
  Astronomy \& Astrophysics, 663, A151

\bibitem[{Diep {et~al.}(2019)Diep, Hoai, Ngoc, Nhung, Phuong, Thai, \&
  Tuan-Anh}]{diep2019}
Diep, P.~N., Hoai, D.~T., Ngoc, N.~B., {et~al.} 2019
  [\eprint{arXiv:1903.11868}]

\bibitem[{Doering {et~al.}(2007)Doering, Meixner, Holfeltz, Krist, Ardila,
  Kamp, Clampin, \& Lubow}]{Doering2007}
Doering, R.~L., Meixner, M., Holfeltz, S.~T., {et~al.} 2007, \aj, 133, 2122

\bibitem[{Dong {et~al.}(2015)Dong, Zhu, Fung, Rafikov, Chiang, \&
  Wagner}]{Dong2015}
Dong, R., Zhu, Z., Fung, J., {et~al.} 2015, \apj, 816, L12

\bibitem[{Dorschner {et~al.}(1995)Dorschner, Begemann, Henning, Jäger, \&
  Mutschke}]{Dorschner1995}
Dorschner, J., Begemann, B., Henning, T., Jäger, C., \& Mutschke, H. 1995,
  \aap, 300, 503

\bibitem[{Doucet {et~al.}(2007{\natexlab{a}})Doucet, Habart, Pantin, Dullemond,
  Lagage, Pinte, Duch{\^e}ne, \& M{\'e}nard}]{doucet2007hd}
Doucet, C., Habart, E., Pantin, E., {et~al.} 2007{\natexlab{a}}, Astronomy \&
  Astrophysics, 470, 625

\bibitem[{Doucet {et~al.}(2007{\natexlab{b}})Doucet, Habart, Pantin, Dullemond,
  Lagage, Pinte, Duch{\^{e}}ne, \& M{\'{e}}nard}]{Doucet2007}
Doucet, C., Habart, E., Pantin, E., {et~al.} 2007{\natexlab{b}}, \aap, 470, 625

\bibitem[{Draine \& Li(2001)}]{Draine2001}
Draine, B.~T. \& Li, A. 2001, \apj, 551, 807

\bibitem[{Draine {et~al.}(2020)Draine, Li, Hensley, Hunt, Sandstrom, \&
  Smith}]{Draine2020}
Draine, B.~T., Li, A., Hensley, B.~S., {et~al.} 2020, Excitation of PAH
  Emission: Dependence on Starlight Spectrum, Intensity, PAH Size Distribution,
  and PAH Ionization

\bibitem[{Dullemond \& Dominik(2004)}]{dullemond2004effect}
Dullemond, C. \& Dominik, C. 2004, Astronomy \& Astrophysics, 421, 1075

\bibitem[{Fairlamb {et~al.}(2015)Fairlamb, Oudmaijer, Mendigut{\'{\i}}a, Ilee,
  \& van~den Ancker}]{Fairlamb2015}
Fairlamb, J.~R., Oudmaijer, R.~D., Mendigut{\'{\i}}a, I., Ilee, J.~D., \&
  van~den Ancker, M.~E. 2015, \mnras, 453, 976

\bibitem[{Feroz {et~al.}(2009)Feroz, Hobson, \& Bridges}]{Multinest}
Feroz, F., Hobson, M., \& Bridges, M. 2009, Monthly Notices of the Royal
  Astronomical Society, 398, 1601

\bibitem[{Follette {et~al.}(2017)Follette, Rameau, Dong, Pueyo, Close,
  Duch{\^{e}}ne, Fung, Leonard, Macintosh, Males, Marois, Millar-Blanchaer,
  Morzinski, Mullen, Perrin, Spiro, Wang, Ammons, Bailey, Barman, Bulger,
  Chilcote, Cotten, Rosa, Doyon, Fitzgerald, Goodsell, Graham, Greenbaum,
  Hibon, Hung, Ingraham, Kalas, Konopacky, Larkin, Maire, Marchis, Metchev,
  Nielsen, Oppenheimer, Palmer, Patience, Poyneer, Rajan, Rantakyr\"{o},
  Savransky, Schneider, Sivaramakrishnan, Song, Soummer, Thomas, Vega, Wallace,
  Ward-Duong, Wiktorowicz, \& Wolff}]{Follette2017}
Follette, K.~B., Rameau, J., Dong, R., {et~al.} 2017, \apj, 153, 264

\bibitem[{Folsom {et~al.}(2012)Folsom, Bagnulo, Wade, Alecian, Landstreet,
  Marsden, \& Waite}]{folsom2012chemical}
Folsom, C., Bagnulo, S., Wade, G., {et~al.} 2012, Monthly Notices of the Royal
  Astronomical Society, 422, 2072

\bibitem[{Fukagawa {et~al.}(2010)Fukagawa, Tamura, Itoh, Oasa, Kudo, Hayashi,
  Kato, Ootsubo, Itoh, Shibai, \& Hayashi}]{Fukagawa2010}
Fukagawa, M., Tamura, M., Itoh, Y., {et~al.} 2010, Publications of the
  Astronomical Society of Japan, 62, 347

\bibitem[{Garufi {et~al.}(2014)Garufi, Quanz, Schmid, Avenhaus, Buenzli, \&
  Wolf}]{Garufi2014}
Garufi, A., Quanz, S.~P., Schmid, H.~M., {et~al.} 2014, \aap, 568, A40

\bibitem[{Geers {et~al.}(2009)Geers, Van~Dishoeck, Pontoppidan, Lahuis, Crapsi,
  Dullemond, \& Blake}]{geers2009lack}
Geers, V., Van~Dishoeck, E., Pontoppidan, K., {et~al.} 2009, Astronomy \&
  Astrophysics, 495, 837

\bibitem[{Geers {et~al.}(2007)Geers, van Dishoeck, Visser, Pontoppidan,
  Augereau, Habart, \& Lagrange}]{Geers2007}
Geers, V.~C., van Dishoeck, E.~F., Visser, R., {et~al.} 2007, \aap, 476, 279

\bibitem[{Gillessen {et~al.}(2006)Gillessen, Perrin, Brandner, Straubmeier,
  Eisenhauer, Rabien, Eckart, Lena, Genzel, Paumard, \&
  Hippler}]{Gillessen2006}
Gillessen, S., Perrin, G., Brandner, W., {et~al.} 2006, in Advances in Stellar
  Interferometry, ed. J.~D. Monnier, M.~Sch\"{o}ller, \& W.~C. Danchi ({SPIE})

\bibitem[{Ginski {et~al.}(2016)Ginski, Stolker, Pinilla, Dominik, Boccaletti,
  de~Boer, Benisty, Biller, Feldt, Garufi, Keller, Kenworthy, Maire,
  M{\'{e}}nard, Mesa, Milli, Min, Pinte, Quanz, van Boekel, Bonnefoy, Chauvin,
  Desidera, Gratton, Girard, Keppler, Kopytova, Lagrange, Langlois, Rouan, \&
  Vigan}]{Ginski2016}
Ginski, C., Stolker, T., Pinilla, P., {et~al.} 2016, \aap, 595, A112

\bibitem[{Gratton {et~al.}(2019)Gratton, Ligi, Sissa, Desidera, Mesa, Bonnefoy,
  Chauvin, Cheetham, Feldt, Lagrange, Langlois, Meyer, Vigan, Boccaletti,
  Janson, Lazzoni, Zurlo, Boer, Henning, D'Orazi, Gluck, Madec, Jaquet, Baudoz,
  Fantinel, Pavlov, \& Wildi}]{Gratton2019}
Gratton, R., Ligi, R., Sissa, E., {et~al.} 2019, \aap, 623, A140

\bibitem[{Guzm{\'a}n-D{\'\i}az {et~al.}(2021)Guzm{\'a}n-D{\'\i}az,
  Mendigut{\'\i}a, Montesinos, Oudmaijer, Vioque, Rodrigo, Solano, Meeus, \&
  Marcos-Arenal}]{guzman2021homogeneous}
Guzm{\'a}n-D{\'\i}az, J., Mendigut{\'\i}a, I., Montesinos, B., {et~al.} 2021,
  Astronomy \& Astrophysics, 650, A182

\bibitem[{Hashimoto {et~al.}(2011)Hashimoto, Tamura, Muto, Kudo, Fukagawa,
  Fukue, Goto, Grady, Henning, Hodapp, Honda, Inutsuka, Kokubo, Knapp,
  McElwain, Momose, Ohashi, Okamoto, Takami, Turner, Wisniewski, Janson, Abe,
  Brandner, Carson, Egner, Feldt, Golota, Guyon, Hayano, Hayashi, Hayashi,
  Ishii, Kandori, Kusakabe, Matsuo, Mayama, Miyama, Morino, Moro-Martin,
  Nishimura, Pyo, Suto, Suzuki, Takato, Terada, Thalmann, Tomono, Watanabe,
  Yamada, Takami, \& Usuda}]{Hashimoto2011}
Hashimoto, J., Tamura, M., Muto, T., {et~al.} 2011, \apj, 729, L17

\bibitem[{{Honda} {et~al.}(2012){Honda}, {Maaskant}, {Okamoto}, {Kataza},
  {Fukagawa}, {Waters}, {Dominik}, {Tielens}, {Mulders}, {Min}, {Yamashita},
  {Fujiyoshi}, {Miyata}, {Sako}, {Sakon}, {Fujiwara}, \&
  {Onaka}}]{2012ApJ...752..143H}
{Honda}, M., {Maaskant}, K., {Okamoto}, Y.~K., {et~al.} 2012, \apj, 752, 143

\bibitem[{Hudgins \& Allamandola(2005)}]{Hudgins2005}
Hudgins, D.~M. \& Allamandola, L.~J. 2005, {ChemInform}, 36

\bibitem[{Isella {et~al.}(2018)Isella, Huang, Andrews, Dullemond, Birnstiel,
  Zhang, Zhu, Guzm{\'{a}}n, P{\'{e}}rez, Bai, Benisty, Carpenter, Ricci, \&
  Wilner}]{Isella2018}
Isella, A., Huang, J., Andrews, S.~M., {et~al.} 2018, \apj, 869, L49

\bibitem[{Jamialahmadi {et~al.}(2018)Jamialahmadi, Ratzka, Pani{\'{c}},
  Fathivavsari, van Boekel, Flement, Henning, Jaffe, \&
  Mulders}]{Jamialahmadi2018}
Jamialahmadi, N., Ratzka, T., Pani{\'{c}}, O., {et~al.} 2018, \apj, 865, 137

\bibitem[{J\"{a}rvinen {et~al.}(2018)J\"{a}rvinen, Carroll, Hubrig, Ilyin,
  Sch\"{o}ller, Castelli, Hummel, Petr-Gotzens, Korhonen, Weigelt, Pogodin, \&
  Drake}]{Jrvinen2018}
J\"{a}rvinen, S.~P., Carroll, T.~A., Hubrig, S., {et~al.} 2018, \apj, 858, L18

\bibitem[{Juh{\'{a}}sz {et~al.}(2010)Juh{\'{a}}sz, Bouwman, Henning, Acke,
  van~den Ancker, Meeus, Dominik, Min, Tielens, \& Waters}]{Juhsz2010}
Juh{\'{a}}sz, A., Bouwman, J., Henning, T., {et~al.} 2010, \apj, 721, 431

\bibitem[{Jäger {et~al.}(1998)Jäger, Molster, Dorschner, Henning, Mutschke,
  \& Waters}]{Jger1998}
Jäger, C., Molster, F., Dorschner, J., {et~al.} 1998, \aap, 339, 904

\bibitem[{Keppler {et~al.}(2018)Keppler, Benisty, M{\"u}ller, Henning,
  Van~Boekel, Cantalloube, Ginski, Van~Holstein, Maire, Pohl,
  {et~al.}}]{keppler2018discovery}
Keppler, M., Benisty, M., M{\"u}ller, A., {et~al.} 2018, Astronomy \&
  Astrophysics, 617, A44

\bibitem[{Keppler {et~al.}(2020)Keppler, Penzlin, Benisty, van Boekel, Henning,
  van Holstein, Kley, Garufi, Ginski, Brandner, Bertrang, Boccaletti, de~Boer,
  Bonavita, Sevilla, Chauvin, Dominik, Janson, Langlois, Lodato, Maire,
  M{\'{e}}nard, Pantin, Pinte, Stolker, Szul{\'{a}}gyi, Thebault, Villenave,
  Zurlo, Rabou, Feautrier, Feldt, Madec, \& Wildi}]{Keppler2020}
Keppler, M., Penzlin, A., Benisty, M., {et~al.} 2020, \aap, 639, A62

\bibitem[{Khalafinejad {et~al.}(2016)Khalafinejad, Maaskant, Mari{\~{n}}as, \&
  Tielens}]{Khalafinejad2016}
Khalafinejad, S., Maaskant, K.~M., Mari{\~{n}}as, N., \& Tielens, A. G. G.~M.
  2016, \aap, 587, A62

\bibitem[{Klarmann {et~al.}(2017)Klarmann, Benisty, Min, Dominik, Berger,
  Waters, Kluska, Lazareff, \& Bouquin}]{Klarmann2017}
Klarmann, L., Benisty, M., Min, M., {et~al.} 2017, \aap, 599, A80

\bibitem[{Kluska {et~al.}(2020)Kluska, Berger, Malbet, Lazareff, Benisty,
  Bouquin, Absil, Baron, Delboulb{\'{e}}, Duvert, Isella, Jocou, Juhasz, Kraus,
  Lachaume, M{\'{e}}nard, Millan-Gabet, Monnier, Moulin, Perraut, Rochat,
  Pinte, Soulez, Tallon, Thi, Thi{\'{e}}baut, Traub, \& Zins}]{Kluska2020}
Kluska, J., Berger, J.-P., Malbet, F., {et~al.} 2020, \aap, 636, A116

\bibitem[{Kokoulina {et~al.}(2021)Kokoulina, Matter, Lopez, Pantin, Ysard,
  Weigelt, Habart, Varga, Jones, Meilland, {et~al.}}]{kokoulina2021first}
Kokoulina, E., Matter, A., Lopez, B., {et~al.} 2021, Astronomy \& Astrophysics,
  652, A61

\bibitem[{Lagage {et~al.}(2006)Lagage, Doucet, Pantin, Habart, Duchene, Menard,
  Pinte, Charnoz, \& Pel}]{Lagage2006}
Lagage, P.-O., Doucet, C., Pantin, E., {et~al.} 2006, Science, 314, 621

\bibitem[{Li \& Draine(2001)}]{Li2001}
Li, A. \& Draine, B.~T. 2001, \apj, 554, 778

\bibitem[{Li \& Lunine(2003)}]{li2003modeling}
Li, A. \& Lunine, J. 2003, The Astrophysical Journal, 594, 987

\bibitem[{Liu {et~al.}(2018)Liu, Jin, Li, Isella, \& Li}]{Liu2018}
Liu, S.-F., Jin, S., Li, S., Isella, A., \& Li, H. 2018, \apj, 857, 87

\bibitem[{Lopez {et~al.}(2022)Lopez, Lagarde, Petrov, Jaffe, Antonelli,
  Allouche, Berio, Matter, Meilland, Millour, {et~al.}}]{lopez2022matisse}
Lopez, B., Lagarde, S., Petrov, R., {et~al.} 2022, Astronomy \& Astrophysics,
  659, A192

\bibitem[{Maaskant {et~al.}(2014)Maaskant, Min, Waters, \&
  Tielens}]{Maaskant2014}
Maaskant, K.~M., Min, M., Waters, L. B. F.~M., \& Tielens, A. G. G.~M. 2014,
  \aap, 563, A78

\bibitem[{Mac{\'{\i}}as {et~al.}(2017)Mac{\'{\i}}as, Anglada, Osorio,
  Torrelles, Carrasco-Gonz{\'{a}}lez, G{\'{o}}mez, Rodr{\'{\i}}guez, \&
  Sierra}]{Macas2017}
Mac{\'{\i}}as, E., Anglada, G., Osorio, M., {et~al.} 2017, \apj, 838, 97

\bibitem[{Mac{\'{\i}}as {et~al.}(2019)Mac{\'{\i}}as, Espaillat, Osorio,
  Anglada, Torrelles, Carrasco-Gonz{\'{a}}lez, Flock, Linz, Bertrang, Henning,
  G{\'{o}}mez, Calvet, \& Dent}]{Macas2019}
Mac{\'{\i}}as, E., Espaillat, C.~C., Osorio, M., {et~al.} 2019, \apj, 881, 159

\bibitem[{Meeus {et~al.}(2012)Meeus, Montesinos, Mendigut{\'{\i}}a, Kamp, Thi,
  Eiroa, Grady, Mathews, Sandell, Martin-Zaïdi, Brittain, Dent, Howard,
  M{\'{e}}nard, Pinte, Roberge, Vandenbussche, \& Williams}]{Meeus2012}
Meeus, G., Montesinos, B., Mendigut{\'{\i}}a, I., {et~al.} 2012, \aap, 544, A78

\bibitem[{Meeus {et~al.}(2001)Meeus, Waters, Bouwman, van~den Ancker, Waelkens,
  \& Malfait}]{Meeus2001}
Meeus, G., Waters, L. B. F.~M., Bouwman, J., {et~al.} 2001, \aap, 365, 476

\bibitem[{Mendigut{\'{\i}}a {et~al.}(2017)Mendigut{\'{\i}}a, Oudmaijer, Garufi,
  Lumsden, Hu{\'{e}}lamo, Cheetham, de~Wit, Norris, Olguin, \&
  Tuthill}]{Mendiguta2017}
Mendigut{\'{\i}}a, I., Oudmaijer, R.~D., Garufi, A., {et~al.} 2017, \aap, 608,
  A104

\bibitem[{Menu {et~al.}(2015)Menu, van Boekel, Henning, Leinert, Waelkens, \&
  Waters}]{Menu2015}
Menu, J., van Boekel, R., Henning, T., {et~al.} 2015, \aap, 581, A107

\bibitem[{Mordasini {et~al.}(2015)Mordasini, Molli{\`e}re, Dittkrist, Jin, \&
  Alibert}]{Mordasini14}
Mordasini, C., Molli{\`e}re, P., Dittkrist, K.-M., Jin, S., \& Alibert, Y.
  2015, International Journal of Astrobiology, 14, 201

\bibitem[{Mulders {et~al.}(2011)Mulders, Waters, Dominik, Sturm, Bouwman, Min,
  Verhoeff, Acke, Augereau, Evans, Henning, Meeus, \& Olofsson}]{Mulders2011}
Mulders, G.~D., Waters, L. B. F.~M., Dominik, C., {et~al.} 2011, \aap, 531, A93

\bibitem[{Muro-Arena {et~al.}(2018)Muro-Arena, Dominik, Waters, Min, Klarmann,
  Ginski, Isella, Benisty, Pohl, Garufi, {et~al.}}]{muro2018dust}
Muro-Arena, G., Dominik, C., Waters, L., {et~al.} 2018, Astronomy \&
  Astrophysics, 614, A24

\bibitem[{Natta \& Kr{\"u}gel(1995)}]{natta1995pah}
Natta, A. \& Kr{\"u}gel, E. 1995, Astronomy and Astrophysics, 302, 849

\bibitem[{Nealon {et~al.}(2020)Nealon, Cuello, Gonzalez, van~der Plas, Pinte,
  Alexander, M{\'{e}}nard, \& Price}]{Nealon2020}
Nealon, R., Cuello, N., Gonzalez, J.-F., {et~al.} 2020, \mnras, 499, 3857

\bibitem[{Okamoto {et~al.}(2017)Okamoto, Kataza, Honda, Yamashita, Fujiyoshi,
  Miyata, Sako, Fujiwara, Sakon, Fukagawa, Momose, \& Onaka}]{Okamoto2017}
Okamoto, Y.~K., Kataza, H., Honda, M., {et~al.} 2017, The Astronomical Journal,
  154, 16

\bibitem[{Osorio {et~al.}(2014)Osorio, Anglada, Carrasco-Gonz{\'{a}}lez,
  Torrelles, Mac{\'{\i}}as, Rodr{\'{\i}}guez, G{\'{o}}mez,
  D{\textquotesingle}Alessio, Calvet, Nagel, Dent, Quanz, Reggiani, \&
  Mayen-Gijon}]{Osorio2014}
Osorio, M., Anglada, G., Carrasco-Gonz{\'{a}}lez, C., {et~al.} 2014, \apj, 791,
  L36

\bibitem[{Pani{\'{c}} {et~al.}(2014)Pani{\'{c}}, Ratzka, Mulders, Dominik, van
  Boekel, Henning, Jaffe, \& Min}]{Pani2014}
Pani{\'{c}}, O., Ratzka, T., Mulders, G.~D., {et~al.} 2014, \aap, 562, A101

\bibitem[{P{\'{e}}rez {et~al.}(2019)P{\'{e}}rez, Casassus, Baruteau, Dong,
  Hales, \& Cieza}]{Prez2019}
P{\'{e}}rez, S., Casassus, S., Baruteau, C., {et~al.} 2019, \aap, 158, 15

\bibitem[{P{\'{e}}rez {et~al.}(2020)P{\'{e}}rez, Casassus, Hales, Marino,
  Cheetham, Zurlo, Cieza, Dong, Alarc{\'{o}}n, Ben{\'{\i}}tez-Llambay,
  Fomalont, \& Avenhaus}]{Prez2020}
P{\'{e}}rez, S., Casassus, S., Hales, A., {et~al.} 2020, \apj, 889, L24

\bibitem[{Perraut {et~al.}(2019)Perraut, Labadie, Lazareff, Klarmann,
  Segura-Cox, Benisty, Bouvier, Brandner, o~Garatti, Caselli,
  {et~al.}}]{perraut2019gravity}
Perraut, K., Labadie, L., Lazareff, B., {et~al.} 2019, Astronomy \&
  Astrophysics, 632, A53

\bibitem[{Petit dit de~la Roche {et~al.}(2021)Petit dit de~la Roche, Oberg,
  van~den Ancker, Kamp, van Boekel, Fedele, Ivanov, Kasper, K{\"a}ufl,
  Kissler-Patig, {et~al.}}]{dit2021new}
Petit dit de~la Roche, D., Oberg, N., van~den Ancker, M.~E., {et~al.} 2021,
  Astronomy \& Astrophysics, 648, A92

\bibitem[{Pi{\'{e}}tu {et~al.}(2005)Pi{\'{e}}tu, Guilloteau, \&
  Dutrey}]{Pitu2005}
Pi{\'{e}}tu, V., Guilloteau, S., \& Dutrey, A. 2005, \aap, 443, 945

\bibitem[{Pineda {et~al.}(2019)Pineda, Szul{\'a}gyi, Quanz, Van~Dishoeck,
  Garufi, Meru, Mulders, Testi, Meyer, \& Reggiani}]{pineda2019high}
Pineda, J.~E., Szul{\'a}gyi, J., Quanz, S.~P., {et~al.} 2019, The Astrophysical
  Journal, 871, 48

\bibitem[{Pinte {et~al.}(2018)Pinte, Price, M{\'{e}}nard, Duch{\^{e}}ne, Dent,
  Hill, de~Gregorio-Monsalvo, Hales, \& Mentiplay}]{Pinte2018}
Pinte, C., Price, D.~J., M{\'{e}}nard, F., {et~al.} 2018, \apj, 860, L13

\bibitem[{Pinte {et~al.}(2019)Pinte, van~der Plas, M{\'{e}}nard, Price,
  Christiaens, Hill, Mentiplay, Ginski, Choquet, Boehler, Duch{\^{e}}ne, Perez,
  \& Casassus}]{Pinte2019}
Pinte, C., van~der Plas, G., M{\'{e}}nard, F., {et~al.} 2019, Nature Astronomy,
  3, 1109

\bibitem[{Poblete {et~al.}(2020)Poblete, Calcino, Cuello, Mac{\'{\i}}as, Ribas,
  Price, Cuadra, \& Pinte}]{Poblete2020}
Poblete, P.~P., Calcino, J., Cuello, N., {et~al.} 2020, \mnras, 496, 2362

\bibitem[{Quanz {et~al.}(2013)Quanz, Amara, Meyer, Kenworthy, Kasper, \&
  Girard}]{Quanz2013}
Quanz, S.~P., Amara, A., Meyer, M.~R., {et~al.} 2013, \apj, 766, L1

\bibitem[{Quanz {et~al.}(2012)Quanz, Birkmann, Apai, Wolf, \&
  Henning}]{Quanz2012}
Quanz, S.~P., Birkmann, S.~M., Apai, D., Wolf, S., \& Henning, T. 2012, \aap,
  538, A92

\bibitem[{Reggiani {et~al.}(2014)Reggiani, Quanz, Meyer, Pueyo, Absil, Amara,
  Anglada, Avenhaus, Girard, Gonzalez, Graham, Mawet, Meru, Milli, Osorio,
  Wolff, \& Torrelles}]{Reggiani2014}
Reggiani, M., Quanz, S.~P., Meyer, M.~R., {et~al.} 2014, \apj, 792, L23

\bibitem[{Rosotti {et~al.}(2019)Rosotti, Benisty, Juh{\'{a}}sz, Teague, Clarke,
  Dominik, Dullemond, Klaassen, Matr{\`{a}}, \& Stolker}]{Rosotti2019}
Rosotti, G.~P., Benisty, M., Juh{\'{a}}sz, A., {et~al.} 2019, \mnras, 491, 1335

\bibitem[{Rosotti {et~al.}(2020)Rosotti, Benisty, Juh{\'a}sz, Teague, Clarke,
  Dominik, Dullemond, Klaassen, Matr{\`a}, \& Stolker}]{rosotti2020spiral}
Rosotti, G.~P., Benisty, M., Juh{\'a}sz, A., {et~al.} 2020, Monthly Notices of
  the Royal Astronomical Society, 491, 1335

\bibitem[{Sanchez-Bermudez {et~al.}(2021)Sanchez-Bermudez, o~Garatti, Lopez,
  Perraut, Labadie, Benisty, Brandner, Dougados, Garcia, Henning,
  {et~al.}}]{2021arXiv210702391S}
Sanchez-Bermudez, J., o~Garatti, A.~C., Lopez, R.~G., {et~al.} 2021, Astronomy
  \& Astrophysics, 654, A97

\bibitem[{Schneider {et~al.}(2020)Schneider, Dougados, Whelan, Eisl{\"o}ffel,
  G{\"u}nther, Hu{\'e}lamo, Mendigut{\'\i}a, Oudmaijer, \&
  Beck}]{schneider2020discovery}
Schneider, P.~C., Dougados, C., Whelan, E., {et~al.} 2020, Astronomy \&
  Astrophysics, 638, L3

\bibitem[{Seok \& Li(2016)}]{Seok2016}
Seok, J.~Y. \& Li, A. 2016, The Astrophysical Journal, 818, 2

\bibitem[{Seok \& Li(2017)}]{Seok2017}
Seok, J.~Y. \& Li, A. 2017, \apj, 835, 291

\bibitem[{Servoin \& Piriou(1973)}]{Servoin1973}
Servoin, J.~L. \& Piriou, B. 1973, physica status solidi (b), 55, 677

\bibitem[{Siebenmorgen \& Heymann(2012)}]{Siebenmorgen2012}
Siebenmorgen, R. \& Heymann, F. 2012, \aaps, 543, A25

\bibitem[{Siebenmorgen \& Kr{\"u}gel(2010)}]{siebenmorgen2010destruction}
Siebenmorgen, R. \& Kr{\"u}gel, E. 2010, Astronomy \& Astrophysics, 511, A6

\bibitem[{{Siebenmorgen} {et~al.}(2000){Siebenmorgen}, {Prusti}, {Natta}, \&
  {M{\"u}ller}}]{Siebenmorgen2000}
{Siebenmorgen}, R., {Prusti}, T., {Natta}, A., \& {M{\"u}ller}, T.~G. 2000,
  \aap, 361, 258

\bibitem[{Skilling(2004)}]{Skilling04}
Skilling, J. 2004, in {AIP} Conference Proceedings ({AIP})

\bibitem[{Spitzer \& Kleinman(1961)}]{Spitzer1961}
Spitzer, W.~G. \& Kleinman, D.~A. 1961, Physical Review, 121, 1324

\bibitem[{Taha {et~al.}(2018)Taha, Labadie, Pantin, Matter, Alvarez, Esquej,
  Grellmann, Rebolo, Telesco, \& Wolf}]{Taha2018}
Taha, A.~S., Labadie, L., Pantin, E., {et~al.} 2018, \aap, 612, A15

\bibitem[{Tang {et~al.}(2017{\natexlab{a}})Tang, Guilloteau, Dutrey, Muto,
  Shen, Gu, ichiro Inutsuka, Momose, Pietu, Fukagawa, Chapillon, Ho, di~Folco,
  Corder, Ohashi, \& Hashimoto}]{Tang2017}
Tang, Y.-W., Guilloteau, S., Dutrey, A., {et~al.} 2017{\natexlab{a}}, \apj,
  840, 32

\bibitem[{Tang {et~al.}(2017{\natexlab{b}})Tang, Guilloteau, Dutrey, Muto,
  Shen, Gu, Inutsuka, Momose, Pietu, Fukagawa, {et~al.}}]{tang2017planet}
Tang, Y.-W., Guilloteau, S., Dutrey, A., {et~al.} 2017{\natexlab{b}}, The
  Astrophysical Journal, 840, 32

\bibitem[{Teague {et~al.}(2018)Teague, Bae, Bergin, Birnstiel, \&
  Foreman-Mackey}]{Teague2018}
Teague, R., Bae, J., Bergin, E.~A., Birnstiel, T., \& Foreman-Mackey, D. 2018,
  \apj, 860, L12

\bibitem[{Telesco {et~al.}(2003)Telesco, Ciardi, French, Ftaclas, Hanna, Hon,
  Hough, Julian, Julian, Kidger, Packham, Pina, Varosi, \&
  Sellar}]{Telesco2003}
Telesco, C.~M., Ciardi, D., French, J., {et~al.} 2003, in Instrument Design and
  Performance for Optical/Infrared Ground-based Telescopes, ed. M.~Iye \&
  A.~F.~M. Moorwood ({SPIE})

\bibitem[{van Boekel {et~al.}(2005)van Boekel, Min, Waters, de~Koter, Dominik,
  van~den Ancker, \& Bouwman}]{vanBoekel2005}
van Boekel, R., Min, M., Waters, L. B. F.~M., {et~al.} 2005, \aap, 437, 189

\bibitem[{van Boekel {et~al.}(2004)van Boekel, Waters, Dominik, Dullemond,
  Tielens, \& de~Koter}]{vanBoekel2004}
van Boekel, R., Waters, L. B. F.~M., Dominik, C., {et~al.} 2004, \aap, 418, 177

\bibitem[{van Der~Plas {et~al.}(2017)van Der~Plas, Wright, M{\'e}nard,
  Casassus, Canovas, Pinte, Maddison, Maaskant, Avenhaus, Cieza,
  {et~al.}}]{van2017cavity}
van Der~Plas, G., Wright, C., M{\'e}nard, F., {et~al.} 2017, Astronomy \&
  Astrophysics, 597, A32

\bibitem[{van~der Plas {et~al.}(2016)van~der Plas, Wright, M{\'{e}}nard,
  Casassus, Canovas, Pinte, Maddison, Maaskant, Avenhaus, Cieza, Perez, \&
  Ubach}]{vanderPlas2016}
van~der Plas, G., Wright, C.~M., M{\'{e}}nard, F., {et~al.} 2016, \aap, 597,
  A32

\bibitem[{Van~Kerckhoven \& Waelkens(2002)}]{van2002nanodiamonds}
Van~Kerckhoven, C. \& Waelkens, C. 2002, Astronomy \& astrophysics, 384, 568

\bibitem[{Varga {et~al.}(2018)Varga, {\'{A}}brah{\'{a}}m, Chen, Ratzka,
  Gab{\'{a}}nyi, \& K{\'{o}}sp{\'{a}}l}]{Varga2018}
Varga, J., {\'{A}}brah{\'{a}}m, P., Chen, L., {et~al.} 2018, Proceedings of the
  International Astronomical Union, 14, 128

\bibitem[{Varga {et~al.}(2021)Varga, Hogerheijde, van Boekel, Klarmann, Petrov,
  Waters, Lagarde, Pantin, Berio, Weigelt, Robbe-Dubois, Lopez, Millour,
  Augereau, Meheut, Meilland, Henning, Jaffe, Bettonvil, Bristow, Hofmann,
  Matter, Zins, Wolf, Allouche, Donnan, Schertl, Dominik, Heininger, Lehmitz,
  Cruzal{\`{e}}bes, Glindemann, Meisenheimer, Paladini, Sch\"{o}ller, Woillez,
  Venema, Kokoulina, Yoffe, {\'{A}}brah{\'{a}}m, Abadie, Abuter, Accardo,
  Adler, Ag{\'{o}}cs, Antonelli, B\"{o}hm, Bailet, Bazin, Beckmann, Beltran,
  Boland, Bourget, Brast, Bresson, Burtscher, Castillo, Chelli, Cid, Clausse,
  Connot, Conzelmann, Danchi, Haan, Delbo, Ebert, Elswijk, Fantei, Frahm,
  Rosas, Gabasch, Gallenne, Garces, Girard, Gont{\'{e}}, Herrera, Graser,
  Guajardo, Guitton, Haubois, Hron, Hubin, Huerta, Isbell, Ives, Jakob,
  Jask{\'{o}}, Jochum, Klein, Kragt, Kroes, Kuindersma, Labadie, Laun, Poole,
  Leinert, Lizon, Lopez, M{\'{e}}rand, Marcotto, Mauclert, Maurer, Mehrgan,
  Meisner, Meixner, Mellein, Mohr, Morel, Mosoni, Navarro, Neumann,
  Nu{\ss}baum, Pallanca, Pasquini, Percheron, Pott, Pozna, Ridinger, Rigal,
  Riquelme, Rivinius, Roelfsema, Rohloff, Rousseau, Schuhler, Schuil, Soulain,
  Stee, Stephan, ter Horst, Tromp, Vakili, van Duin, Vinther, Wittkowski, \&
  Wrhel}]{Varga2021}
Varga, J., Hogerheijde, M., van Boekel, R., {et~al.} 2021, \aap, 647, A56

\bibitem[{Verhoeff {et~al.}(2010)Verhoeff, Min, Acke, van Boekel, Pantin,
  Waters, Tielens, van~den Ancker, Mulders, de~Koter, \&
  Bouwman}]{Verhoeff2010}
Verhoeff, A.~P., Min, M., Acke, B., {et~al.} 2010, \aap, 516, A48

\bibitem[{Visser {et~al.}(2007)Visser, Geers, Dullemond, Augereau, Pontoppidan,
  \& van Dishoeck}]{Visser2007}
Visser, R., Geers, V.~C., Dullemond, C.~P., {et~al.} 2007, \aap, 466, 229

\bibitem[{Wagner {et~al.}(2015)Wagner, Apai, Kasper, \& Robberto}]{Wagner2015}
Wagner, K., Apai, D., Kasper, M., \& Robberto, M. 2015, \apj, 813, L2

\bibitem[{Walsh {et~al.}(2016)Walsh, Juh{\'{a}}sz, Meeus, Dent, Maud, Aikawa,
  Millar, \& Nomura}]{Walsh2016}
Walsh, C., Juh{\'{a}}sz, A., Meeus, G., {et~al.} 2016, \apj, 831, 200

\bibitem[{Woitke {et~al.}(2009)Woitke, Kamp, \& Thi}]{woitke2009radiation}
Woitke, P., Kamp, I., \& Thi, W.-F. 2009, Astronomy \& Astrophysics, 501, 383

\bibitem[{{Yu} {et~al.}(2021){Yu}, {Teague}, {Bae}, \&
  {{\"O}berg}}]{2021ApJ...920L..33Y}
{Yu}, H., {Teague}, R., {Bae}, J., \& {{\"O}berg}, K. 2021, \apjl, 920, L33

\bibitem[{Zeidler {et~al.}(2015)Zeidler, Mutschke, \& Posch}]{Zeidler2015}
Zeidler, S., Mutschke, H., \& Posch, T. 2015, \apj, 798, 125

\end{thebibliography}
%

\appendix

\color{black}

\section{Target sample: Detailed introduction} \label{app_targets}

\subsection{AB Aurigae} \label{ABAUR}


The disk around this object, extending up to $\approx$550 au \citep{Hashimoto2011} is extremely rich in substructures: gaps at distances of $\approx$30, 70 au were detected in, for example, millimeter CO isotopologue observations with the Institut de radioastronomie millimetrique (IRAM) interferometer \citep{Pitu2005}, near-infrared polarimetry with the Subaru telescope \citep{Hashimoto2011} and dust continuum and $^{12}$CO observations with ALMA \citep{Tang2017}. These observations also detected spiral structures on different scales. In addition, exquisite images of the circumstellar disk as observed with SPHERE showcase complex spiral structures, as recently reported by \citet{Boccaletti2020}, and further constrain a "bump" in an inner spiral, possibly induced by a $\approx$4-13~\Mjup \  companion at a distance of $\approx$40~au \citep{Poblete2020}.

The object's spatially unresolved N-band spectrum was modeled by, for example, \citet{vanBoekel2005}. While abundant silicate emission was observed and modeled, PAH emission was not detected. Later, \cite{Juhsz2010} detected several PAH bands in the Spitzer spectrum of the object.

\subsection{HD95881} \label{HD95881}

\citet{Verhoeff2010} performed robust fitting of a radiative-transfer model to reproduce multiple observables of the protoplanetary disk around this object across different bands--spanning from Q-band imaging, infrared spectroscopy, and K- and N-band interferometry. The derived density and temperature structure of the disk indicates that it consists of a thick puffed-up inner rim and an outer region that has a flaring gas surface and is relatively void of “visible” dust grains, and conclude that the disk is in a transition phase from a gas-rich flaring disk to a gas-poor self-shadowed disk. 

More recently, \citet{Jrvinen2018} detected weak mean longitudinal stellar magnetic fields using the High Accuracy Radial velocity Planet Searcher polarimeter (HARPSpol) attached to the ESO 3.6 m telescope at the La Silla observatory.

\subsection{HD~97048} \label{HD97048}
The well-known circumstellar disk around HD\,97048 has been thoroughly investigated. This young object ($\sim$5 Myr according to \citet{Fairlamb2015}) exhibits a particularly extended disk (up to $\sim$\,1000\ AU), which was observed at an early stage with the \textit{Hubble} Space Telescope \citep{Doering2007} and since then on multiple additional occasions.\ Its extended emission \citep[e.g.,][]{Lagage2006, Ginski2016}, substructure \citep[e.g.,][]{Ginski2016, vanderPlas2016}, and chemistry \citep[e.g.,][]{Siebenmorgen2000, Meeus2012, Booth2019} have been probed.


Observations of, for example, millimeter-sized continuum emission observed with ALMA \citep{vanderPlas2016, Walsh2016}, tracing the continuum emission up to $\approx$355 au, and polarized scattered light observations with SPHERE \citep{Ginski2016} reveal gaps on distance scales of $\approx$40-350 au, as well as tracing the continuum emission to $\approx$355 au. 

A study by \citet{Quanz2012} focusing on the inner region of the disk ($\sim$15-160 au) with the VLT/NACO polarimetric differential imager was able to detect and resolve dust continuum emission in the H and K bands. Studies aimed at specifically probing the properties of PAHs \citep{vanBoekel2004, Lagage2006} were able to resolve the distribution thereof spatially. Finally, \cite{Pinte2019} report a planet discovered by its kinematic signature in the ALMA line observations data.

\subsection{HD~100453} \label{HD100453}

The disk around this object possesses ample and diverse substructures, such as spiral arms and shadows \citep[e.g.,][]{Benisty2016}, a gap approximately of the size of $\sim$1-20 au \citep{Wagner2015, Khalafinejad2016}, and finally--an M dwarf companion at a distance of $\approx$120 au \citep[e.g.,][]{Wagner2015, Dong2015}, with which binary-disk interaction (namely the spiral arms of the outer disk) is evident \citep{Rosotti2019}. Later, \cite{rosotti2020spiral} presented clear evidence for that hypothesis.

While no direct detection of planets embedded in the disk has been made, \citet{Nealon2020} posit a potential existence thereof as an explanation for the mutual inclination between the inner and outer disk components \citep{bohn2022probing}. Modeling the near-infrared intensity distribution of the disk based on interferometric observations, \cite{Klarmann2017} found evidence for stochastically heated particles.

\subsection{HD~100546} \label{HD100546}

Previous observations and modeling of the disk around this object traditionally divide it into an inner and outer component \citep{Bouwman2003, Pani2014}, separated by a gap between approximately 1 and 10 au \citep[e.g.,][]{Menu2015, Jamialahmadi2018}), with a possible misalignment of the two \citep{Kluska2020}. Additional more delicate substructure features, such as spirals \citep{Boccaletti2013, Follette2017} and inflow or jets \citep{Mendiguta2017} had been postulated and later found by \citet{schneider2020discovery} to be most likely a jet. The same is true for posited existence of companions at different distances throughout the disk \citep[e.g.,][]{Quanz2013, Brittain2014, Cugno2019, Casassus2019, Prez2020}. \citet{vanBoekel2004} were able to spatially resolve the N-band PAH emission of this object with the TIMMI2 mid-infrared instrument on the ESO 3.6 m telescope at La Silla Observatory, albeit with a seeing-limited spatial resolution of only $\approx$0\farcs85.






\subsection{HD~163296} \label{HD163296}

Substructure in the disk of this object had been identified across a wide spectral range of observations, predominantly in the submillimeter with ALMA \citep[e.g.,][]{Isella2018} and in near-infrared polarimetry \citep[e.g.,][]{Garufi2014}. These constrained the presence of gaps on distances around $\approx$140, 80, 50, and 10 au. Subsequently, companions were suggested as a potential mechanism to induce these gaps \citep[e.g.,][]{Liu2018, Teague2018, Pinte2018}. More recently, evidence for a temporally variable sub-structure on sub-au scales was found in VLT Interferometer (VLTI) observations using Multi AperTure mid-Infrared Spectroscopic Experiment (MATISSE) \citep{Varga2021} and GRAVITY \citep{2021arXiv210702391S}. Finally, \citet{dit2021new} recently observed this object with VISIR-NEAR in imaging mode and spatially resolved images in two different PAH narrowband filters.


\subsection{HD~169142} \label{HD169142}

This object has been studied and observed with various methods across a wide swath of the electromagnetic spectrum--from the submillimeter and millimeter observations with ALMA \citep{Macas2019}, near-infrared \citep{Fukagawa2010} and mid-infrared \citep{Okamoto2017}. Additionally, scattered light polarimetry surveys with SPHERE \citep{Bertrang2018} produced images of the multiple rings (up to four \cite{Macas2017,Prez2019}), and interferometric surveys in the near-infrared were also conducted with the VLTI-The Precision Integrated-Optics Near-infrared Imaging Experiment (PIONIER) instrument in order to constrain the inner-region substructure of the disk \citep{Chen2018, Chen2019}. The innermost ring (inner disk rim, \cite{Bertrang2018}) and the outer three rings are located at distances of $\approx$20 and $\approx$29, 55, 74, and 87 au, respectively. While planets were not directly detected in the disk, their potential presence at different distance scales or that of ongoing planet formation was postulated by, for example, \citet{Osorio2014}, \citet{Reggiani2014}, and \citet{Biller2014}, but are inconclusive. More recently, a scattered light angular differential imaging study by \cite{Gratton2019} with SPHERE suggested the presence of a $\approx$2~\Mjup \  planet at a distance of $\sim$100-150 au. Strong spatially unresolved PAH emission features at 3.3, 6.2, 7.7, 8.6, 11.3, and 12.6~$\mu$m were detected by \citet{Seok2016} and were recently spatially resolved by \citet{dit2021new} with VISIR-NEAR in the imaging mode.


\subsection{HD~179218} \label{HD179218}

While a less studied object, its circumstellar disk manifests mainly through its infrared excess \citep{Meeus2001} and is subsequently classified as a group-Ia source (i.e., of possible flared geometry), of which PAHs are an optimal tracer. Indeed, \citet{Meeus2001}  and \citet{Bouwman2001} both detect PAH emission at the $\approx$8.6 and 11.3 $\mu$m regions, which was later confirmed by \citet{Juhsz2010}. Additionally, \citet{Meeus2001} and \citet{Bouwman2001} detected silicate emission, which was later confirmed and modeled by \citet{vanBoekel2005}. More recently, \citet{Taha2018} observed the object with the CanariCam \citep{Telesco2003} high-resolution mid-infrared imager and spectrograph and were able to resolve the N-band PAH emission spatially, concluding that similarly to HD\,97048--PAHs extend far across the flared disk's surface layer in a predominantly ionized state. Recent MATISSE-VLTI preliminary low spectral resolution observations of this object could not infer the presence or not of a PAH emission in the inner disk region ($<$10 au). Stochastically heated carbon dust particles are, however, probably populating this inner disk region \citep{kokoulina2021first}.

\color{black}

\section{Parametric model: Template plots} \label{APP_Param_Model}





\begin{table*}
\small
    \centering
    \begin{tabular}{c  c  c  c } 
    \hline\hline
   Dust Component & Chemical Formula & Shape & Reference \\ 
    \hline
    Amorphous olivine & Mg$_{2x}$Fe$_{2-2x}$SiO$_4$ & Homogenous Spheres & \citet{Dorschner1995} \\

    Amorphous pyroxene & Mg$_x$Fe$_{1x}$SiO$_3$ & Homogenous Spheres & \citet{Dorschner1995} \\

    Crystalline forsterite & Mg$_2$SiO$_4$ & Irregular (DHS) & \citet{Servoin1973} \\

    Crystalline enstatite & MgSiO$_3$ & Irregular (DHS) & \citet{Jger1998} \\

    Amorphous silica & SiO$_2$ & Irregular (DHS) & \citet{Spitzer1961} \\
    \hline
    \end{tabular}
    \caption{Characteristics of silicate components and their provenances, adopted from \citet{vanBoekel2005}. Here an $x$  in the chemical formulae denotes the magnesium content of the material and ranges between $0 \lesssim x \lesssim 1$. The templates used in our study are similar to those in \citet{vanBoekel2005}, who assumed $x = 0.5$ for amorphous olivine and pyroxene, and $x = 1$ for crystalline olivine and pyroxene (i.e., forsterite and enstatite, respectively).}
    \label{Tab_silicateSpecies}
\end{table*}

\begin{table*}
\small
    \centering
    \begin{tabular}{c c c c} 
    \hline\hline
    Emission Band [$\mu$m] & Charge State When Dominant & Emission Source \\ 
    \hline
     7.9 & ionized &  CC stretching\\
     8.6 & ionized & CH in-plane bending  \\
     11.3 & Neutral & CH out-of-plane bending  \\
     12.6 & Neutral & CH out-of-plane bending \\
     \hline
    \end{tabular}
    \caption{N-band PAH emission features. Characteristics are adopted from \citet{Hudgins2005}.}
    \label{Tab_PAHTemplates}
\end{table*}
   
\begin{figure*}[h]
\centering
\begin{tabular}{cc}
    \hspace{-1.0cm}
        \includegraphics[scale = 0.7]{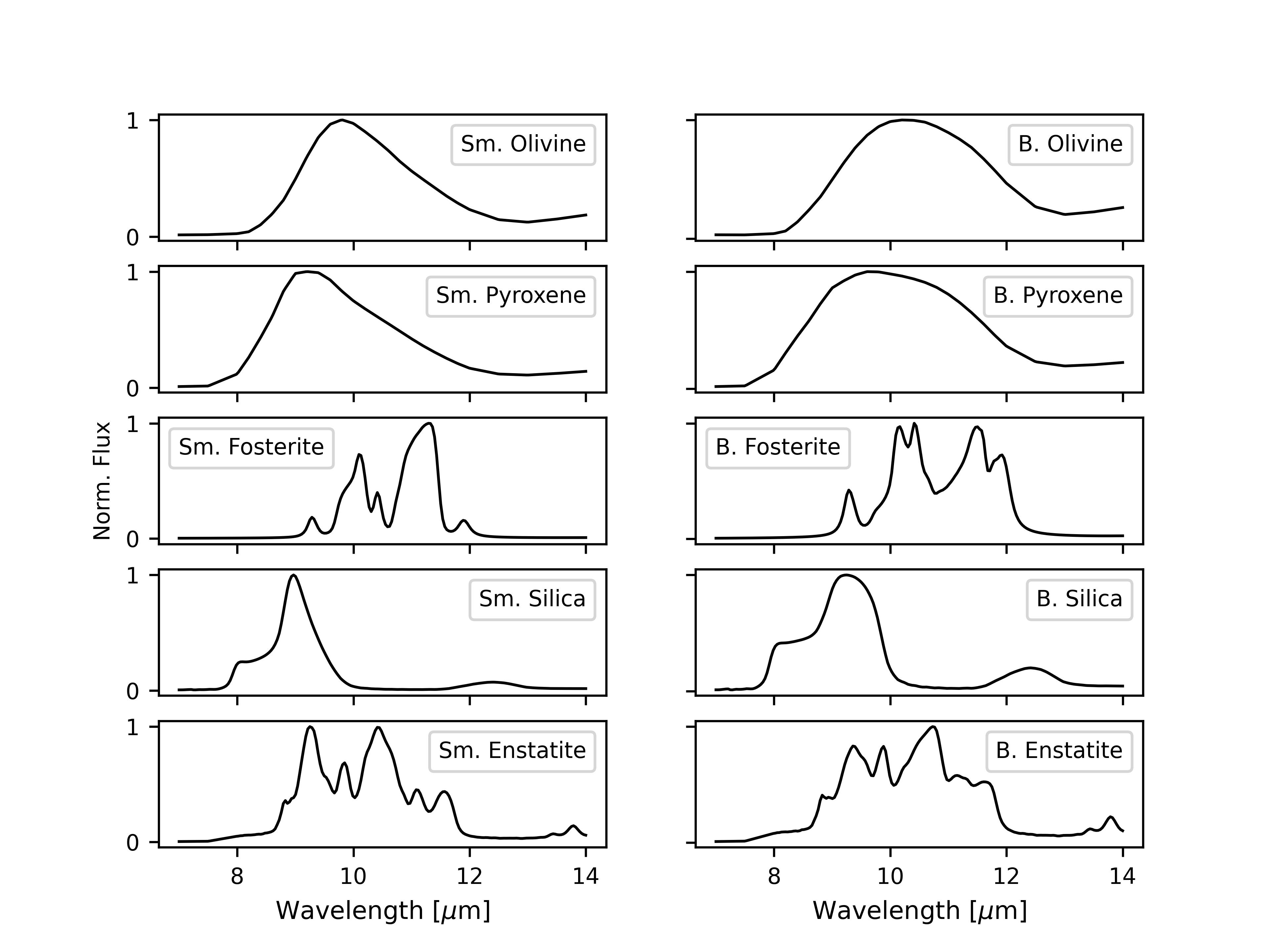}\\
    \hspace{-1.0cm}
  \includegraphics[scale = 0.7]{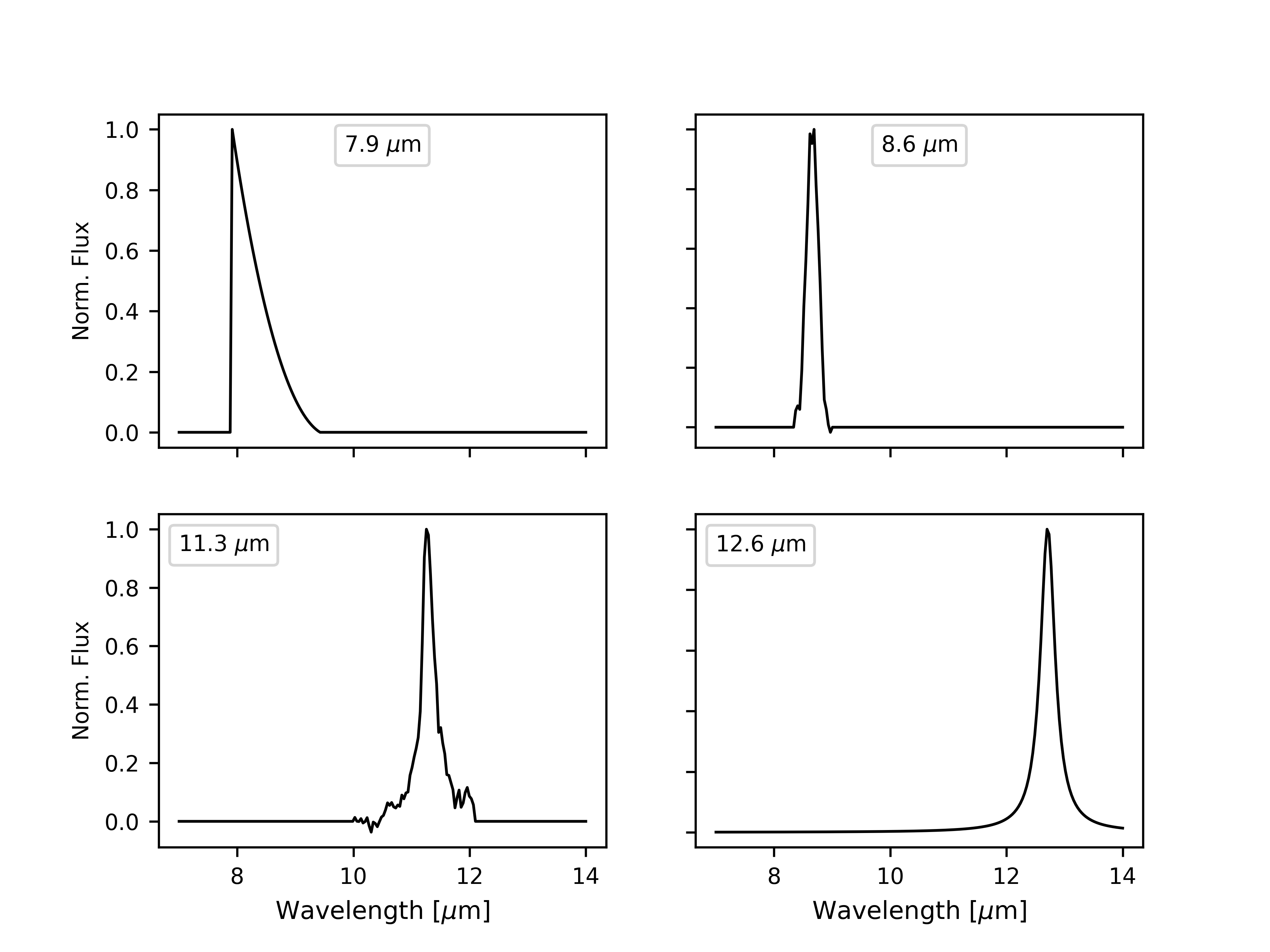} \\
\end{tabular}
\caption{Templates used in the parametric model. \textbf{Upper Panels}: Silicate species templates ($S_i$) of five silicate species (Table \ref{Tab_silicateSpecies}), where for each specie there exit two templates--corresponding to two prescribed grain-sizes of 0.1 $\mu$m ("small") and 1.5~$\mu$m ("big"). \textbf{Bottom Panels}: PAH component templates ($P_j$).}
\label{Fig_Templates}
\end{figure*}

\section{Statistical model}
\label{APP_Stat_Model}

We evaluated the goodness of fit to the reduced N-band spectra for each simulated model with parameter set $\Theta$ assuming a normal distribution, whose log is
\begin{equation}
\log(L) = -0.5\sum_{i=1} ^m \ln\left(2\pi\sigma_i \right) - 0.5\sum_{i=1} ^m \ln\left(\frac{(y_{i,\Theta} - x_i)^2}{\sigma_i ^2} \right),
\label{Eq_LogLike}
\end{equation}
where there are $m$ observed spectral measurements (1024 pixels) $x_i$ with respective uncertainties \textbf{$\sigma_i$}, and \textbf{$y_i$} is the simulated N-band emission spectrum associated with the model parameters \textbf{$\Theta$}. As the measurement uncertainty of the observed flux values, $\sigma_i$, we adopt the standard deviation of the flux over all objects' exposures. 

Our derived best-fit values of the model parameters are the median values of the posterior distribution, and the uncertainties are the $50\pm 34.1$ percentiles ranges of the posterior distribution, computed after removal of the MCMC "burn-in" phase (points with relative likelihood $<10^{-3}$ times that of the best-fit). We refer to the range of uncertainties as $\sigma$, defined irrespective of the distribution. For a normal distribution, $\sigma$ corresponds to the standard deviation. We note that the uncertainty derived for each parameter in the model from its respective posterior distribution, which we refer to as $\sigma$, is not the same as the $\sigma_i$ in Eq.  \ref{Eq_LogLike}, which represents the uncertainty in the observed {fluxes}.

\color{black}

\section{Unresolved spectra posterior distributions} \label{APP_posteriors} 
\begin{figure*}
\centering
\includegraphics[scale = 0.18]{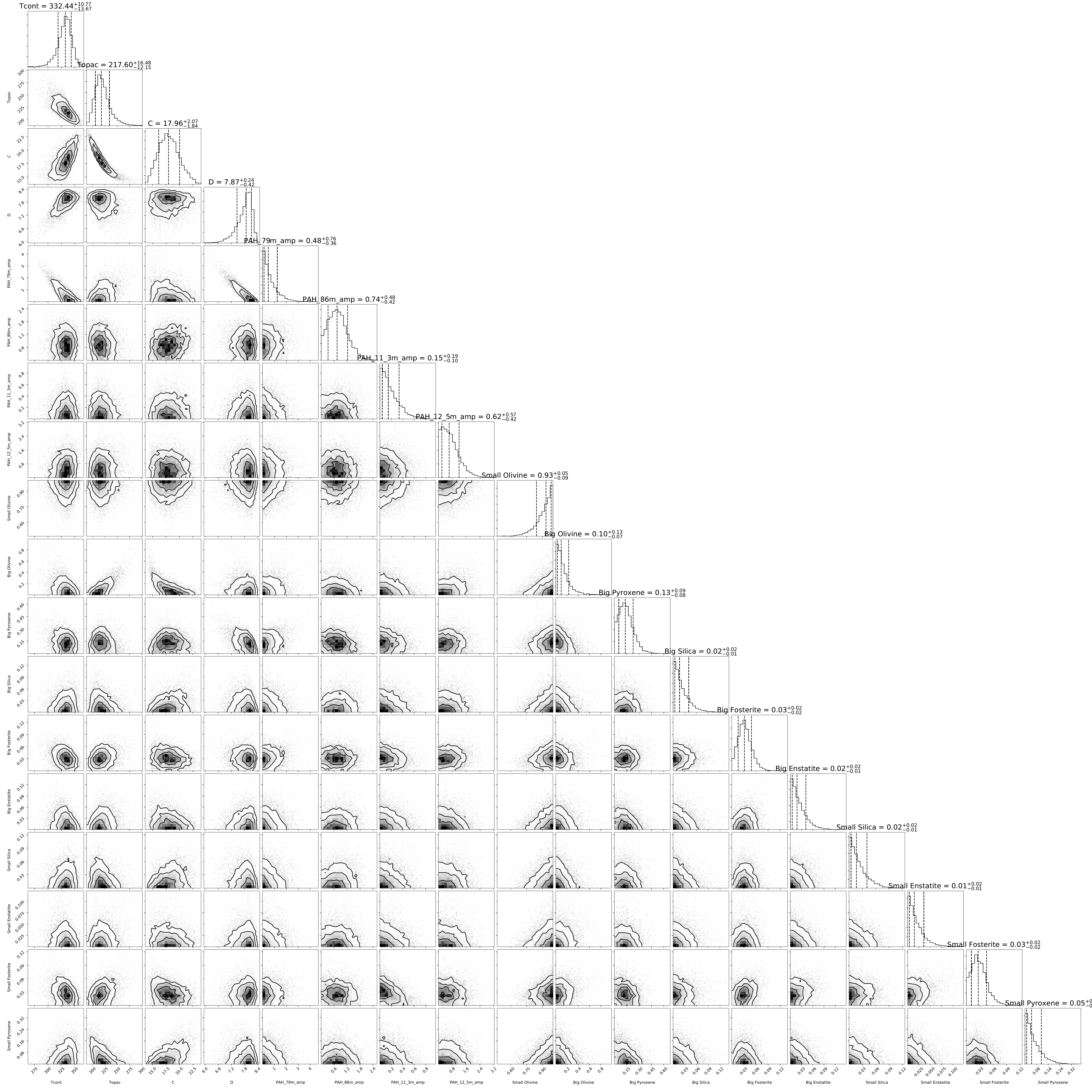}
\caption{\color{black} AB~Aurigae: unresolved fitting posterior distributions. \color{black}}
\label{Fig_corner_ABAUR}
\end{figure*}

\begin{figure*}
\centering
\includegraphics[scale = 0.18]{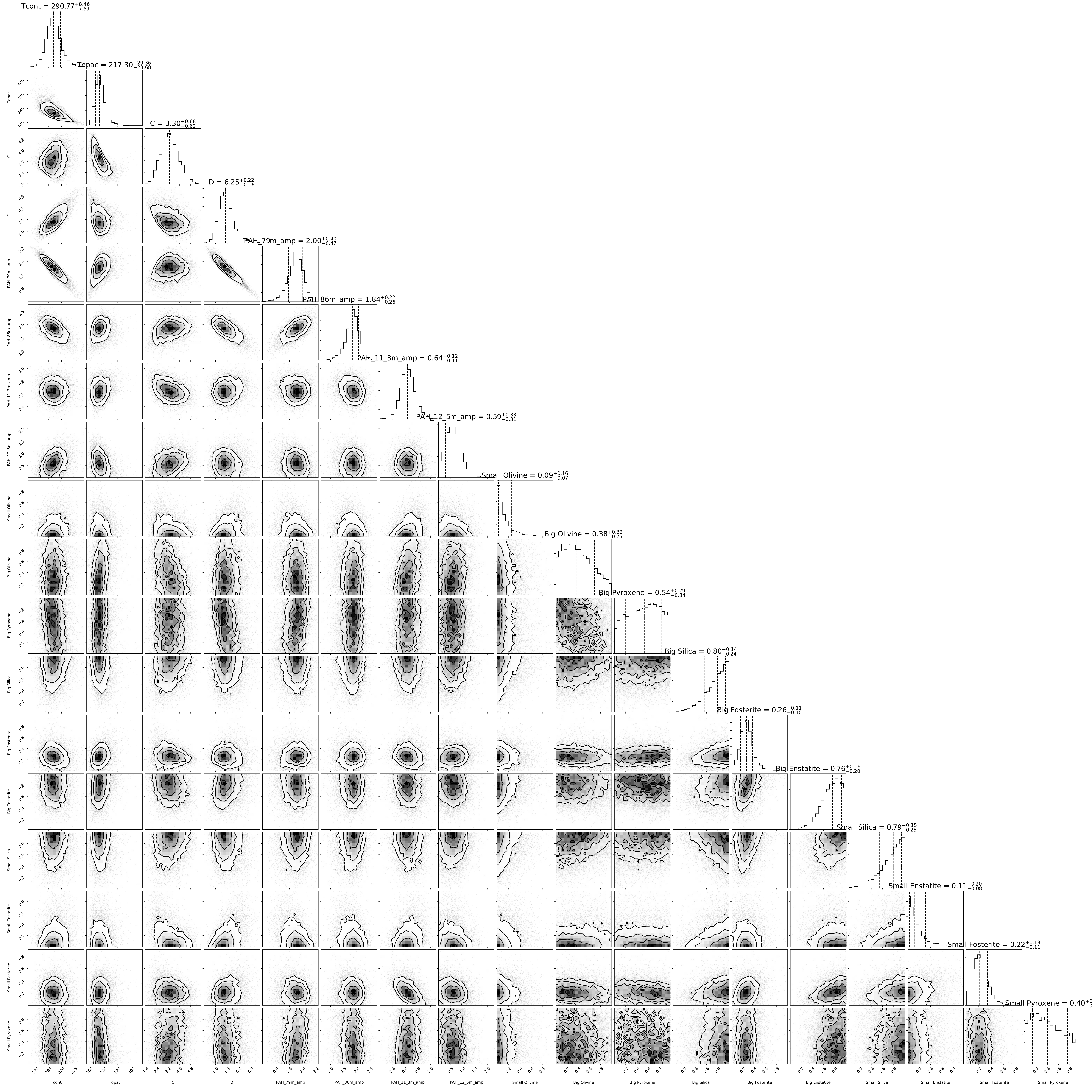}
\caption{\color{black} HD~95881: unresolved fitting posterior distributions. \color{black}}
\label{Fig_corner_HD95881}
\end{figure*}

\begin{figure*}
\centering
\includegraphics[scale = 0.18]{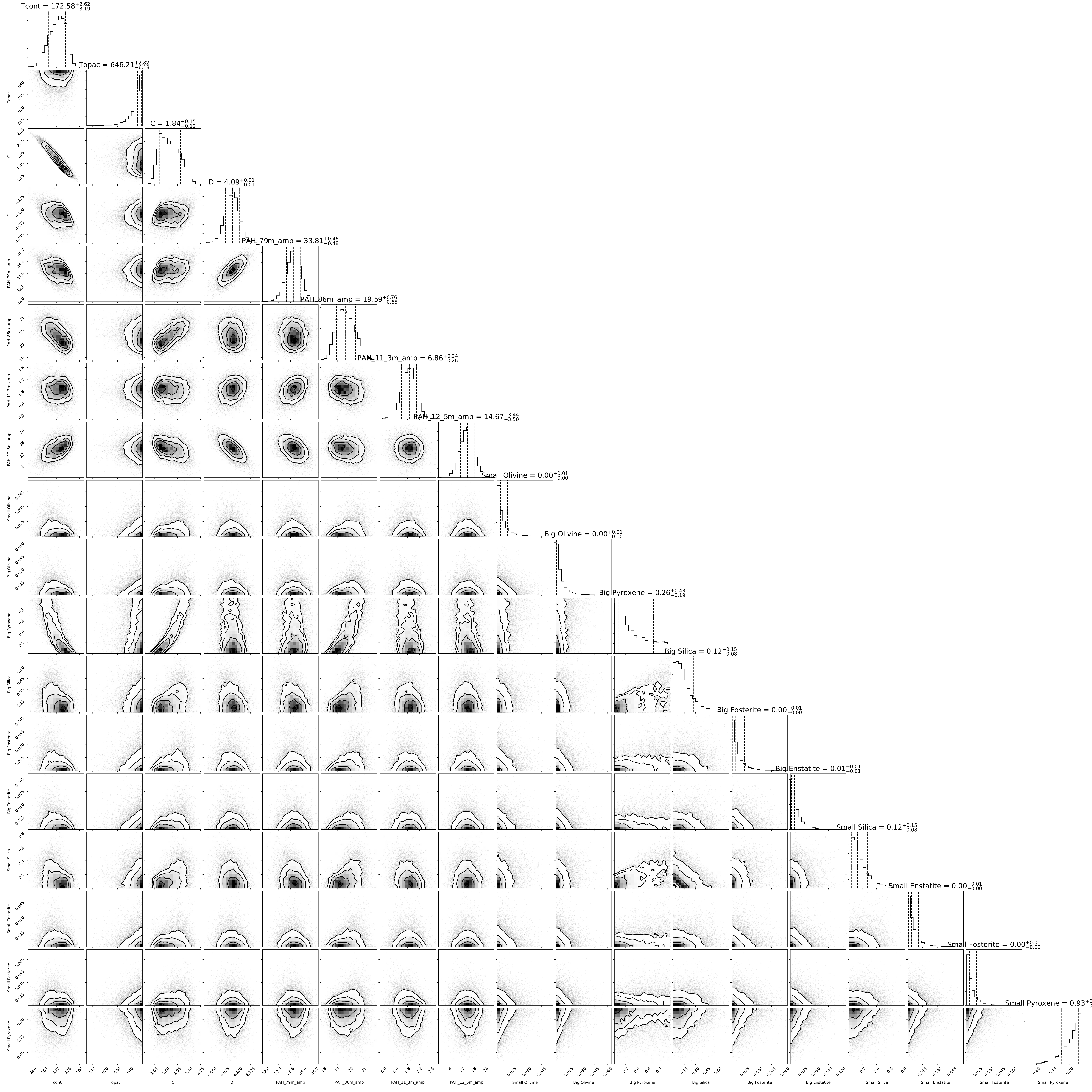}
\caption{\color{black} HD~97048: unresolved fitting posterior distributions. \color{black}}
\label{Fig_corner_HD97048}
\end{figure*}

\begin{figure*}
\centering
\includegraphics[scale = 0.18]{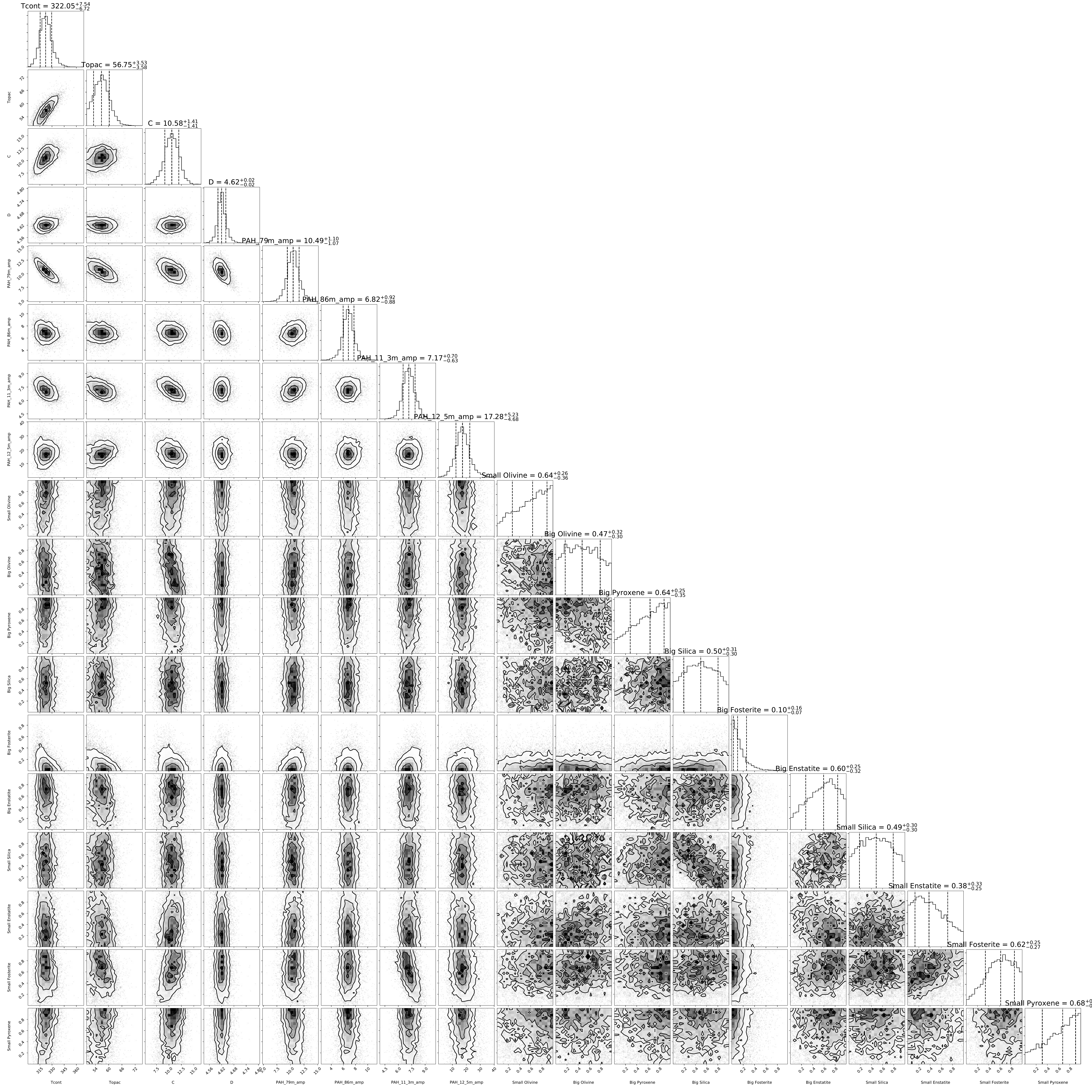}
\caption{\color{black} HD~100453: unresolved fitting posterior distributions. \color{black}}
\label{Fig_corner_HD100453}
\end{figure*}

\begin{figure*}
\centering
\includegraphics[scale = 0.18]{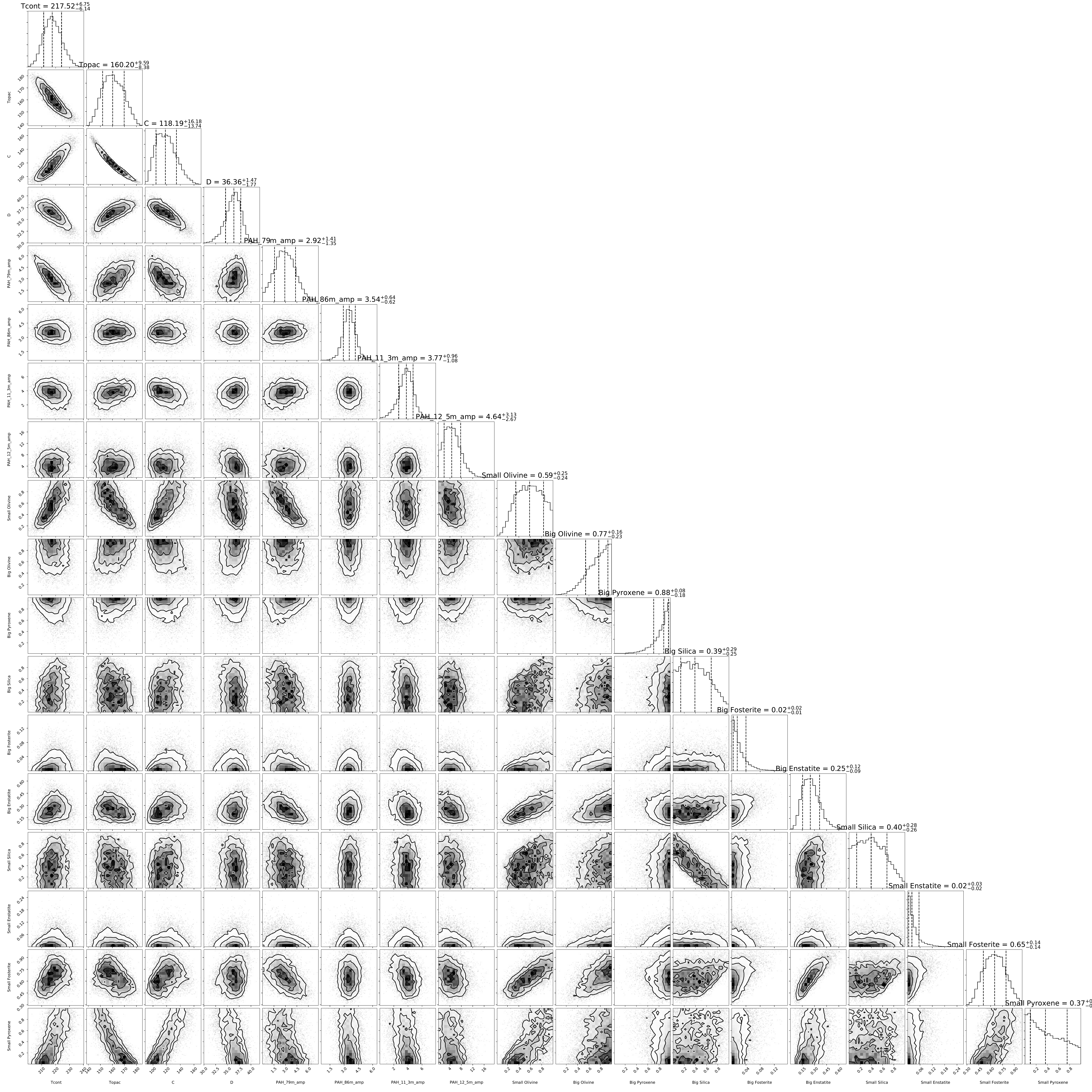}
\caption{\color{black} HD~100546: unresolved fitting posterior distributions. \color{black}}
\label{Fig_corner_HD100546}
\end{figure*}

\begin{figure*}
\centering
\includegraphics[scale = 0.18]{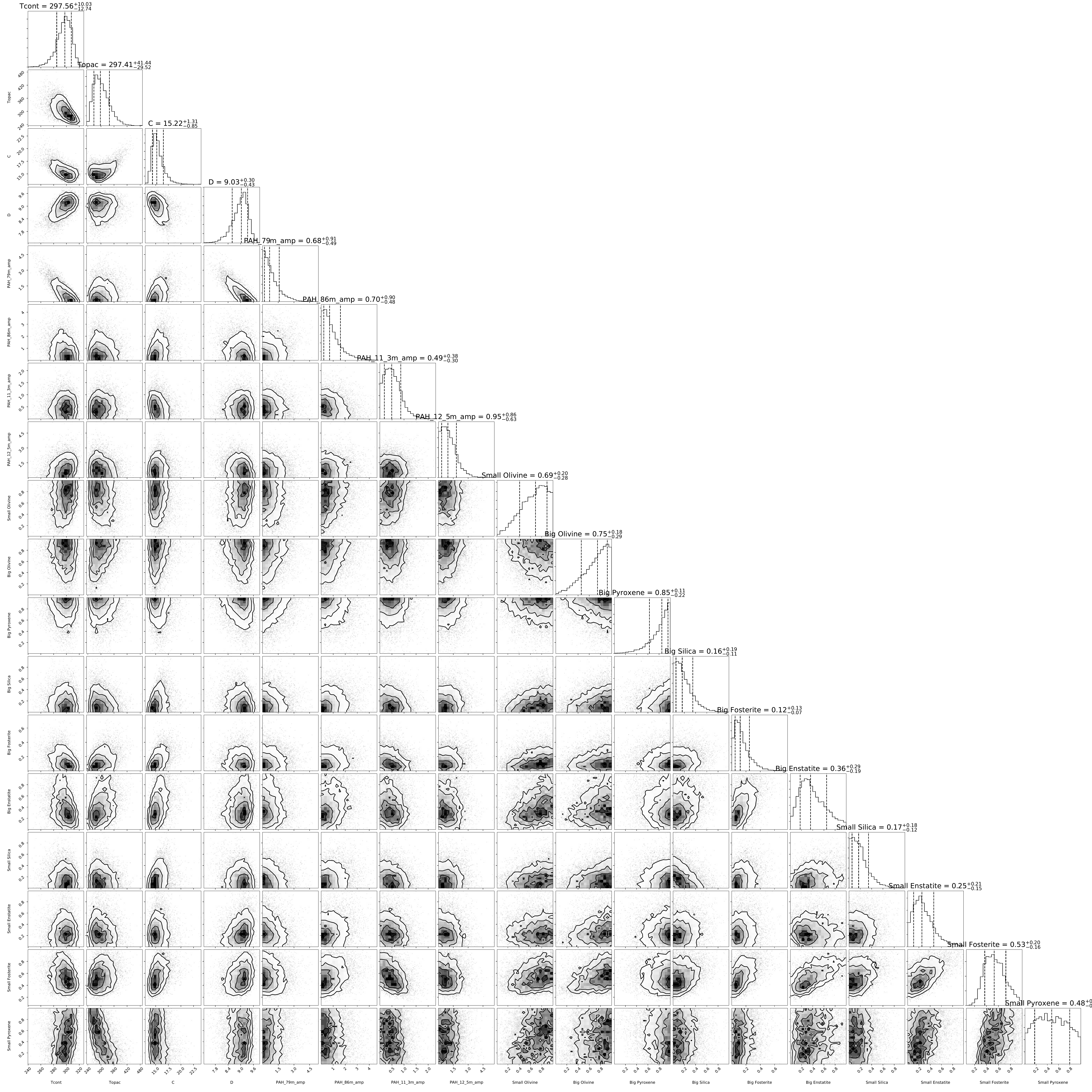}
\caption{\color{black} HD~163296: unresolved fitting posterior distributions. \color{black}}
\label{Fig_corner_HD163296}
\end{figure*}

\begin{figure*}
\centering
\includegraphics[scale = 0.18]{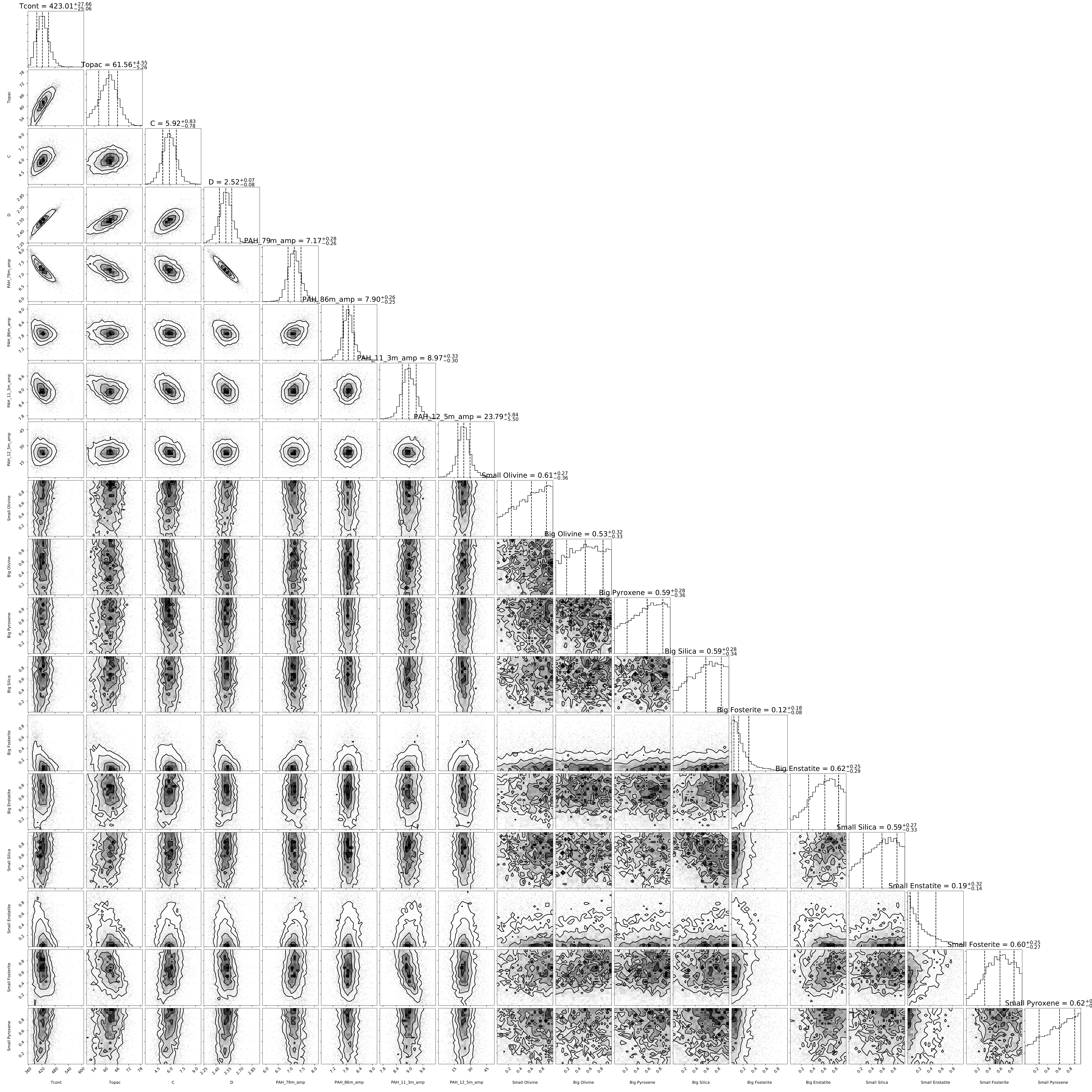}
\caption{\color{black} HD~169142: unresolved fitting posterior distributions. \color{black}}
\label{Fig_corner_HD169142}
\end{figure*}

In Figs. \ref{Fig_corner_ABAUR}-\ref{Fig_corner_HD169142}, we plot the posterior distribution of all model constituents (see Sect. \ref{Param_Model}) of the unresolved spectra fitting procedure (see Sect. \ref{Results_1D}) for all objects in our sample.


\begin{figure*}
\centering
\includegraphics[scale = 0.18]{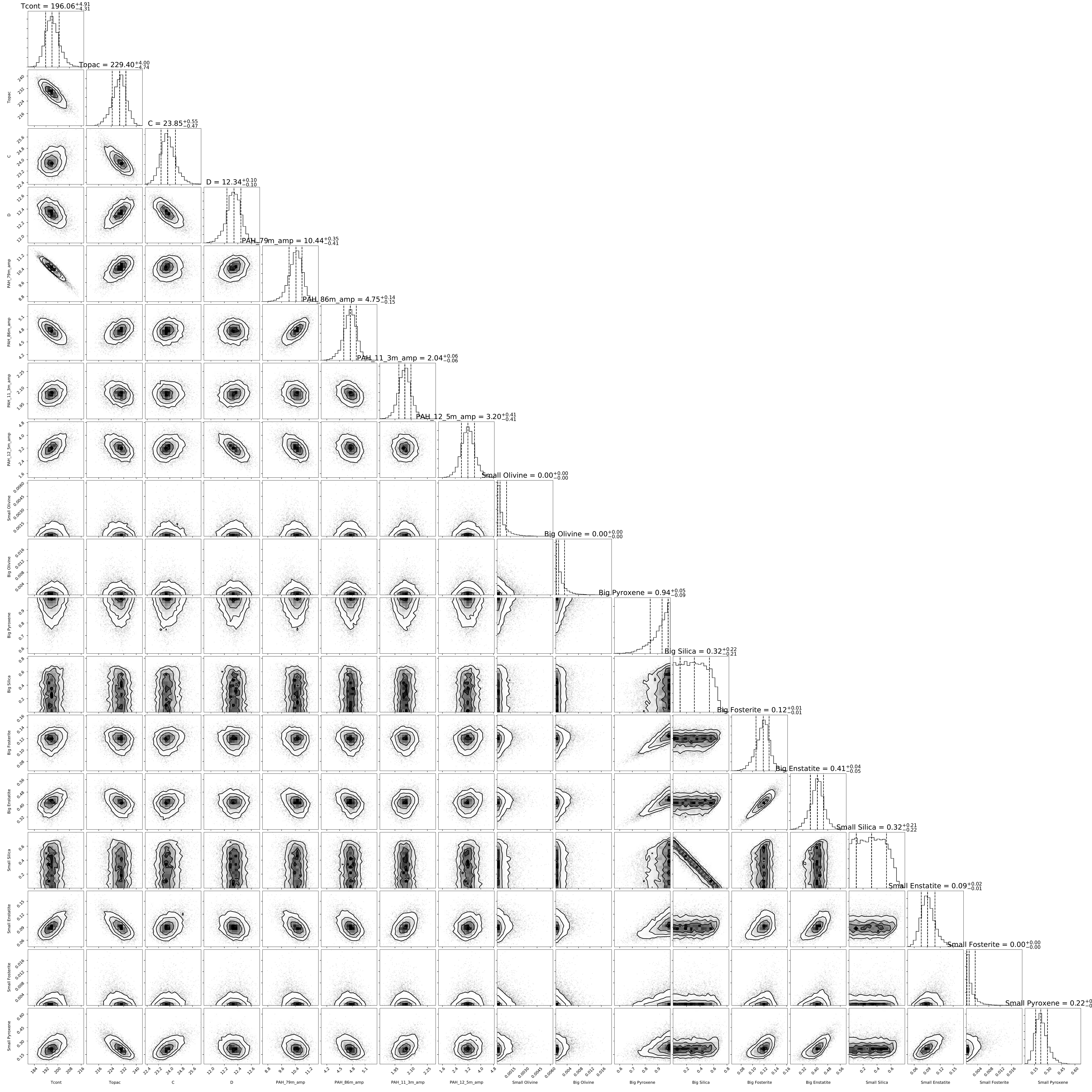}
\caption{\color{black} HD~179218: unresolved fitting posterior distributions. \color{black}}
\label{Fig_corner_HD179218}
\end{figure*}

\section{Temperature probability distribution and PAH model emission spectra} \label{APP_Li plots}

\begin{figure*}
\centering
\includegraphics[scale = 0.8]{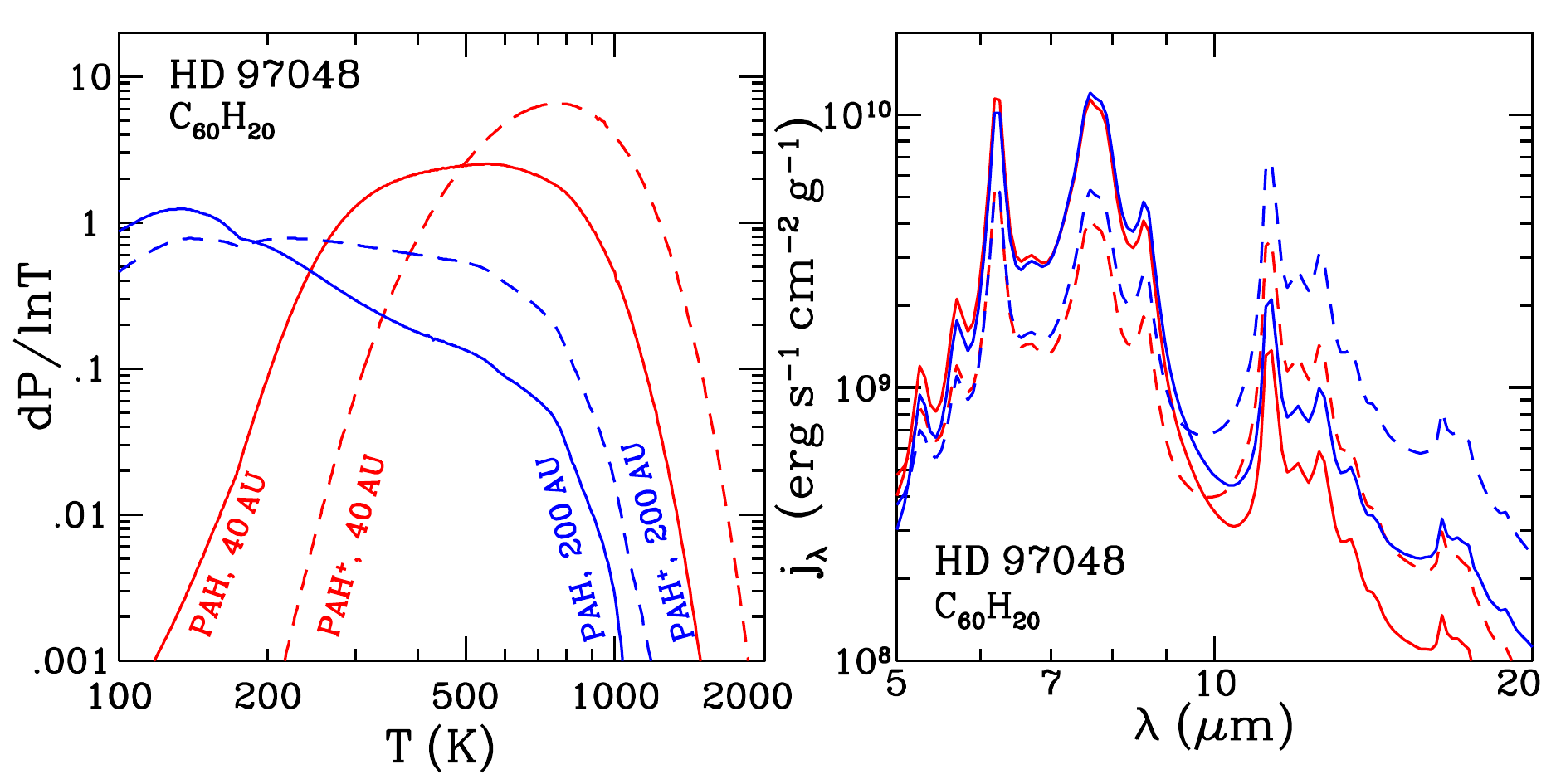}
\caption{HD~97048: temperature probability distributions (left panel) and emission spectra (right panel) of neutral C$_{60}$H$_{20}$ (similar to Fig. \ref{Fig_Li_PAH_ABAUR}).}
\label{Fig_Li_PAH_HD97048}
\end{figure*}

\begin{figure*}
\centering
\includegraphics[scale = 0.8]{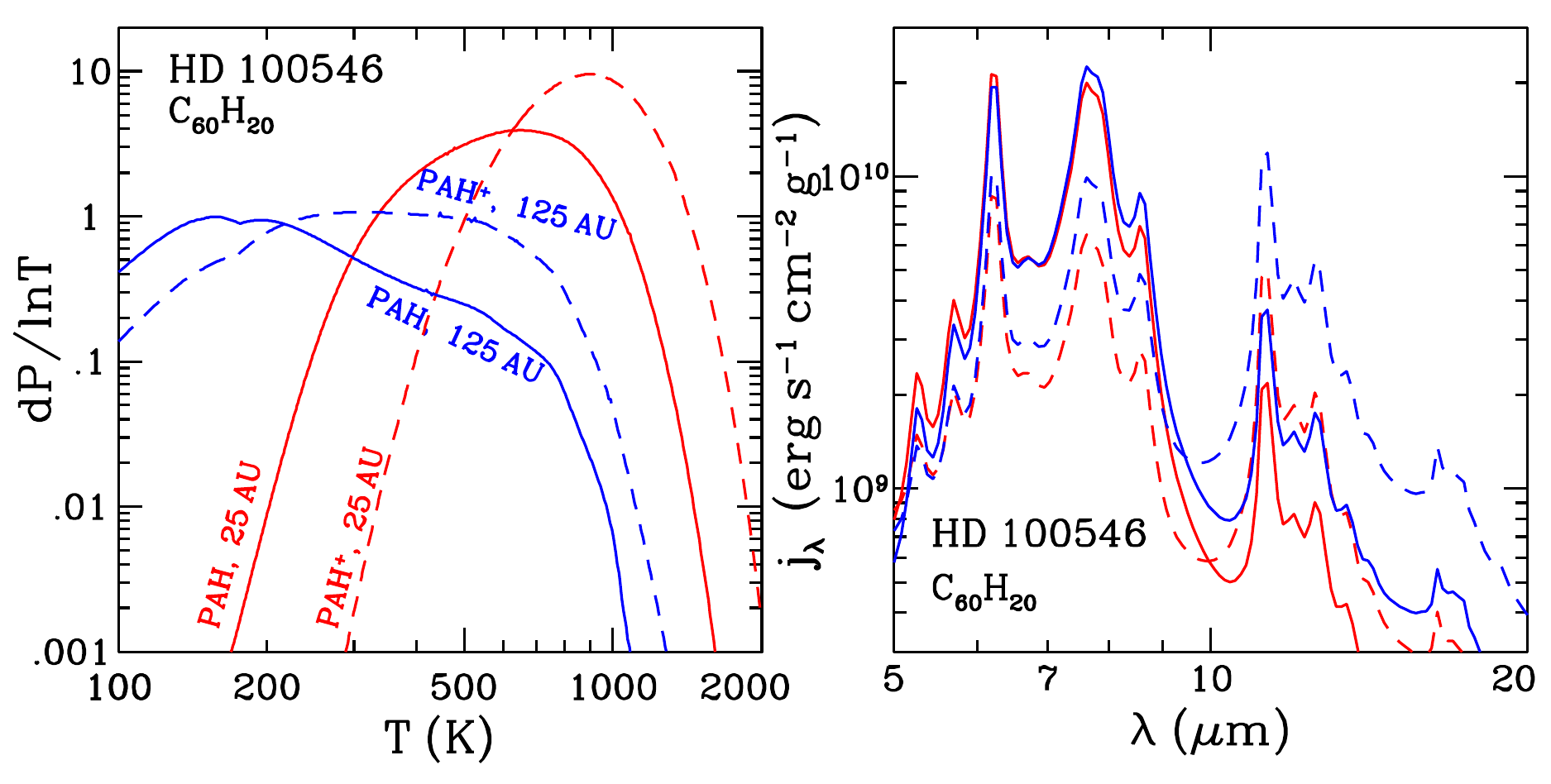}
\caption{HD~100546: temperature probability distributions (left panel) and emission spectra (right panel) of neutral C$_{60}$H$_{20}$ (similar to Fig. \ref{Fig_Li_PAH_ABAUR}).}
\label{Fig_Li_PAH_HD100546}
\end{figure*}

\begin{figure*}
\centering
\includegraphics[scale = 0.8]{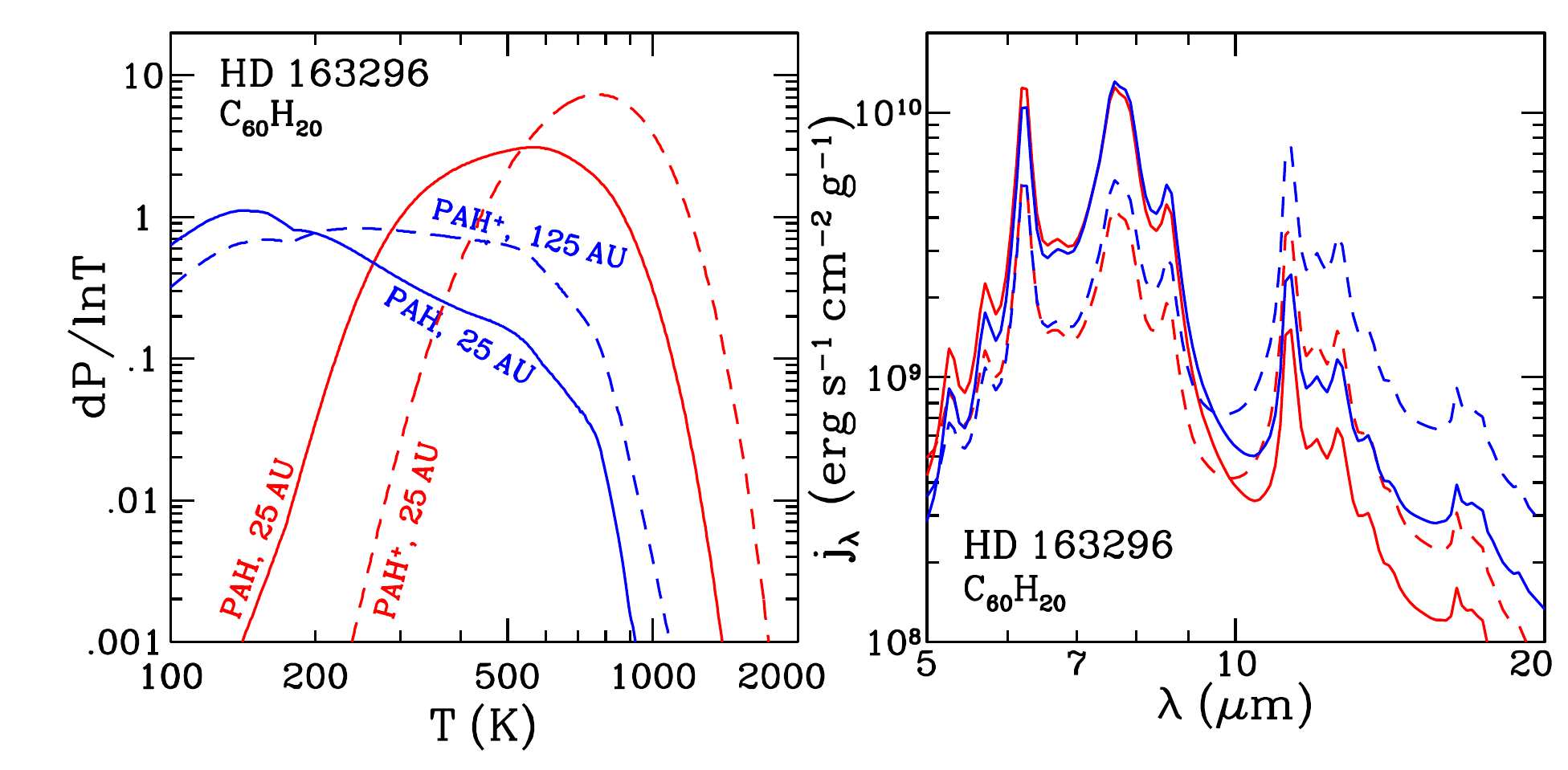}
\caption{HD~163296: temperature probability distributions (left panel) and emission spectra (right panel) of neutral C$_{60}$H$_{20}$ (similar to Fig. \ref{Fig_Li_PAH_ABAUR}).}
\label{Fig_Li_PAH_HD163296}
\end{figure*}

\begin{figure*}
\centering
\includegraphics[scale = 0.8]{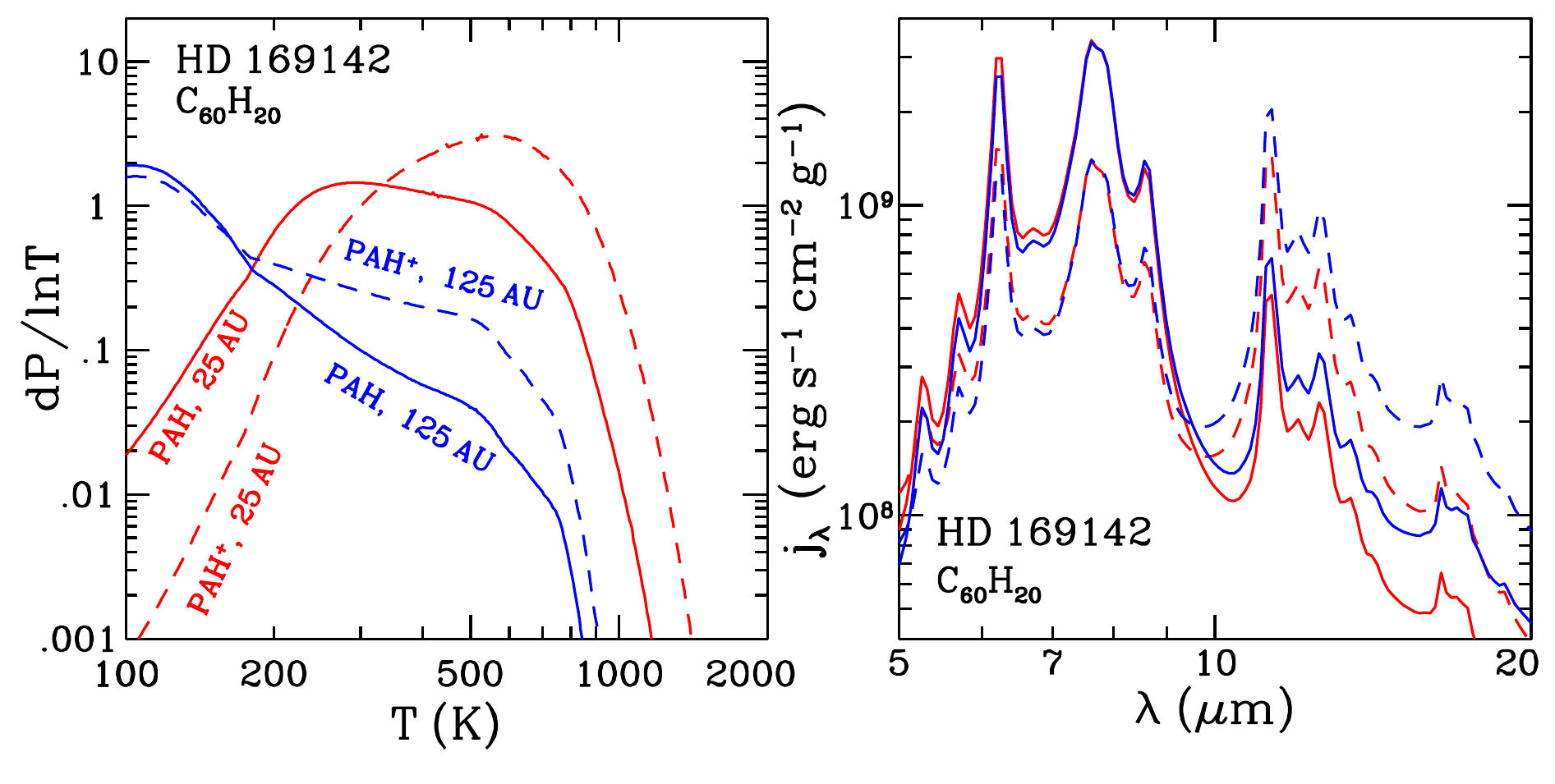}
\caption{HD~169142: temperature probability distributions (left panel) and emission spectra (right panel) of neutral C$_{60}$H$_{20}$ (similar to Fig. \ref{Fig_Li_PAH_ABAUR}).}
\label{Fig_Li_PAH_HD169142}
\end{figure*}

In Figs. \ref{Fig_Li_PAH_HD97048}-\ref{Fig_Li_PAH_HD169142}, we plot the temperature probability distribution and PAH model emission spectra (see Sect. \ref{Disc_SPHERE}) for HD~97048, HD~100546, HD~163296, and HD~169142, respectively.

\color{black}

\section{PSF model}
\label{sec:PSF_model}

The inner disk regions dominate the total flux of our targets but are either spatially unresolved or marginally resolved with VISIR-NEAR. The PSF redistributes the flux arising from these regions in the VISIR focal plane, leading to a small fraction of the light landing in the off-axis apertures. In order to estimate this fraction  we create a high-S/N PSF model as follows. 

For each of our calibrator observations, we created a 2D spectrum (dimensions: across-slit (dispersion) and along-slit (spatial) direction), which we normalized such that the flux in each spectral bin is unity. The spectral resolution of VISIR in the mode employed here is approximately $R=300$; hence, we measured a quasi-monochromatic PSF at each wavelength.

We over-sampled the 2D spectra of the calibrators by a factor of 2 in the spatial dimension, aligned them, and combined them into one high-S/N 2D spectrum. We then resample the spatial profiles in each spectral bin to a common grid in terms of $\lambda/D$ (where $D$ is the VLT primary mirror diameter of 8.2~m), over-sampling by a factor of 2.

\begin{figure}
   \begin{center}
    \includegraphics[width=0.50\textwidth, trim=0 0 0 0]{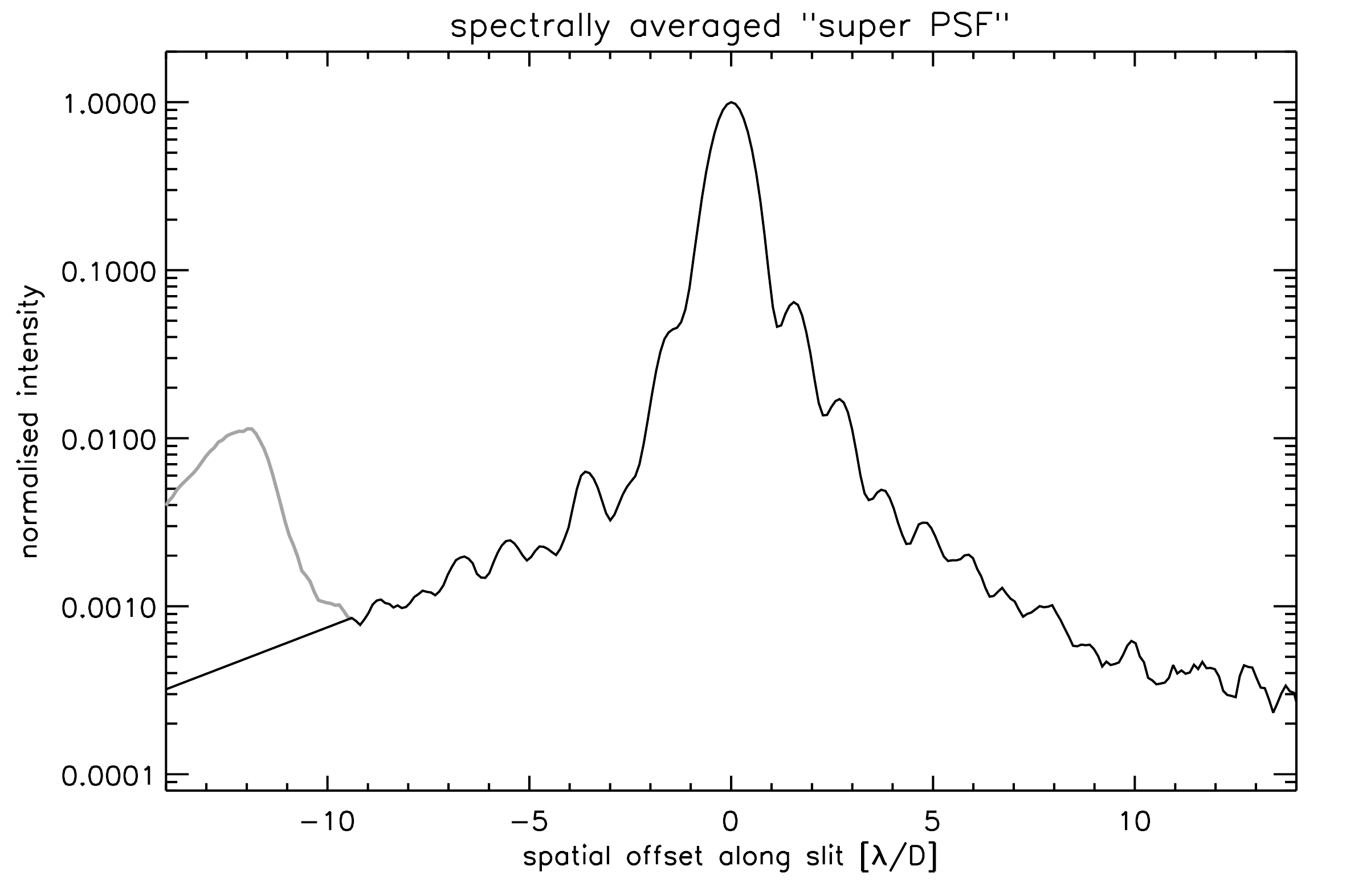}

    \caption{VISIR-NEAR quasi-monochromatic PSF as recorded on our calibrators.}
    \label{fig:super_PSF}
    \end{center}
\end{figure}

We spectrally averaged these quasi-monochromatic PSFs to produce a single, 1D "super PSF" along the spatial dimension, which is displayed in Fig.~\ref{fig:super_PSF}. More than ten diffraction rings are visible.
    
    There is an optical ghost approximately 4 arcsec below the target on the VISIR detector (toward the left of the profile in Fig.~\ref{fig:super_PSF}, as the gray part of the curve left of $\approx-10\lambda/D$). We replaced this part of our PSF profile with the continuation of a linear fit between $9\lambda/D$ and $4\lambda/D$. We note that the ghost is far enough away from the target to not interfere with any potential spatially extended emission that is the topic of this study.
Finally, we resampled the super PSF at each wavelength back onto the instrumental spatial grid of the VISIR detector.

\section{Potential PSF-widening due to atmospheric conditions} \label{APP_PSF_WIDENING}

\begin{figure*}
\centering
\includegraphics[scale = 0.6]{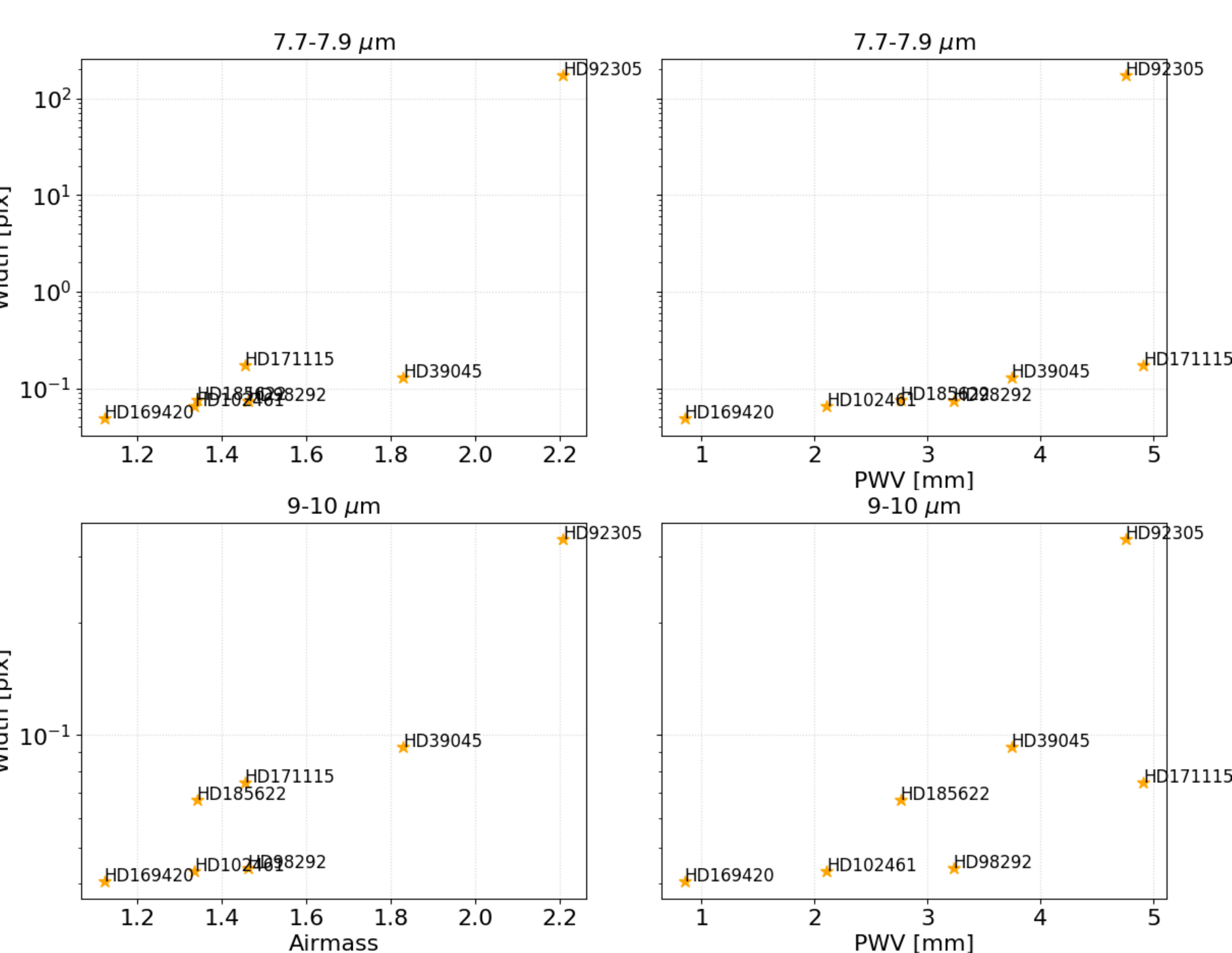}
  \caption{Calibrator PSF residual width (in pixels) vs. atmospheric conditions. The residual widths are calculated as follows: (\textbf{1}) each column (i.e., spectral coordinate; see Fig. \ref{Fig_2D_Spectrum}) of the detector is fitted with a Gaussian to estimate the width of the PSF. (\textbf{2}) The fitted widths are then fitted with a first-order polynomial (ideally, $\lambda/D$). (\textbf{3}) The fitted polynomial is then subtracted from the fitted widths to generate a residual widths vector -- which represents the nonlinear deviations of the fitted widths. \textbf{Left Panel}: PSF residual width vs. airmass during observation for the water vapor and ozone bands. \textbf{Right Panel}: PSF residual width vs. precipitable water vapor (PWV) during observation for the water vapor and ozone bands.}
     \label{Fig_Calib_width_vs_Atmos}
\end{figure*}

\begin{figure*}
\centering
\includegraphics[scale = 0.6]{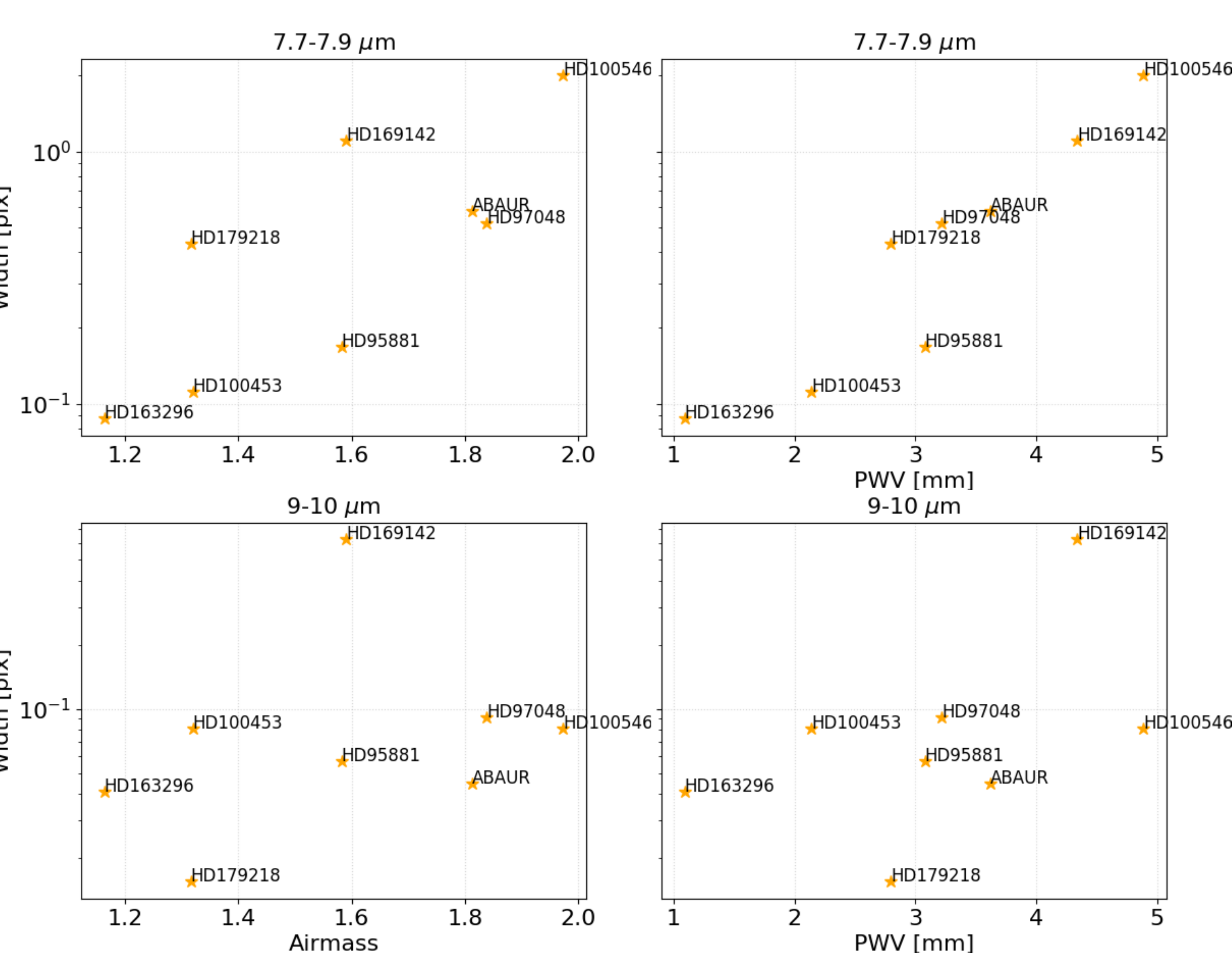}
  \caption{Science target PSF residual width (in pixels) vs. atmospheric conditions (similar to Fig. \ref{Fig_Calib_width_vs_Atmos}).}
     \label{Fig_Sci_width_vs_Atmos}
\end{figure*}

To affirm the assumption that our detected spatial distributions of the PAHs represent the intrinsic structure of the observed objects, rather than a misinterpretation of a PSF widened due to sub-optimal atmospheric conditions, we conducted a series of tests to detect potential nonlinear widening of the science targets' PSF (i.e., not scaled as $\approx\lambda/D$) at particular spectral bins where such widening might be expected, for example, the 9-10 $\mu$m ozone band and the 7.7-7.9~$\mu$m water vapor band. In the former, no PAH features are present, so a widening of the PSF in that band is not necessarily indicative of the intrinsic distribution of PAHs within the object, whereas in the latter the opposite is true. Thus, if widening as a function of the PSF is detected in both bands, one would have reason to suspect the derived PAH spatial distributions. However, the {absence} of such widening in the ozone band would serve to show reinforce our conviction in the data. Additionally, we perform these tests on the calibrators as well, which should ideally exhibit no widening of the PSF at all, if it indeed reflects the intrinsic structure of the observed objects. We plot our results for the calibrators in Fig. \ref{Fig_Calib_width_vs_Atmos} and for the science targets in Fig. \ref{Fig_Sci_width_vs_Atmos}.

Fig. \ref{Fig_Calib_width_vs_Atmos} indicates that with the exception of a single faulty observation of HD\,92305 (which was not used as a calibrator in the processing stage), the PSF residual widths of calibrators do not seem to exhibit a strong correlation with the airmass nor the PWV in either band. The same can be said about the residual widths of the science targets' PSF in the ozone band (HD~169142 is spatially resolved in the continuum and therefore deviates). By observing the water-vapor bands of the ozone band, one might assume, however, a strong correlation and be reluctant to attribute those widths to intrinsic energy distributions within the objects themselves. However, the first three cases serve to show that there seems to be {no} correlation between the given atmospheric conditions under scrutiny and the PSF residual widths, which serves to reinforce the assumption that the observed intensity distributions in the water-vapor (or 7.9~$\mu$m PAH) band of the science targets is not an atmospheric effect, but rather an intrinsic property of their structure.

\end{document}